\documentclass{aa}  
\usepackage{graphicx}
\usepackage{txfonts}
\usepackage{hyperref}
\hypersetup{
    colorlinks=true,
    linkcolor=blue,
    filecolor=magenta,  
    citecolor=cyan,
    urlcolor=cyan,
}
\usepackage{natbib}

\begin{document} 

   \title{Characterization of Low Surface Brightness structures in annotated deep images}
  
   \author{Elisabeth Sola
          \inst{1}
          \and
          Pierre-Alain Duc\inst{1}
          \and
          Felix Richards\inst{2}
          \and
          Adeline Paiement \inst{3}
          \and 
          Mathias Urbano \inst{1}
          \and
          Julie Klehammer \inst{1}
          \and
          Michal B\'ilek \inst{4,5,6} 
          \and
          Jean-Charles Cuillandre \inst{7}
          \and 
          Stephen Gwyn \inst{8}
          \and
          Alan McConnachie \inst{9}
          }

   \institute{Université de Strasbourg, CNRS, Observatoire astronomique de Strasbourg (ObAS), UMR 7550, F-67000 Strasbourg, France\\
              \email{elisabeth.sola@astro.unistra.fr} 
            \and
                Department of Computer Science, Swansea University, UK
                \and 
                Universit\'e de Toulon, Aix Marseille Univ, CNRS, LIS, Marseille, France
                \and 
                European Southern Observatory, Karl-Schwarzschild-Stra{\ss}e 2, 85748 Garching bei M{\"u}chen, Germany
                \and 
                LERMA, Observatoire de Paris, CNRS, PSL Univ., Sorbonne Univ., 75014 Paris, France
                \and 
                Coll\`ege de France, 11 place Marcelin Berthelot, 75005 Paris, France
                \and
                AIM, CEA, CNRS, Universit\'e Paris-Saclay, Universit\'e de Paris, F-91191 Gif-sur-Yvette, France
                \and
                National Research Council, Canadian Astronomy Data Centre, Victoria, BC, V9E 2E7, Canada
                \and NRC Herzberg Astronomy and Astrophysics, 5071 West Saanich Road, Victoria, British Columbia, Canada, V9E2E7
             }

   \date{Received : November 16, 2021; accepted : March 04, 2022}

  \abstract  
   {The identification and characterization of Low Surface Brightness (LSB) stellar structures around galaxies such as tidal debris of on-going or past collisions is essential to constrain models of galactic evolution. So far most efforts have focused on the numerical census of samples of varying sizes, either through visual inspection or more recently with  deep learning. Detailed analysis including photometry has been carried out for a small number of objects, essentially because of the lack of convenient tools able to precisely characterize tidal structures around large samples of galaxies.}
   {Our goal is to characterize in detail, and in particular  obtain quantitative measurements, of LSB structures identified in deep images of samples consisting of hundreds of galaxies.}
   {We developed an online annotation tool that enables contributors to delineate with precision the shapes of diffuse extended stellar structures, as well as artefacts or foreground structures. All parameters are automatically stored in a database which may be queried to retrieve quantitative measurements.
   We annotated LSB structures around 352 nearby massive galaxies with deep images obtained with the Canada-France-Hawaii Telescope as part of two large programs: MATLAS and UNIONS/CFIS. Each LSB structure was delineated and labeled according to its likely nature: stellar shells, streams associated to a disrupted satellite, tails formed in major mergers, ghost reflections or cirrus.}
   {From our database containing 8441 annotations, the area, size, median surface brightness and distance to the host of 228 structures were computed. The results confirm the fact that tidal structures defined as streams are thinner than tails, as expected by numerical simulations. In addition, tidal tails appear to exhibit a higher surface brightness than streams (by about 1 mag), which may be related to different survival times for the two types of collisional debris. We did not detect any tidal feature fainter than 27.5 mag\,arcsec$^{-2}$, while the nominal surface brightness limits of our surveys range between 28.3 and 29 mag\,arcsec$^{-2}$, a difference that needs to be taken into account when estimating the sensitivity of future surveys to identify LSB structures. 
   }
    {We compiled an annotation database of observed LSB structures around nearby  massive galaxies including tidal features that may be used for quantitative analysis and as a training set for machine learning algorithms.
    }
    

   \maketitle

\section{Introduction}
\subsection{Detection of tidal features}
According to hierarchical models of galactic evolution, galaxies assemble through successive mergers, accretion of smaller systems and smooth accretion of gas \citep[e.g.,][]{Kauffman_et_al_1993, Cole_et_al_2000, Baugh_et_al_2003, Stringer_and_Benson_2007}. These interactions between galaxies leave collisional debris, in particular tidal features such as tidal tails, streams or shells, that have different morphologies and survival lifetime \citep[e.g.,][]{Arp_1966, Toomre_and_Toomre_1972, Quinn_1984, Mancillas_et_al_2019}. Since the different types of features are produced by different types of collisions, their identification and classification give valuable information about the mass assembly history of the host galaxy and more in general about models of galactic evolution. However, the detection of collisional debris is complicated by their Low Surface Brightness (LSB), challenging detection by traditional methods.

Detecting LSB tidal features in the Local Group is possible by stellar count even using ground-based missions, such as SDSS with which \cite{Belokurov_et_al_2006} revealed streams and stellar structures in the Sagittarius dwarf spheroidal. The resolved stellar populations photometric survey PAndAS \citep{Martin_et_al_2014} also studied this "Field of streams", as well as the surrounding of M31 where signs of galactic interactions were studied \citep{McConnachie_et_al_2009}. Substructures in M31 halo were detected in other surveys, such as \cite{Ibata_et_al_2001} or \cite{Ferguson_et_al_2002} with the Isaac Newton Telescope or \cite{Gilbert_et_al_2012} with the SPLASH survey. A wealth of new structures have been disclosed in the Milky Way by the Gaia mission \citep[e.g.,][]{Malhan_et_al_2018, Antoja_et_al_2020}.

However, for systems with increasing distance from the Local Group, stars are less and less resolved individually so that the detection of tidal features relies mainly on the study of the combined diffuse light. Overcoming the observational challenge of detecting faint and extended structures has been made possible by the development of dedicated instruments and/or data reduction pipelines able to produce deep images (i.e. images with a depth sufficient to detect faint structures) with limited undesirable artefacts. To that extent, both professional and small-sized telescopes enabled the discovery of a wealth of LSB structures around nearby galaxies. For instance, \cite{Martinez_Delgado_et_al_2010} and \cite{Javanmardi_et_al_2016} have respectively discovered streams and LSB systems around spiral galaxies with several small-size telescopes, while the Dragonfly Telephoto Array \citep{van_Dokkum_et_al_2014} produced very deep images of nearby galaxies. The Huntsman telescope \citep{Spitler_et_al_2019} is based on the same principle as Dragonfly. In addition, LSB studies were conducted using professional telescopes such as the Burrell Schmidt telescope \citep[e.g.,][]{Mihos_et_al_2015}, the CFHT \citep[e.g.,][]{Ferrarese_et_al_2012,Duc_2020}, the VLT Survey Telescope (VST) \citep[e.g.,][]{Venhola_et_al_2017,Iodice_et_al_2021}, or the Subaru telescope \citep[e.g.,][]{Alabi_et_al_2021,Jackson_et_al_2021}.

\subsection{Identification of tidal features}
Once the deep images are acquired, the LSB structures need to be identified, a task  traditionally done with a visual inspection by one or a handful of contributors. This inspection can  be performed on the images resulting from the basic data reduction,  possibly  adjusting  the scaling and dynamics of the image intensity to enhance the LSB structures,  or on images with an advanced processing, such as residual images obtained by subtracting the  light profile model of the target  (using softwares like GALFIT \citep{Peng_et_al_2002} for example). This technique discloses the inner most tidal debris  as asymmetries that are not well fitted by symmetric models of galactic light \citep[e.g.][]{Bell_et_al_2006,McIntosh_et_al_2008,Tal_et_al_2009}, but generates multiple artefacts. 

A number of surveys of tidal features made by a direct visual inspection of the images by a few expert contributors have been carried out.   For instance,   \cite{Jackson_et_al_2021} have inspected Subaru Hyper Suprime-Cam  images of a sample of 118 low redshift, massive central galaxies;
\cite{Kluge_et_al_2020},  Wendelstein Telescope Wide Field Images of the  170 local brightest cluster galaxies;  \cite{Bilek_et_al_2020},  CFHT MATLAS images \citep{Duc_2020} of 177 massive nearby early-type galaxies; \cite{Morales_et_al_2018} SDSS images of  297 Local Volume galaxies;  
\cite{Atkinson_Abraham_Ferguson_2013}, CFHTLS-Wide images of luminous 1781 galaxies  at a redshift of $0.04<z<0.2$;
\cite{Bridge_et_al_2010} the 2 square degrees CFHTLS-Deep survey images of 27,000 galaxies.

As the sample size of galaxies increases, the classification by a limited team of experts becomes progressively infeasible. Galaxy features identification   may then be done by a  crowd of citizen scientists.  Their potential  lack of expertise is compensated by the higher number of individual annotations per galaxy. The reliability of the classification increases with  the number of participants. This is exploited by the Galaxy Zoo project. For instance, \cite{Casteels_et_al_2012} used the Galaxy Zoo 2 classifications made by 80,000 volunteers of Sloan Digital Sky Survey images to study tidal features of about 150,000 galaxies of similar mass.
Similarly, the Galaxy Cruise\footnote{Galaxy Cruise, \url{https://galaxycruise.mtk.nao.ac.jp/en/index.html}} project aims at classifying the presence of shells, rings, streams, and distorted halos from Subaru-HSC images. 
However, a method relying on the participation   of volunteers  with limited science knowledge can only be reliable for simple  tasks.   
Precise classifications require prior knowledge of the field. It is for example the case when trying to disentangle the tidal tails associated with major mergers (i.e. mergers between two similar-mass galaxies) and the streams which trace minor mergers (i.e. mergers with a lower-mass galaxy):  just focusing on their shape may not be enough as both collisional debris exhibit only subtle differences, as discussed in this paper.  A prior knowledge on galaxy evolution and mergers is required to separate them. Hence, most studies on mergers relying on  citizen-science efforts have focused on the presence of tidal distortions or collisional debris,  without making any attempt to distinguish sub-classes of objects.

As an alternative approach for large samples, fully automated methods have  been developed. 
The level  of the morphological asymmetry of galaxies can reveal tidal disturbances associated to on-going or past mergers \citep{Pawlik_et_al_2016}. More in general  non-parametric methods such as the Gini-$M_{20}$ parameter  \citep[e.g.,][]{Abraham_et_al_2003,Lotz_2004} and/or the CAS system \citep[e.g.,][]{Abraham_et_al_1994,Abraham_et_al_1996,Conselice_et_al_2003,Conselice_et_al_2008,Conselice_2009}, that do not assume a particular function for the galactic light distribution, have been often used. 
However, such parameters are not very sensitive to LSB structures, as they are flux-weighted and dominated by the contribution of the most luminous parts of a galaxy. On the contrary, \cite{Wen_et_al_2014} developed the $A_O-D_O$ method, which is efficient to select asymmetric galaxies with faint features like tidal tails.
In addition, \cite{Mantha_et_al_2019} proposed a new tool to extract and quantify galactic morphological substructures from residual images, including plausible tidal features, along with a measure of their surface brightness.
Automated tidal feature identification can also be performed using algorithms to separate the high and low spatial frequencies in the image \citep{Kado-Fong_et_al_2018}, respectively corresponding to tidal features and galaxy light, allowing a quantitative analysis of their properties.
However, like for the citizen scientist approach, these methods to not allow subtle classification and disentangling between various types of tidal features.

Another promising approach towards identifying faint tidal features on large number of objects is machine learning, as it offers the possibility to work with large samples of galaxies. In particular, convolutional neural networks (CNNs) have been used to classify the morphologies of galaxies \citep[e.g.,][]{Huertas-Company_et_al_2015,Dieleman_et_al_2015,Ferrero_et_al_2020,Dominguez-Sanchez_et_al_2018,Tohill_et_al_2021}.
CNNs are also able to capture disturbed galactic morphologies that can be the hint of mergers and interactions \citep[e.g.][]{Pearson_et_al_2019,Ferreira_et_al_2020}. They can be used to identify LSB tidal features in observational images: \cite{Walmsley_et_al_2018} and \cite{Bickley_et_al_2021} used CNNs to identify tidally-disrupted galaxies and classify tidal features. They obtained high accuracy and low contamination, and in overall performed better than other automated techniques. In addition, \cite{Pearson_et_al_2019} were able to classify merger features in SDSS observations from a CNN trained on snapshots from the EAGLE simulation. However, currently the deep learning approach is unable to precisely classify the different types of tidal features, unless it is trained with large sample of images that have been previously precisely annotated. Unsupervised techniques could offer a solution to this problem \citep[e.g.,][]{Martin_et_al_2020,Uzeirbegovic_et_al_2020,Spindler_et_al_2021,Cheng_et_al_2021}, although there may be less control over the output.

The efforts to  classify the morphology of galaxies and detecting surrounding tidal perturbations  is not restricted to images obtained with telescopes. More and more detailed, realistic, images are produced by numerical simulations. They have the main advantage of providing the ground truth when interpreting the results. Indeed, it is possible to track the merger trees of LSB hosting galaxies and their 3D information to follow the LSB structures wrapping around galaxies \citep[e.g.,][]{Hendel_and_Johnston_2015, Pop_et_al_2018, Mancillas_et_al_2019,  Ebrova_et_al_2021, Bilek_et_al_2021}.

Several types of simulations aim at predicting the formation history of galaxies, including interactions and mergers along with their tidal debris \citep[e.g.,][]{Helmi_and_White_1999,Cooper_et_al_2010,Bullock_and_Johnston_2005,Johnston_et_al_2008, Pillepich_et_al_2018, Schaye_et_al_2015, Crain_et_al_2015}.
In order to be comparable with those observed in the real Universe, simulated galaxies must not be idealized \citep{Bottrell_et_al_2019}, and should include realistic sky, resolution, contamination sources and surface brightness limit. Identification of the tidal features have been carried out on these realistic mock images: for instance, \cite{Mancillas_et_al_2019} performed a census of tidal features around galaxies from a hydrodynamical simulation, while \cite{Martin_et_al_2022} investigated their nature, frequency and visibility around galaxies from the NewHorizon cosmological simulation.

\subsection{Annotation of tidal features}
One should note that most of the techniques mentioned above, whether applied to observational images or simulated ones,  focus on qualitative aspects such as the probability of presence of one or several tidal features. Quantitative morphological and photometric measurements of tidal debris have so far been made for very limited numbers of galaxies \citep[e.g.,][]{SSLS} or restricted to on-going tidally interacting systems. Systematic measurements for larger samples are needed to fully characterize the various types of LSB structures in order to make quantitative comparisons with numerical simulation-based models of galaxy evolution. Such quantitative properties will offer important calibrations towards understanding the physical nature of mergers causing the observed LSB structures and more in general to reconstruct the late assembly histories of individual galaxies. To that end, dedicated tools providing detailed annotations are needed.

In this paper, to systematically characterize tidal structures in deep observations, we have developed a tool that allows users to annotate large samples of galaxies via an accessible and intuitive on-line interface. The shapes of tidal features  may be drawn   with precision directly on the displayed images. Although other tools such as the Zooniverse platform also enable citizen scientists to delineate morphological features, such as spiral arms or bars \citep{Masters_et_al_2021}, our interface provides new functionalities. It offers a larger variety in the annotation shapes used to delineate features, and the flexibility to switch between bands, which are facilities that are well suited for LSB structures annotations.
 
Given the complexity of the precise annotation task and focus, our tool is more adapted to expert users. The delineated tidal features are stored in a database, from which we can then determine their distribution of shapes, sizes and surface brightness.
Delineation is the first step toward exploiting the full 2D profiles of the individual features, which gives additional constraints on the merger that created them.

The paper is organized as follows. In section \ref{section:deep-images} of this paper, we present the data we used for the deep images. We introduce the annotation server, its features and the annotation process in section \ref{section:annotation-tool}. In section \ref{section:analysis-tools}, we detail the analysis tools that were used to retrieve quantitative measurements. Then, in section \ref{section:results} we present the results obtained from the annotations, and we discuss them in section \ref{section:discussion}. Finally, we outline the conclusions in section \ref{section:conclusion}.

\section{Deep images} \label{section:deep-images}
The images and surveys we used in this paper are briefly described here.
We used data from the 3.6-meter Canada-France-Hawaii Telescope (CFHT), with the wide-field optical imager, MegaCam. In particular, we utilised images from two CFHT Large Programs: the Canada-France Imaging Survey (CFIS\footnote{CFIS, \url{https://www.cfht.hawaii.edu/Science/CFIS/}}) and the Mass Assembly of early-Type GaLAxies with their fine Structures survey (MATLAS\footnote{MATLAS, \url{http://obas-matlas.u-strasbg.fr}}).
MegaCam offers a wide field of view of 1°$\times$1° with a resolution of 0.18 arcsecond per pixel. Images were processed by the Elixir-LSB dedicated pipeline optimized for the detection of LSB structures \citep[Cuillandre, private communication, ][]{Duc_et_al_2015}. The limiting surface brightness reaches 28.3 mag\,arcsec$^{-2}$ for CFIS (Cuillandre, private communication) and 28.9 mag\,arcsec$^{-2}$ for MATLAS in the $r$-band.

While MATLAS targets nearby massive galaxies, especially of early-type  
 \citep{Duc_et_al_2015, Duc_2020, Bilek_et_al_2020}, CFIS is a blind survey that will cover 5,000 square degrees in the Northern hemisphere in the $u$ and $r$ band \citep{Ibata_et_al_2017}, with additional bands available from observations made with other telescopes as part of the Ultraviolet Near Infrared Optical Northern Survey (UNIONS) project.
 \footnote{The UNIONS project is a collaboration of wide field imaging surveys of the northern hemisphere. UNIONS consists of the Canada-France Imaging Survey (CFIS), conducted at the 3.6-meter CFHT on Maunakea, members of the Pan-STARRS team, and the Wide Imaging with Subaru HyperSuprime-Cam of the Euclid Sky (WISHES) team. CFHT/CFIS is obtaining deep u and r bands; Pan-STARRS is obtaining deep i and moderate-deep z band imaging, and Subaru/WISHES is obtaining deep z band imaging. 
 }

 The galaxies annotated for the work presented in this paper are located within the 3,600 square degrees that were covered by CFIS in March 2021. Annotation was done for the $r$-band images from MATLAS and CFIS.

In these CFHT deep images, we selected massive and nearby galaxies belonging to the reference ATLAS$^{3D}$ (main and parent) samples \citep{Atlas3D}. They include objects with distances smaller than 42 Mpc, an absolute $K$-band magnitude brighter than -21.5 mag and a stellar mass higher than $6\times 10^9 M_\odot$. In order to study LSB features as a function of the morphology of the host galaxies, we selected 2 sub-samples of comparable sizes consisting of 186 Early-Type Galaxies (ETGs) and 166 Late-Type Galaxies (LTGs) hence a total of 352 galaxies. 
If a galaxy was present in both surveys, we used the MATLAS image, because this survey is deeper. By doing so, we are biased toward finding faint features in ETGs, because they were the primary targets of MATLAS. Table \ref{table:nb-gal-annotated} summarizes the number of galaxies per survey.
\begin{table}
\caption{Number of galaxies surveyed in this work, detailed by their survey and morphological type}             
\label{table:nb-gal-annotated}     
\centering                         
\begin{tabular}{c c}        
\hline\hline                
Type & Number of galaxies studied  \\   
\hline                        
    MATLAS ETGs & 179  \\
    CFIS ETGs & 7  \\ 
    MATLAS LTGs & 53  \\ 
    CFIS LTGs & 113  \\ 
    Total  & 352  \\
\hline                                  
\end{tabular}
\end{table}

The CFIS 30 arcmin wide tiles were combined using SWarp \citep{SWarp} and then cropped in order to center the final image on the galaxy of interest. Users annotated structures in images with a field of view (FoV) of 31$\times$31 arcmin.
This FoV corresponds to an average physical size of 250$\times$250 kpc and is equivalent to an average size of 50 effective radii around the target galaxies (with a minimum of 50$\times$50 kpc, or 6.5 effective radii), enough  to visualize the entire galaxy, its neighborhood and potentially most of its extended tidal features. Inspecting larger areas would have been too time consuming given the number of stellar structures and instrumental artefacts to annotate, and would not have been relevant to this study. The images were downsized by a factor of 3 (i.e. binned 3$\times$3), both to decrease the size of the files on the web server  and to enhance very faint structures.

To further enhance the visual identification of fine structures, a transformation from linear scale to a slightly modified inverse hyperbolic sine, asinh, was applied, with the following formula: 
\begin{equation}
    \text{asinh}(ADU) = \log \left( \alpha \times (ADU-b) + \sqrt{ \alpha^{2} \times (ADU-b)^{2} +1} \right) 
\end{equation}
where asinh is the value of the pixel in asinh scale; $ADU$ is the pixel value in linear scale; $b$ is the background value and $\alpha$ is a parameter to tune, chosen here as 1.
The background level was set to a fixed value of  0 in our case. This is motivated by the fact that the Elixir-LSB pipeline precisely processes the images in order to achieve a flat background over a given field of view, after correcting for residual instrumental or large scale sky artefacts. However some local contamination of sources, such as star halos or Galactic cirrus, remain after this processing. Therefore the real  background may locally be non-zero. Local determination of the background is required to get a precise photometry, but for this paper we fixed it to the standard fixed value  as a reasonable approximation.

In addition to asinh scaled images, surface brightness (SB) maps scaled in mag\,arcsec$^{-2}$ were produced and used to characterize the tidal features. The relation from linear to surface brightness scale is the following:
\begin{equation}
    \mu = -2.5\log_{10}\left(\frac{ADU-b}{pixsize^{2}}\right) + 30
\end{equation}
where $\mu$ is the surface brightness value of the pixel in mag\,arcsec$^{-2}$ in AB magnitude; $ADU$ is the value of the pixel in the original linear image; $b$ is the value of the background (chosen here as 0), $pixsize$ is the size of one pixel in arcsecond and 30 is the value of the zero point.

As it will be explained in \ref{general_description}, the on-line annotation tool requires the images to be in a particular format \citep[HiPS, see][]{hips} in order to display them. 
Hence, we created the HiPS after the asinh and surface brightness scalings. Having a single image does not enable the user to adjust on-line the image dynamics\footnote{The original MATLAS server available at \url{https://obas-matlas.u-strasbg.fr/WP/} allows the user to adjust on line the contrasts and cuts, as discussed in \cite{Bilek_et_al_2020}} such as the contrast or the cuts, however this homogeneity turns out to be an asset since the consistency of the images makes it easier to understand the differences between the annotations of several users.

In addition, $g-r$ colormaps were computed from the surface brightness maps for a sub-sample of 177 ETGs and 53 LTGs from MATLAS with available $g$-band images, and exploited to further characterize the LSB strutures,  as described  in section \ref{section:g-r_color}.

Finally, we also considered for our annotations  shallower true-color images from the Data Release 1 (DR1) of  PanSTARRS   \footnote{PanSTARRS, \url{https://panstarrs.stsci.edu/}} (see Section \ref{section:annotation-tool}), an imaging survey that covered the entire sky north of Dec=-30 deg in five bands ($g$, $r$, $i$, $z$ and $y$) \citep{Chambers_et_al_2016}.

\section{LSB structures annotation tool} \label{section:annotation-tool}
The annotation tool we developed is applicable to any imaging survey. However we focus here specifically  on its use for the study of LSB structures. 

\subsection{General description} \label{general_description}

Visual classification methods are much faster and efficient with a web browser based tool that provides online  facilities such as image visualization, navigation as well as immediate structure  identification  and labeling.  Simple and clear interfaces are needed when numerous collaborators or citizen scientists are asked to review the data, such as the Zooniverse platform\footnote{ Zooniverse \url{https://www.zooniverse.org/}}. Although the latter enables collaborators to record quantitative information through an annotation tool, most platforms simply offer the possibility to assess the presence of a given feature.

In this paper, we present a web-based annotation tool that enables collaborators (referred to as 'users' in this paper) to draw with precision the shapes of LSB structures superimposed on deep images and label them,  allowing a quantitative analysis of the LSB structures of various types. In comparison with the Zooniverse interface, our tool offers several drawing options, an easy navigation through the image, the possibility to display images from other surveys and a simple way to verify the annotations once they have been drawn. The annotation process, described in detail in section \ref{annotation-process}, relies on the visualization of astronomical images thanks to an online tool that uses  the Aladin Lite\footnote{Aladin Lite \url{https://aladin.u-strasbg.fr/AladinLite/}} facility  developed by the Centre de Données astronomiques de Strasbourg (CDS).  It enables the visualization of sky regions, overlaid with object information from astronomical databases such as SIMBAD. 
Data from various surveys, such as PanSTARRS DR1, can be displayed and explored, but custom images can also be added, provided that the images are in the Hierarchical Progressive Surveys \citep[HiPS, see][]{hips} format.
The HiPS format enables the representation of large astronomical datasets as the resolution increases when the users zoom on a part of the image. It relies on the hierarchical partitioning of a sphere into smaller and smaller diamonds as the order of the partitioning increases, each diamond being identified by a unique index and order. Hence, we added our own HiPS images, whose origin is detailed in section \ref{section:deep-images}.

\subsection{Annotation process} \label{annotation-process}
The annotation process can be divided in several steps: the selection of the galaxy to annotate, the annotation itself, its classification and the verification or modification of the annotation. All  users are previously identified and logged in to record their annotations.

The  annotation is made directly from the image navigation interface. The latter allows us to zoom in and out in the images, navigate through them and switch between different layers, i.e. images with different intensity scalings or from various surveys. Catalogues from for instance the SIMBAD database can be overlaid to display pieces of information about the objects in the image, including their velocity when known. 
 
\begin{figure*}
   \centering
   \includegraphics[width=\linewidth]{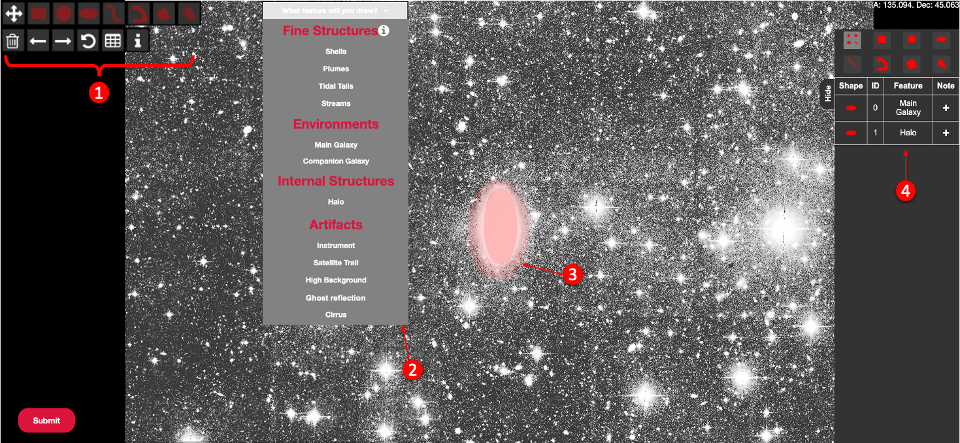}
      \caption{The annotation interface with its main facilities: drawing buttons (label 1), classification menu (label 2), examples of already drawn annotations (label 3)  and summary table (label 4). The background  is a CFIS asinh-scaled image.}
         \label{interface-dessin}
\end{figure*}

\begin{figure}
   \centering
   \includegraphics[width=\linewidth]{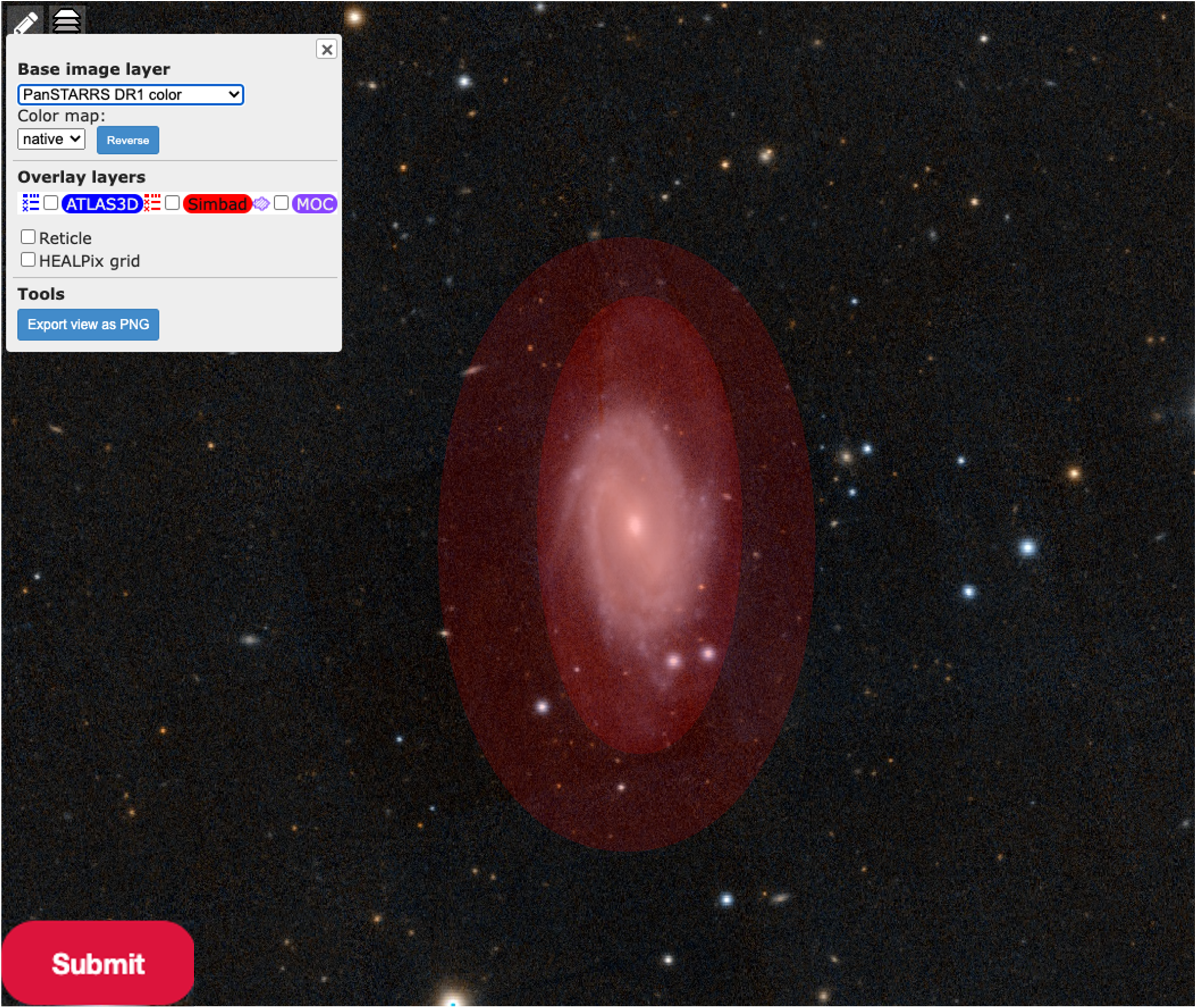}
      \caption{Zoom on a shallower version of Figure \ref{interface-dessin} with the same annotated galaxy. The Pan-STARRS DR1 RGB-color image layer that was used to delineate the 'main galaxy' (inner red ellipse) is shown as background.}
         \label{fig:interface-panstarrs}
\end{figure}

Drawing tool buttons are used to make the annotations. The interface is displayed in Figure \ref{interface-dessin} with a CFIS asinh-scaled image, while a Pan-STARRS DR1 color image layer of the same galaxy is shown in Figure \ref{fig:interface-panstarrs}. The user is asked to draw the external boundary of the features present in the image, as defined later in this section. 
To do so, the most appropriate type of shapes among \textit{Circles}, \textit{Ellipses}, \textit{Rectangles}, \textit{Polygons} or \textit{Curved lines} are selected. \textit{Curved lines} are cubic Bézier curves defined with four control points. The shapes, superimposed on the images with a semi-transparent red color, may be adjusted with precision. Afterwards, the user needs to associate the drawn shape with a label from a menu. All annotations may be checked and further updated from a Summary table. A tutorial explains how to draw the annotations with specific tools and then how to label them.

 The aim is to annotate every feature visible on the image relevant for our study. This means that several annotations are drawn on the same image. It includes the stellar structures of interest and contaminants, among which:
\begin{itemize}
    \item \textbf{Main galaxy}: the target galaxy, encompassing its more luminous features, such as spiral arms for late-type galaxies. Such annotation is made on standard shallow images, in particular the PanSTARRS DR1 color images.\footnote{ We delineate the main galaxy on shallow images as they best highlight the bright inner structures. Indeed, currently, our Aladin-based tool does not offer the possibility to adjust the contrast and dynamics of the images, but allows us to switch between different sets of images through a convenient layer interface.  Among the existing  surveys made available through the interface, PanSTARRS DR1 was chosen as it covers the entire sky north of Dec=-30 deg and thus encompasses  all our galaxies, contrary to SDSS for instance.}
    The outer elliptical isophote of the galaxy is delineated. A posteriori, with the tool presented later, this isophote roughly corresponds to a surface brightness of 24 mag\,arcsec$^{-2}$.
    \item \textbf{Halo}: the extended low surface brightness stellar halo around the galaxy, as seen on the deep CFHT images. Its outer isophote - generally elliptical but sometimes disturbed - is traced\footnote{Note that the extended presumably thick non-starforming  disks that may be present  around LTGs are considered  here as a halos as there is no way to probe their 3D shape, especially when seen face-on.}.
    \item \textbf{Tidal Tail}: defined here and in the previous papers of this series \citep[e.g.,][]{Duc_et_al_2015,Bilek_et_al_2020} as elongated stellar features whose stellar material likely comes from the target galaxy, and which then should have formed during major mergers \citep[e.g.][]{Arp_1966,Toomre_and_Toomre_1972,Mihos_1995}. If the shape differs from the standard, antennae-like elongated structure, it is labeled as a \textit{Plume}.
    \item \textbf{Streams}: generally thin and elongated tidal features, whose material does not seem to originate from the target galaxy. The stream is either attached to a companion galaxy progenitor whose mass is much lower than that of the target galaxy, or it is defined as an orphan stream if no progenitor is visible. In all cases, such streams likely trace past or on-going minor mergers \citep[e.g.][]{Bullock_and_Johnston_2005,Belokurov_et_al_2006,Martinez_Delgado_et_al_2010}.
    \item \textbf{Shells}: arc-shaped features, that are often present in groups and are concentric. They are typically formed during intermediate mass encounters (with a mass ratio around 1:10) with specific orbital parameters \citep[e.g.,][]{Prieur_1990, Ebrova_2013,Duc_et_al_2015, Pop_et_al_2018}. 
    \item \textbf{Companion}: a nearby massive galaxy with a known velocity
    close to that of the target galaxy. We considered a difference of velocities of about 200 km$\,s^{-1}$ for the threshold.  This value is  more conservative  than the larger velocity threshold often  used to identify companions. Indeed we want to make sure that  the two galaxies are currently involved in a tidal interaction able to produce visible tidal debris. The outer envelope of the companion on deep imaging is delineated. 
    \item \textbf{Ghost reflections}: artificial and extended round halos around bright stars caused by internal reflections on the detector and optical elements of the camera.
    \item \textbf{Instrument}: remaining instrument signature (CCD gap).
    \item \textbf{Satellite Trail}: trail of any satellite passing in the image.
    \item \textbf{Cirrus}: dust clouds in our Galaxy, scattering the optical light, and showing up as diffuse but structured (usually filamentary) features on the deep images. Regions likely contaminated by cirrus emission are delineated.
    \item \textbf{High Background}: regions in the deep images, with background levels higher than the blank sky values, not clearly identified as structured cirrus. These regions may trace dust illuminated by bright objects. They are not flat field defects but are really due to higher foreground emissions in these regions.
\end{itemize}
The visual classification of tidal features among tails, streams and shells, is necessarily subjective and may be ambiguous. It then makes sense to have several users annotating and making the classification. One of the goals of this paper is to precisely characterize each type of structures and retrospectively assess the relevance of the classification.

\begin{figure*}
   \centering
   \includegraphics[width=\hsize]{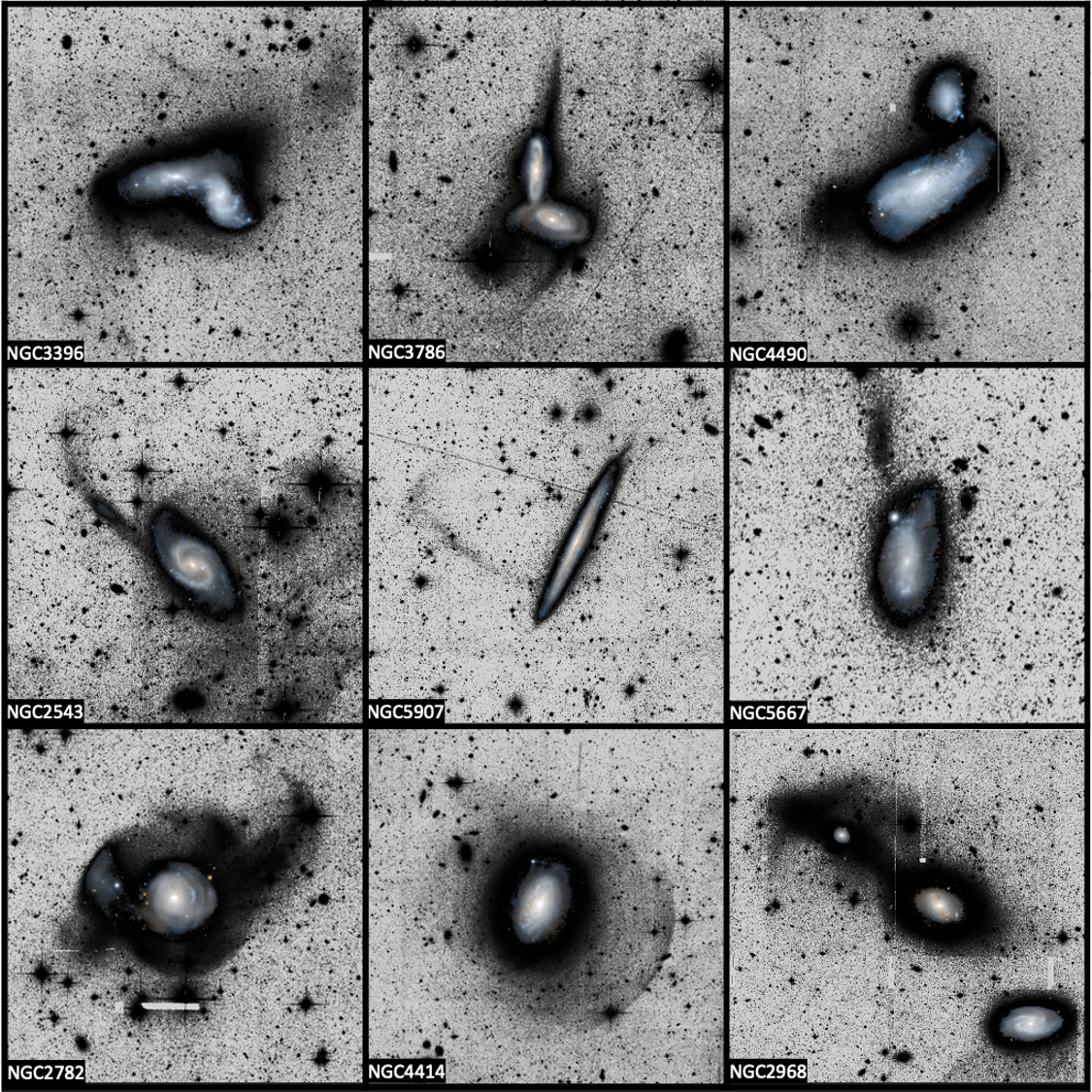}
      \caption{Examples of  tidal features detected in CFIS $r$-band  images displayed with a  asinh scale. A true color image from the PanSTARRS DR1  survey is overlaid at the center of the target galaxy. The first row shows tidal tails and plumes, the middle row streams and the bottom one shells.}
         \label{example-LSB-features}
\end{figure*}

The annotation of contaminants such as cirrus, ghost reflections or high background is essential as they might pollute the stellar structures of interest and either make complicated their detection or skew their annotation. Examples of galaxies with tidal features from CFIS are presented in Figure \ref{example-LSB-features}, while Figures C1, C2 and C3 from \cite{Bilek_et_al_2020} present tidal features identified in MATLAS images.

\begin{figure*}
   \centering
   \includegraphics[width=17cm]{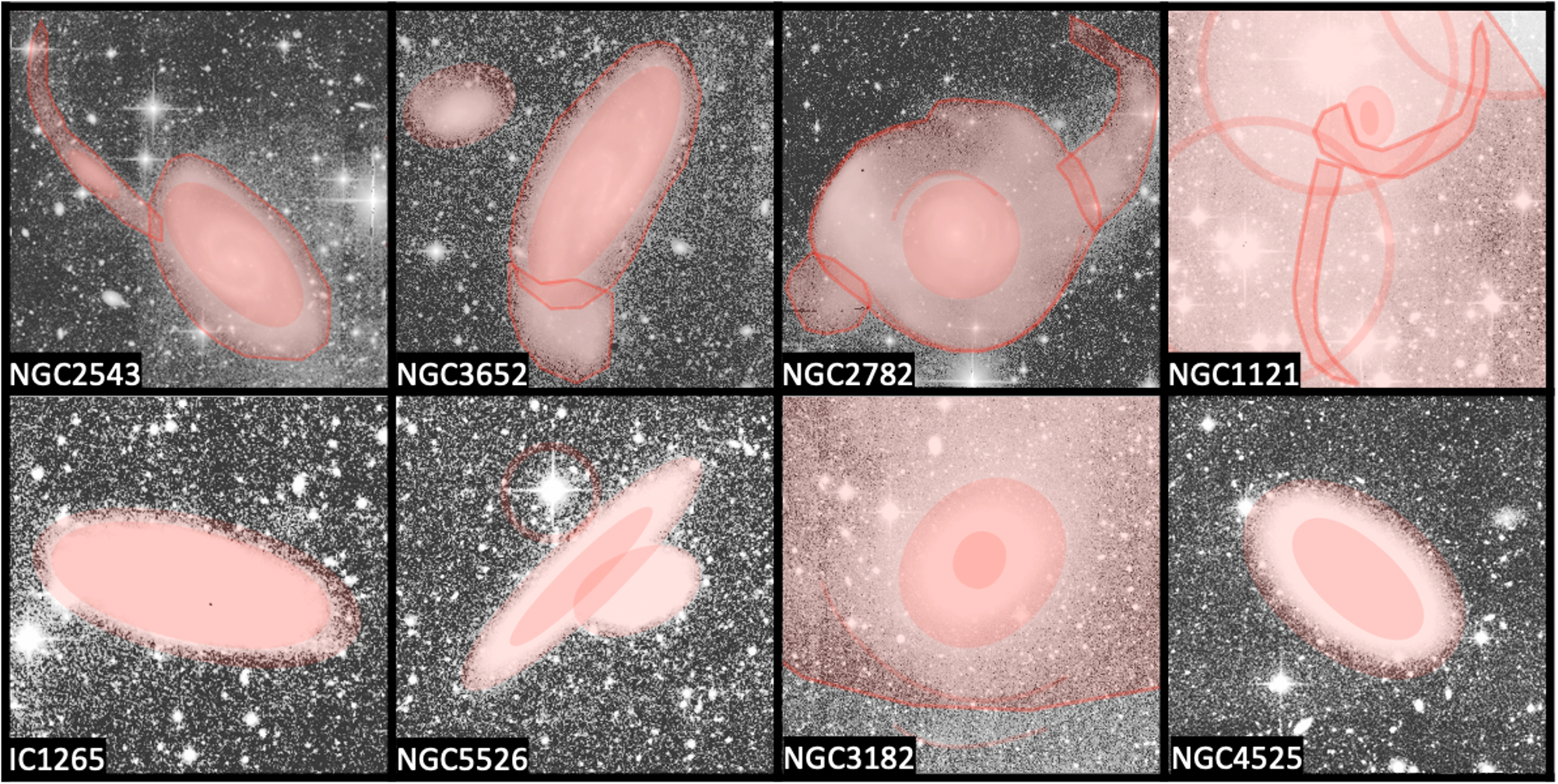}
      \caption{Examples of annotated galaxies. Some of them present tidal features and pollutants or a companion galaxy while others only present the halo and the center of the galaxy.}
         \label{examples_annotate_undergoing}
\end{figure*}

Once the user has finished annotating a galaxy, the shape parameters, positions (in pixel coordinates and in right ascension and declination) and labels of each annotation are stored in a database hosted by a server. These can be used to redraw the annotations on the images or plot them as thumbnails, as it will be seen later. Examples of annotated galaxies are visible in Figure \ref{examples_annotate_undergoing}.

The annotations are made by several users. In this work, all the users are considered to have the same level of expertise in the identification of tidal features. Note however that our  tool  allows us to take into account possible different levels of expertise. This must be taken into account, as it reflects the degree of confidence one can have in the annotations. To that end, weights can in principle be attributed to the users; this is discussed in section \ref{section:discussion_weights}.

\subsection{Thumbnails} \label{section:thumbnails}
In addition to the annotation tool itself, several analysis tools were developed and integrated into the server. One of the main features is the possibility to see the annotations that have been done, through thumbnails. Indeed, quickly visualizing annotations made by a user may be difficult when many features have been annotated. Hence, we developed a fast way to draw the contours of the main features to have an overview of the shapes of the annotations and to detect errors or missing features such as the \textit{Main Galaxy} or the \textit{Halo}. The thumbnails only contain the external boundaries of the annotations, not the images themselves. Thumbnails can also be used to visualize annotations of several users on a same page, for comparison purposes.

\begin{figure}
   \centering
   \includegraphics[width=\linewidth]{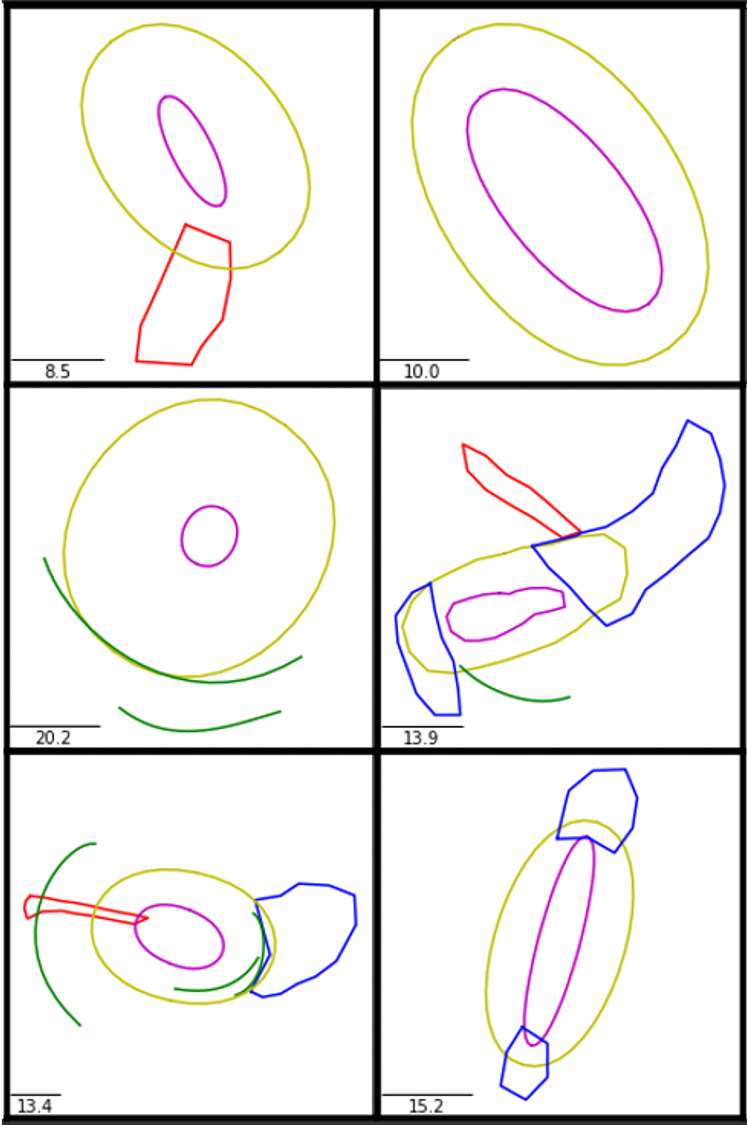}
      \caption{Examples of annotation thumbnails. The center of the target galaxy is plotted in purple, its halo in yellow, tidal tails in blue, streams in red and shells in green. A scalebar in kiloparsecs is shown at the bottom left of each thumbnail.}
         \label{examples_thumbnails_30Re}
\end{figure}

The \textit{Thumbnails} page enables users to choose the galaxies to be plotted, the size of the thumbnail box (in arcseconds ("), kiloparsec (kpc) or effective radius $R_{e}$ of the target galaxy) and the type of annotations: e.g. main galaxy, halo, tidal tails, streams and/or shells. Furthermore, the possibility to represent all features with thumbnails enables a global visual comparison of their shapes and sizes. Such thumbnails, representing only the shapes of tidal features and free of any contaminant (host galaxy, image artefacts, cirrus), could be used to train an algorithm to classify structures based on their shape. This could complement other machine learning algorithms that would be trained on the original images. Once generated, the thumbnails can be downloaded and displayed as webpages. A few examples are illustrated in Figure \ref{examples_thumbnails_30Re}. 

\section{Analysis tools} \label{section:analysis-tools}

In this section, we will present the analysis tools developed to characterize the LSB structures. As previously mentioned, retrieving quantitative measurements about the LSB tidal features is essential to infer the history and mass assembly of a galaxy. Yet for large samples of galaxies, except for fine structure numerical censuses, detailed analyses have not been systematically performed.

The annotation tool we developed offers new possibilities. Indeed, we now have access to the projected shape of the structures, since the users are invited to delineate  the contours with precision. This allows us to retrieve the coordinates of the boundaries of the structures, making possible the determination of the area they cover and their length among other properties. For instance, this will be useful to determine retrospectively whether our criterion to separate streams from tidal tails is relevant. 

\subsection{Area}\label{section:area}
The first step was to determine what is the area covered by each structure. As mentioned in section \ref{annotation-process}, there are different annotation shapes that must be considered. Since \textit{Curved lines} are cubic Bézier curves, it is not possible to determine their area. For the other shapes, the coordinates of all the points forming the contour of the annotation are retrieved in right ascension (RA) and declination (DEC). The distance between two coordinates was obtained using the on-sky separation\footnote{Astropy SkyCoord \url{https://docs.astropy.org/en/stable/api/astropy.coordinates.SkyCoord.html}}, which enabled us to compute the areas of \textit{Circles}, \textit{Rectangles} and \textit{Ellipses}.

To compute areas of simple polygons, we used the shoelace formula  which is given by 
\begin{displaymath}
    A_{polygon}= \frac{1}{2} \displaystyle\left\lvert \left(\sum_{i=1}^{n-1} x_{i}y_{i+1} \right) +x_{n}y_{1} - \left(\sum_{i=1}^{n-1} x_{i+1}y_{i} \right) -x_{1}y_{n} \right\rvert 
\end{displaymath}
where $A_{polygon}$ is the area, $n$ is the number of sides of the polygon and $(x_{i},y_{i}), i=1,...,n$ are the ordered planar coordinates of the vertices of the polygon.

\subsection{Length} \label{section:length}
In addition to the area, the computation of the projected longest length in tidal structures is also important to characterize them.
The definition of the longest length depends on the shape of the annotation. For \textit{Ellipses}, the longest length corresponds to the length of the major axis; for \textit{Circles} it is the diameter; for \textit{Rectangles} and \textit{Curved lines} it is simply the length.

\begin{figure}
   \centering
   \includegraphics[width=\linewidth]{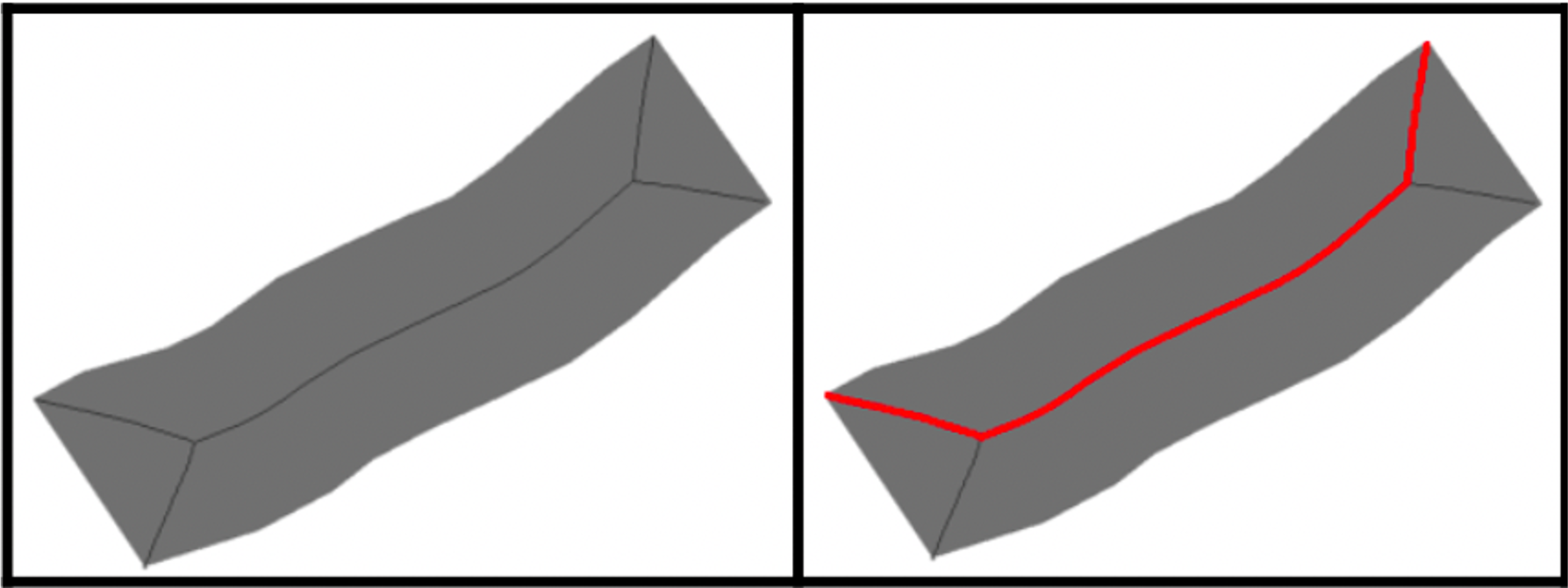}
      \caption{Example of the longest length in a polygon (in gray). \textit{Left}: medial axis (black line) of the polygon obtained during the skeletonization. \textit{Right}: longest length kept (red line) in the polygon}
         \label{skeleton}
\end{figure}

The process is more complicated when dealing with \textit{Polygons}, as they can have various complicated shapes. In these cases, the medial axis was used as the longest length: it can be seen as the topological skeleton and it is defined as the set of points that have at least two closest points on the polygon (i.e. it is the set of points that are equidistant to the contour of the polygon).
The medial axes were obtained using scikit-image \citep{scikit-image} through skeletonization. At the end of this step, the skeletons can have several branches but only the longest possible path was kept. To this purpose, the FilFinder package \citep{filfinder} was used. An example of the previous steps is shown in Figure \ref{skeleton}.

A limit of this method arises for round polygons, i.e. whose shape tends to be circular. In such cases, the topological skeleton tends to be shorter and shorter, up to the limit case of a circle where the skeleton is reduced to the center of the circle. This leads to underestimate the estimated lengths. To tackle this issue, we adapted our method for round polygons and approximate them by the maximum volume inscribed  ellipse. The length is then taken as that of  the major axis. We also modify the value of the area defined in section \ref{section:area} to keep the area of the computed ellipse. 
One way to evaluate how round a polygon is is to compute the isoperimetric ratio $q$ :
\begin{equation}
    q= 4\pi \frac{area}{perimeter^2}
\end{equation}

The value of the isoperimetric ratio $q$ is between 0 and 1, it reaches 1 in the case of a disk. In our case, we had to determine a threshold value for $q$ above which a polygon would be considered round enough for our skeleton method to be incorrect. We determined this threshold to be 0.8, based on our measurements and shapes of the polygons (see section \ref{section:results}). Hence, for polygons with $q<0.8$ we kept the topological skeleton defined above; and for $q\ge0.8$, we approximated them by an ellipse.

\subsection{Distance between shells and the primary galaxy}

\cite{Wilkinson_et_al_1987} and \cite{Prieur_1990} divided shells into several types according to their  position and orientations with respect to the host galaxy.
Type I shells are concentric and centered on the host galaxy; type II shells are circular and randomly distributed around the galaxy;  type III shells appear irregular and not concentric, either because there are very few shells or because of their untypical shape.

Hence, the study of the position of a shell relatively to the center of its primary galaxy is important, as comparisons with simulations can give us hints about the merger that triggered the shells, as well as its age \citep[e.g.,][]{Ebrova_et_al_2021,Bilek_et_al_2021}. Shell radii are useful also for investigating the gravitational fields of galaxies \citep[e.g.,][]{Hernquist_et_al_1987,Bilek_et_al_2013}.

With our annotations, we computed the projected distance between the center of the galaxy and the point located in the middle of the \textit{Curved line} defining the shell. Here we make the assumption that the \textit{Curved lines} are circular arcs. 

\subsection{Progenitor of tidal tails}
Our current annotation server does not allow the user to manually assign an annotated feature to a specific galaxy, for instance the primary galaxy or the companion. By default all  annotations are  tied to the primary galaxy. This is especially an issue for  systems made of two interacting target galaxies, present in the same image. 
In this case, one wishes to attach tidal tails to their real progenitor. There are four different possibilities when determining the progenitor of tidal tails:
\begin{itemize}
    \item \textit{Case 1}: The tail has an overlap with the halo of the primary galaxy but not with that of the companion:  the progenitor is then the main galaxy.
    \item \textit{Case 2}: The tail has an overlap with the halo of the companion but not with that of the primary galaxy: the progenitor is then the companion.
    \item \textit{Case 3}: The tail has no apparent overlap with the target galaxy or the companion. In that case, the position of the center of mass of the tail is computed, as well as the positions of the centers of the primary galaxy and its companion. Then, the distance from each galactic center to the center of the tail is computed, and the tail is associated to the galaxy with the shortest distance. If the difference in distances is small and inferior to an arbitrary threshold, then the progenitor is set to \textit{Unsure}.
    \item \textit{Case 4}: The tail has an overlap with both the primary galaxy and its companion: it looks like a bridge. In this case, the area of the intersection of the tail with both galaxies is computed, and the galaxy with the biggest intersection area is kept as the progenitor. If the difference of areas is small and inferior to an arbitrary threshold, then the same computation on distances than for \textit{Case 3} is performed.
\end{itemize}
For the results, only the tidal tails associated to the primary galaxy were kept, in order to count each structure only once. For streams, by definition they are not attached to the primary galaxy so all the annotations of streams are kept.

\subsection{Surface brightness measurements} \label{section:sb_measurements}
The assessment of the surface brightness (SB) values of each type of tidal feature is important to make comparison with simulations but also  determine whether they will be detectable in surveys to come, such as Euclid or the Vera Rubin Observatory. With our annotation tool, such measurements can be indirectly retrieved.

Indeed, since the coordinates of the contours of the annotations are available, it is possible to retrieve them to create masks that may then be attached to the surface brightness maps. To do so, the RA, DEC coordinates of the contours are converted into pixel coordinates of the FITS image on which we want to create the mask, using Astropy World Coordinate System functions\footnote{Astropy WCS\,\url{https://docs.astropy.org/en/stable/wcs/index.html}} . The interior of that boundary in pixel coordinates is then filled to have the mask, using OpenCV-Python\footnote{ OpenCV-Python\,\url{https://pypi.org/project/opencv-python/}} functions. 
Then, we apply this mask to the SB FITS file and we read the SB values of all the pixels inside the mask\footnote{For shells, the SB values of the pixels along the \textit{Curved line} are retrieved.}. For each annotation of a given type, the median value within the mask area is computed. This way, contribution of light coming from contamination sources such as foreground stars or background galaxies is removed, provided the field is not too crowded. This provides a representative (instead of an average) value of the SB of the structure. A direct aperture photometry would require a proper masking of all contaminant sources, which is beyond the scope of this initial study.

Another potential issue could arise if there is a SB gradient along the structure: deeper imaging will make the median SB value fainter because more LSB pixels are considered. However, since we do not want to get a precise value but to compare trends between tidal features, this issue is not a major one.
So our estimate of the SB value of a given structure is quite uncertain. Our measurements are however useful to compare trends between the different classes of structures. For all the annotations of the same type, the median value of the previously computed medians was computed. 

Finally, it is also possible to retrieve the SB value along the contour of the annotation (by transforming the RA, DEC coordinates of the contours into into pixel coordinates of the SB FITS files, then reading the corresponding SB values and keeping the median value). This is useful especially for stellar halos (see section \ref{section:sb_halos}), as the corresponding ellipse annotation approximates the outer isophote, whose SB value can be compared to the limiting SB of the survey.

\subsection{Color measurements}\label{section:g-r_color}

As mentioned in section \ref{section:deep-images}, $g$ and $r$ band images were available for  all MATLAS galaxies, so we were able to construct the corresponding $g-r$ colormaps. For CFIS, this was not possible as only the $r$-band was available. We created the MATLAS SB files in the $g$-band using the same process as the one described in section \ref{section:deep-images} for the $r$-band images. The $g-r$ colormaps were computed from the $g$ and $r$-band SB files.

Afterwards, we applied the masks of our annotations on the $g-r$ colormap FITS file. For each annotation, we retrieved the $g-r$ values of the pixels inside the mask, and we kept the median value to remove the contribution of outlier pixels. Finally, we estimated the median $g-r$ value for each annotation type (tidal feature, halo or main galaxy). However, these values are only estimates as the colors were computed on polluted images with sources of contamination such as bright reflections or cirri that can overlap with the LSB features of interest.

\subsection{Level of contamination}\label{section:level_contamination}
The contamination of the images by pollutants such as ghost reflections, cirrus, high background, satellite trails or artefacts coming from the instrument can be high in deep images and lead to biased annotations. For instance, if the primary galaxy is embedded in a bright ghost reflection from a nearby star, the user is likely to underestimate the real size of the halo. In order to quantify this degree of contamination, we automatically assigned a reliability index based on the intersection between the halo of the primary galaxy and pollutants as follows. The higher the reliability index, the cleaner the image. 

\begin{itemize}
   \item None: The annotation of the halo is impossible (e.g. due to a high contamination by bright sources)
    \item 1: The entire halo is embedded in a ghost reflection or a high background region
    \item 2: The halo has an overlap with a ghost reflection or a high background region
    \item 3: The entire halo is embedded in a companion galaxy
    \item 4: The entire halo is embedded in a ghost reflection coming from the core of the galaxy 
    \item 5: The halo has an overlap with a companion galaxy 
    \item 6: The entire halo is embedded in cirrus
    \item 7: The halo has an overlap with cirrus
    \item 8: The halo has an overlap with a satellite trail 
    \item 9: There is no pollutant overlapping with the halo
\end{itemize}

As the reliability index can take several values for a given halo, it is stored in a list. To have an average reliability index per galaxy, it was necessary to compute a weighted average of the values. In order to penalize strong pollutants, the values [1,2,3,4,5,6,7,9] were associated to the weights [8,7,6,5,4,3,2,1] respectively. The satellite trails are not taken into account as their impact on the classification is very low. The weighted reliability index in our images is discussed in section \ref{section:discuss-level-contamination}.

In addition, in section \ref{section:similarity_annotations}, we define a similarity index that assesses the similarity between two users' annotations of the same physical type (e.g. halos, pollutants). This similarity index is compared as a function of the level of contamination we defined in this section.

\subsection{Annotations kept} \label{section:selection_process}

By construction, several users have annotated the same structures for any given galaxy. 
We therefore faced the difficulty of keeping  the most representative annotations for any specific galaxy. 
We  present in this section our  selection process. 
It is relatively simple for the diffuse halos since they were systematically annotated by all users. We made the basic assumption that more expert users tend to see fainter features in deep images \citep{Bilek_et_al_2020}. Therefore, we decided, for a given galaxy and structure, to only keep the annotation with the largest area.
For the annotations corresponding to the  brightest part of the galaxy (and made based on shallow images, like PanSTARRS-DR1), the extent is not a relevant criterion and we chose the annotation that represents the median area as our final annotation.

The process is a bit more complicated for tails and streams, as they were not always identified by all users, or might have been delineated in various ways.  
For these tidal features, we proceeded in two steps. First, we considered streams and tails together, and we paired the annotations of the two first users. We used the same method as for the reliability index: two paired structures with a percentage of intersection \footnote{The intersection score is computed as follows: each annotation is defined by its contour coordinates and is considered as a filled polygon (using the shapely package (\url{https://shapely.readthedocs.io}). The area of each polygon is computed, as well as the area of the region of the intersection between the two polygons. The intersection score is then the area of the intersection region divided by the area of the larger polygon.} higher than a given threshold (namely 25\%) are considered as being the same structure. A same unique identifier is attributed to them. We repeated the process with the annotations of the other users: the new feature was paired with the previous ones, the percentage of intersection was computed and then the feature was either associated with an already existing unique identifier or as a new one. 
This iterative process is illustrated on the panels \textit{a)} to \textit{d)} of Figure \ref{fig:unique_id}.

Afterwards, for the features sharing the same unique identifier,  the one with the largest area  is kept following the procedure used for the stellar halos. Hence, at the end each galaxy will have tidal features with different unique identifiers, as visible on the panel \textit{e)} of Figure \ref{fig:unique_id}.
In the following, only the annotations that were kept after the selection process are taken into account (except if explicitly mentioned otherwise).

\section{Results} \label{section:results}

In this section, we present statistical results based on our annotations of tidal features, including their geometrical properties and surface brightness. We opted for median values rather than mean ones in order to get the most representative values. The database contains 8441 annotations. The number of annotations per feature type is detailed in Table \ref{table:nb_annotations}. The annotations have been made by four users: two of them have annotated all the galaxies, while 30\% and 58.5\% of the galaxies were delineated by the two other users.

\begin{table}
\caption{Number of annotations stored in the database as a function of their type. In parenthesis is indicated the number of annotations kept after our selection process on the main galaxy, halo, tidal tails and streams.}   
\centering  
\label{table:nb_annotations}  
\begin{tabular}{cc} 
    \hline \hline
    Annotation type & Number \\
    \hline
    Main Galaxy & 1013 (352) \\
    Halo & 962 (340)  \\
    Tidal Tails & 433 (223) \\
    Streams & 171 (84) \\
    Shells & 260  \\
    Companion Galaxy & 808 \\ 
    High Background & 1121\\
    Ghosted Halo & 3238 \\
    Cirrus & 283 \\
    Satellite Trail & 30 \\
    Instrument & 122 \\
    Total & 8441 (6861) \\ 
    \hline
\end{tabular}
\end{table}

\subsection{Tidal tails and streams}
A total of 223 tidal tails and 84 streams in our database  were kept after the selection process.\footnote{We do not make a distinction between tidal tails (with typical, antennae-like shape) and plumes.} Here we present several geometrical and surface brightness analyses of these features. Based on these analyses, we provide a discussion on the criterion used to differentiate tidal tails from streams in section \ref{section:difference_tails_streams}.

\subsubsection{Qualitative interpretation of global shapes}
\begin{figure*}
   \centering
   \includegraphics[width=\linewidth]{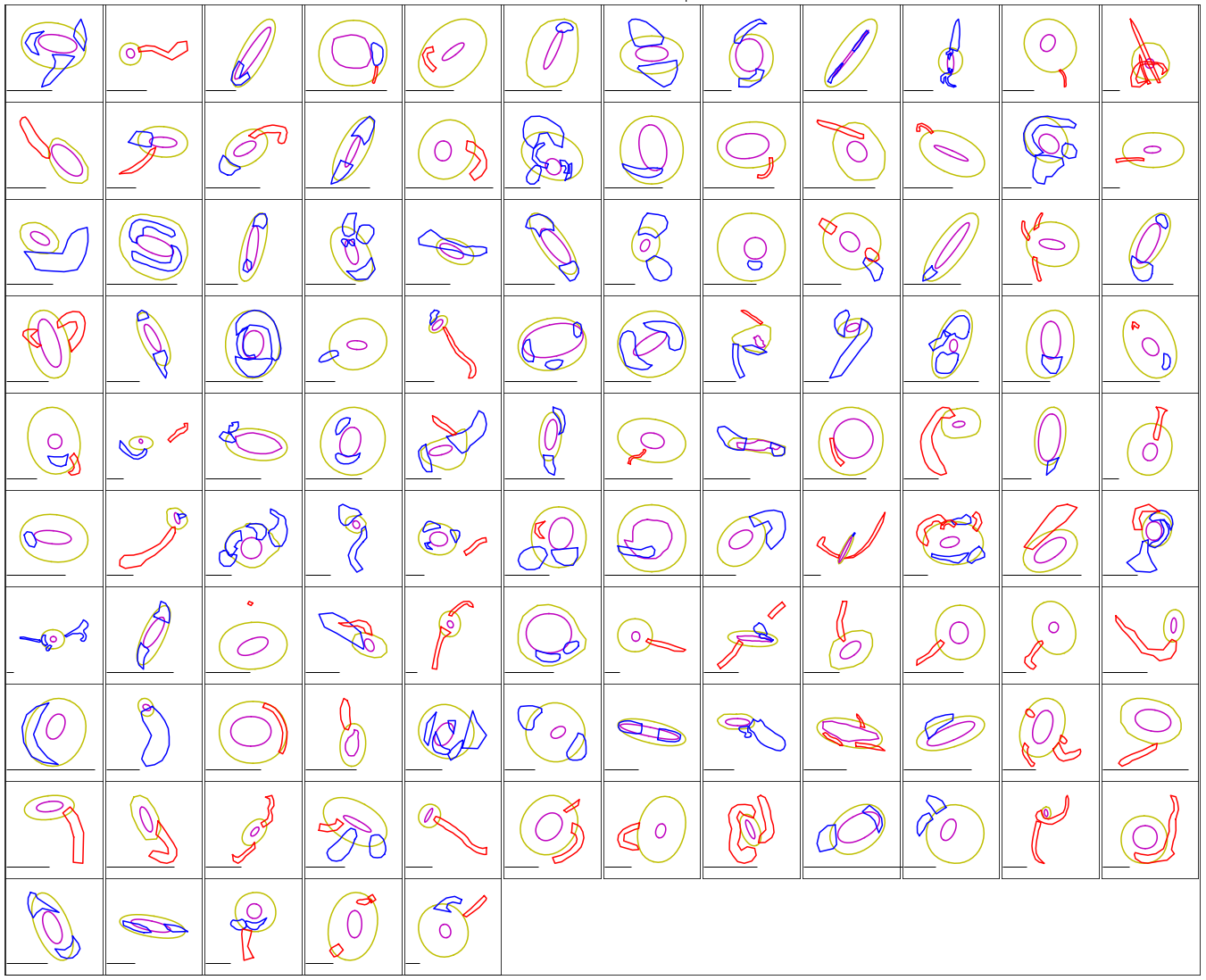}
      \caption{Thumbnails of all the galaxies having tidal tails or streams. The main galaxy is represented in magenta, the halo in yellow, tidal tails in blue and streams in red. A scalebar representing 30 kpc at the bottom left gives an indication of the size of the structures.}
         \label{fig:thumbnails_all_galaxies}
\end{figure*}

 To have an initial overview of their morphology and of their location with respect to the target galaxy, we inspected their thumbnails as represented in Figure \ref{fig:thumbnails_all_galaxies}. 
Their shapes and sizes  show a great variability. For a given galaxy, tidal tails often appear in pairs and seem to be rather located near the foci of the ellipse defining the halo, whereas streams seem more isolated. 
Globally,  tidal tails appear rounder and broader than streams, while streams seem more elongated and thinner. Note that the thickest tails were referred as \textit{Plumes} in our study.

To better compare the individual shapes of each type of tidal structure, we present in Appendix \ref{section:appendix_tail_streams} the footprint of the tails and streams without their host galaxy as a function of the morphological type: tidal features for LTGs are shown in Figure \ref{fig:indidivual_stream_tails_LTG} and in Figure \ref{fig:indidivual_stream_tails_ETG} for ETGs. All thumbnails  have  the same physical size (namely $50\times 50$ kpc) and they are sorted by increasing mass of the host galaxy. The mean galaxy mass in each row is detailed in the text of Appendix \ref{section:appendix_tail_streams}.

From these figures, more massive galaxies tend to host larger or more extended tidal tails. For ETGs, tidal tails seem slightly rounder and larger than for LTGs, but ETGs are on average more massive than LTGs. 
For streams, there is no clear trend neither as a function of the mass of the galaxy nor of its morphological type, which is expected as the material does not originate from the primary galaxy but from a companion.

One important point to note is the fact that tidal tails and streams look relatively similar. Although streams globally seem more elongated than tails, there is no obvious visual difference between them: some tails look like streams and conversely, for all mass range.

Though global trends are observed, clearly the large variability of structures (that might be partly due to delineation errors) does not allow us to make a sharp distinction between tails and streams simply based on a visual inspection. 
In the following sections, we use a more quantitative approach to assess whether a statistically significant distinction between these two types of features can be found.

\subsubsection{Quantitative interpretation: area}
\begin{figure}[h!]
   \centering
   \includegraphics[width=\linewidth]{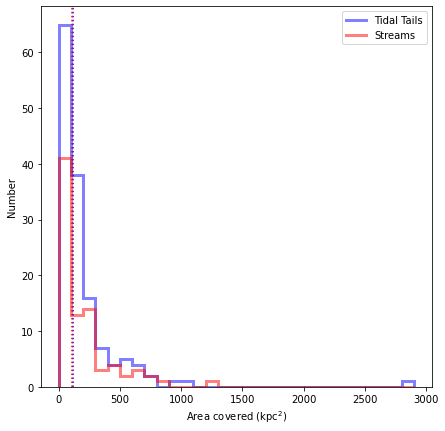}
      \caption{Histogram of the area (in square kiloparsecs) covered by tidal tails (in blue) and streams (in red), in bins of width 100 kpc$^2$. The median of each distribution is represented by the dotted lines. 
      }
         \label{fig:hist_area_tails_streams}
\end{figure}

The distributions of the areas of each type of structure is plotted in Figure \ref{fig:hist_area_tails_streams}. Table \ref{table:area-streams-tails} summarizes the results, making a distinction between LTGs and ETGs. 

\begin{table*}[h!]
\caption{Median areas covered by tidal tails and streams for each type of galaxy, associated with their standard deviation. Values are expressed in square kiloparsecs.}            
\label{table:area-streams-tails}     
\centering                          
\begin{tabular}{ccccc}        
\hline \hline
Galaxy type & \multicolumn{2}{c}{Tidal Tails}  & \multicolumn{2}{c}{Streams}\\
    &  Median area (kpc²) & std &  Median area (kpc²) &std\\
\hline
All galaxies & 116 & 295 & 108 & 227\\
ETGs & 161 & 445 & 139 & 252\\
LTGs & 102 & 149 & 86 & 171\\
\hline
\end{tabular}
\end{table*}

The  distributions of areas hardly differ for both tidal structures: they are peaked around 115 kpc$^{2}$ for tidal tails and 110 kpc$^{2}$ for streams. 
Most structures cover areas smaller than 300 kpc$^{2}$, and few between 300 and 800 kpc$^{2}$. Note that the  structures with the largest areas are associated to systems showing on-going interactions.

From the table, structures surrounding ETGs seem more extended than structures surrounding LTGs, but this difference is not statistically significant. Indeed, we applied Mood's statistical test (testing the null hypothesis that two samples come from populations with the same median) on the areas of the structures as a function of the morphological type: the p-value is 0.08 (for tidal tails) and 0.18 (for streams), which are higher than 0.05, so we cannot reject the null hypothesis that the medians are the same at a confidence level of 5\%.

The  values of the standard deviations may be reduced when taking into account  trends with the mass of the host and environment.   This will be studied in detail in another paper.

\subsubsection{Quantitative interpretation: length and width} \label{section:length_width_tails_streams}
\begin{figure*}
   \centering
\includegraphics[width=\linewidth]{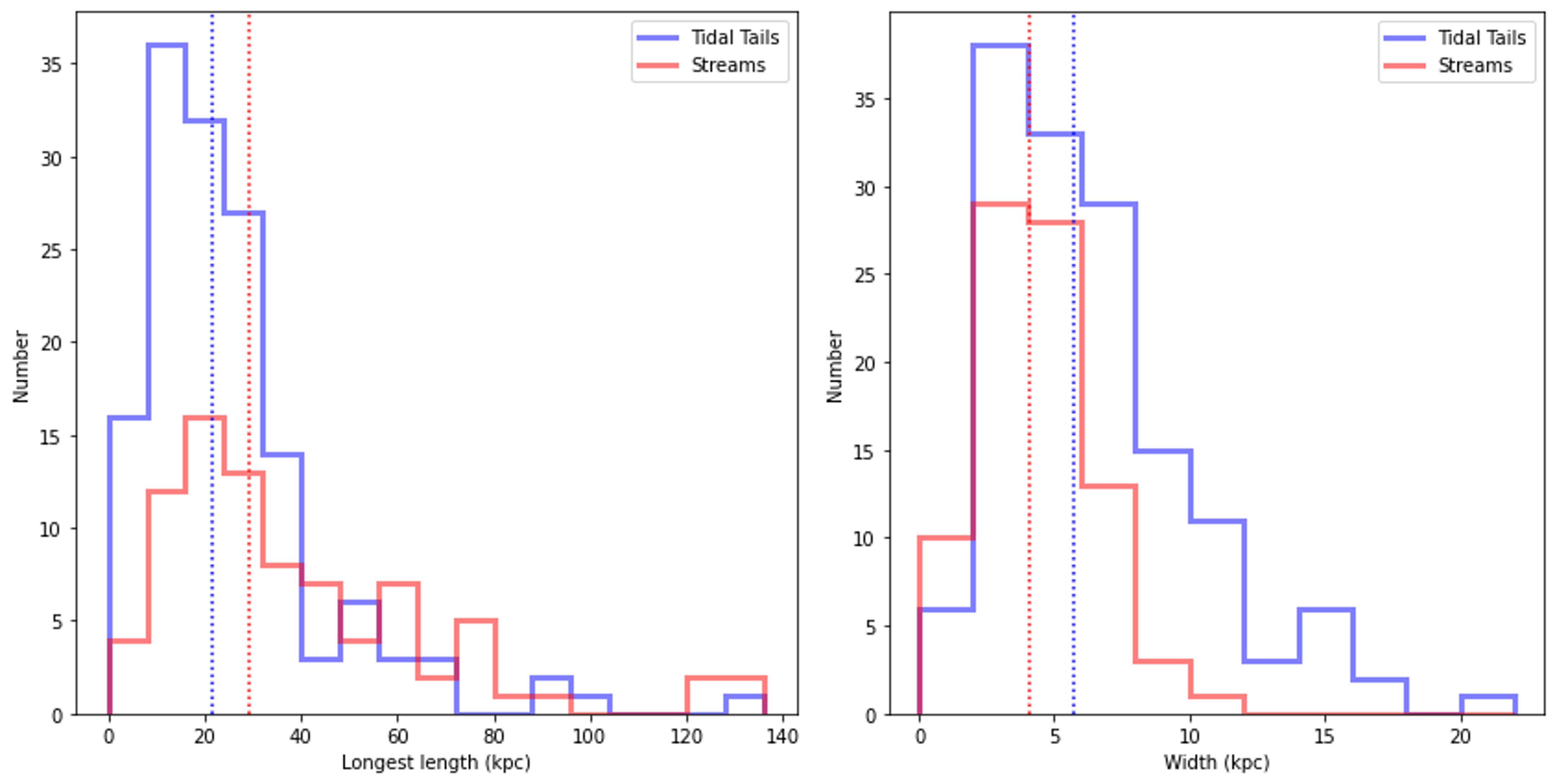}
      \caption{Histograms of the length in kiloparsecs of tidal tails (in blue) and streams (in red), in bins of width 8 kpc (\textit{left}); and of their width in bins of width 2 kpc (\textit{right}). The median of each distribution is represented by the dotted lines. 
      }
         \label{fig:hist_width_length_tails_streams}
\end{figure*}

Median values of the measured length are summarized in Table \ref{table:length-streams-tails} while their histograms are shown in Figure \ref{fig:hist_width_length_tails_streams}. They show that streams are longer than tidal tails (with respectively a median value of 29 kpc and 22 kpc when combining all galaxies), for all galaxy types. The computation of Mood's test gives a p-value of 0.009, which is smaller than 0.05 so we can conclude that the medians of the length for tidal tails and streams are not the same at a significance of 5\%. The distribution of the length of streams is more extended and flatter than for tails.
A few structures reach a length longer than 80 kpc.

\begin{table*}
\caption{Median values of the longest length in tidal tails and streams for each type of galaxy, associated with their standard deviation. Values are expressed in kiloparsecs.}             
\label{table:length-streams-tails}     
\centering                         
\begin{tabular}{ccccc}       
\hline\hline                
Galaxy type & \multicolumn{2}{c}{Tidal Tails}  & \multicolumn{2}{c}{Streams}\\
& Median length (kpc) & std & Median length (kpc) & std\\
\hline
All galaxies & 22 &20 & 29 &29\\
ETGs & 23 &27 & 34 &31\\
LTGs & 20 &14 & 25&25 \\
\hline
\end{tabular}
\end{table*}

An  estimate of the width of tidal tails and streams is obtained by dividing the area they cover by their longest length. This relies on the assumption that these features have ribbon or rectangular shapes. For round polygons, they are approximated by an ellipse (as explained in section \ref{section:length}) so the width corresponds to the minor axis (while the length corresponds to the major axis).
The results are presented in Table \ref{table:width} while the histogram of the width of tidal features for all the galaxies is shown in Figure \ref{fig:hist_width_length_tails_streams}.

\begin{table*}
\caption{Median values of the width in tidal tails and streams for each type of galaxy, associated with their standard deviation. Values are expressed in kiloparsecs.} 
\label{table:width}      
\centering                        
\begin{tabular}{ccccc}       
\hline\hline                 
Galaxy type & \multicolumn{2}{c}{Tidal Tails}  & \multicolumn{2}{c}{Streams}\\
     & Median width (kpc) & std & Median width (kpc) &std\\
\hline
All galaxies & 5.7& 3.8 & 4.1 &2.1 \\ 
ETGs  & 6.6 &4.5 & 4.3 &2.1 \\
LTGs & 5.0 &3.1 & 3.9&2.0\\ 

\hline
\end{tabular}
\end{table*}

From this histogram, one can see that tidal tails are on average wider than streams. This difference is statistically significant, as Mood's test on the medians of the width for tidal tails and streams gives a p-value of 0.0016.
Almost all the streams have a width less than 10 kpc with a peak around 4 kpc, while for tails the peak is around 6 kpc and the distribution is more extended. A few tails are very wide (higher than 14 kpc). 

The fact that tidal tails are wider than streams was expected from models : indeed, the width of a tail or stream increases with the velocity dispersion of the stars that form this structure \citep[e.g.,][]{Johnston_et_al_1996,Johnston_1998}. Yet, the velocity dispersion of a galaxy depends both on its morphological type and of its mass \citep[e.g.,][]{Bernardi_et_al_2010,Bezanson_et_al_2012}. The more massive the galaxy, the higher the velocity dispersion and therefore the wider the tail. Hence, since streams originate from low-mass companions, their velocity dispersion is smaller. This contributes to their widths being smaller than for tails. This is consistent with Figure 6 from \cite{Hendel_and_Johnston_2015} who found in their simulation that if the infalling galaxy satellite had a higher mass, the width of the debris increases. Therefore, our results comfort our approach to make a distinction between tidal tails and streams.

\begin{figure*}
   \centering
   \includegraphics[width=\linewidth]{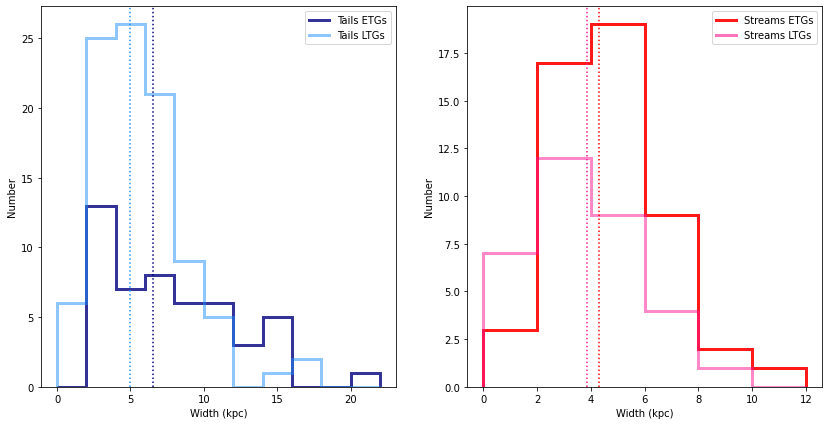}
      \caption{Histograms of the width in kiloparsecs of tidal tails (\textit{left}) and streams (\textit{right}) in bins of 2 kpc as a function of the morphological type of the galaxies: ETGs are represented by darker shades than LTGs. The median of each distribution is represented by the dotted lines.}
         \label{fig:hist_width_feature_morphtype}
\end{figure*}

To explore the results even further, the histograms of the width of tidal tails and streams as a function of the morphological type of their host galaxy is shown in Figure \ref{fig:hist_width_feature_morphtype}.
From it, one can see that the distributions of the width of tidal tails are very different for ETGs and LTGs: the distribution is flatter and more extended for ETGs, with a median value of 6.6 kpc, while for LTGs the distribution is peaked around 5 kpc, with few tails having a width between 10 and 18 kpc. This was also expected, as the velocity dispersion is higher for ETGs than for LTGs, producing wider tails. Contrary to the significant difference for tidal tails, there is no real difference for streams between the distributions for ETGs and LTGs, which are relatively similar. This was also expected, as the morphological type of the primary galaxy is not related to the one of its small companion producing streams.

We can also represent the length and width of tails normalized by the effective sizes of their host galaxy, and as a function of the morphological type, as visible in Figure \ref{fig:tails_width_length_ETG_LTG} of Appendix \ref{section:normalized_length_width_tails}. The same trends than previously mentioned for tails are visible, i.e. a flatter distribution for the width of ETGs, and a slightly longer length, than for LTGs. We did not normalize the length and width of streams as they do not originate from the primary galaxy.

To summarize this section, we found statistical differences between tidal tails and streams: from our measures, streams are more elongated and thinner than tidal tails, which was already hinted by the visual inspection. The agreement with theoretical arguments gives credibility to our classification based on visual impression.

\subsubsection{Surface brightness}
\begin{figure}
   \centering
   \includegraphics[width=\linewidth]{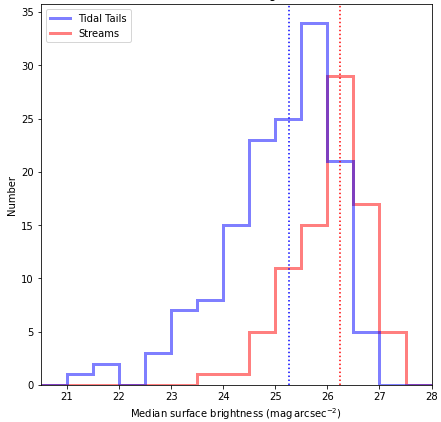}
      \caption{Histogram of the median surface brightness value in magnitudes per square arcsecond for tidal tails (in blue) and streams (in red), in bins of width 0.5 mag\,arcsec$^{-2}$ . The median of each distribution is represented by the dotted lines. 
      }
         \label{fig:hist_SB_tails_streams}
\end{figure}

In Table \ref{table:sb-measurements-tail-streams}, we give the overall median inner SB value for each type of structure and host galaxy type.
 The distributions of these values are visible in Figure \ref{fig:hist_SB_tails_streams}.
\begin{table*}
\caption{Median inner SB measurements for tidal tails and streams, expressed in magnitudes per square arcsecond, associated with their standard deviation.} 
\label{table:sb-measurements-tail-streams}     
\centering                          
\begin{tabular}{cccccc}
\hline \hline
Galaxy type & \multicolumn{2}{c}{Tidal Tails}  & \multicolumn{2}{c}{Streams}\\
 & Median inner SB &std& Median inner SB  &std\\
\hline
All galaxies & 25.3 &1.1 & 26.2 & 0.7\\
ETGs & 25.4 & 0.9 & 26.1 &0.7 \\
LTGs & 25.1 & 1.1 & 26.3 & 0.7  \\
\hline
\end{tabular}
\end{table*}

Having a median SB of 26.2 mag\,arcsec$^{-2}$, streams are fainter than tails by 0.9 mag. This difference is statistically significant, as Mood's test applied on the SB values of tidal tails and streams returns a p-value of $8.6\times10^{-8}$.
As seen in Figure \ref{fig:hist_SB_tails_streams},  the SB distribution of streams is narrower than that of  tidal tails. 
None of them have SB fainter than 27.5 mag\,arcsec$^{-2}$. 

Several reasons could explain why tidal tails are on average brighter than streams. First, our sample includes on-going or recent mergers. Their tails are too young to face the evaporation process that fade collisional debris. Besides, those formed from gas-rich LTGs can contain young and luminous stellar populations \citep[e.g.,][]{Elmegreen_et_al_1993,star_formation_tails}. In addition, by definition tails form from massive objects and should therefore be more massive and  luminous than the streams which emanate from lower-mass satellites. A second explanation could be related to the different survival times of each type of  structures  \citep[see e.g.,][]{Mihos_1995,Ji_et_al_2014,Mancillas_et_al_2019}:  streams tend to be visible for a longer time than tails.
Tails could be more fragile than streams and therefore could disappear faster, i.e. their typical morphology would start to be lost when the structure orbits the galaxy more than once. The old tail would rather resemble multiple streams. We might be able to detect tails only when they are young enough to keep their typical morphology, hence bright enough.

\subsubsection{Overall bending}
As seen in numerical simulations of galaxy mergers \citep[e.g.,][]{Bullock_and_Johnston_2005, Cooper_et_al_2010, Lux_et_al_2013,warp_stream}, tidal streams follow approximately  the orbit of their progenitors: wrapping  around the primary galaxy, they appear as strongly curved. On the other hand, the shape of tails that emanate from the primary galaxy is mainly driven by tidal forces. Depending on their orientation and until their material falls back on the primary, these structures may appear as relatively straight. Therefore curvature may be another criterion to disentangle streams and young tails. 

To obtain a basic estimate of the curvature, we fitted the topological skeleton (as defined in section \ref{section:length}, it is the medial axis and can be seen as a thinner version of the shape that is equidistant to its boundaries) of tidal features by a linear function, using a least-squares  regression. Note that we performed this computation only for not-round polygons, as the skeleton for round polygon does not represent properly the shape. We compared the $R^2$ correlation  coefficient determined for streams and tails: the higher $R^2$, the closer the feature is to a straight curve. 

For streams, the mean and median $R^2$ are respectively 0.62 and 0.79 with an associated standard deviation of 0.37. For tails, the mean and median $R^2$ values are respectively 0.59 and 0.64, with a standard deviation of 0.34. It appears that there is a large variability, especially for streams where the difference between the mean and the median value of $R^2$ is more important. In addition, the linear fit for streams is better than for tails. Hence, streams do not appear more curved than tails as we would have expected from simulations.

One possible explanation, besides the projection effects, the uncertainties of the method and/or possible confusions between streams and tails, is that the depth of the survey may not be sufficient to follow the structures over large distances. They must be long enough to get a reliable estimate of the  curvature, which is often not the case (see Figures \ref{fig:indidivual_stream_tails_LTG} and \ref{fig:indidivual_stream_tails_ETG}). Besides, the bending of stream might be invisible if it is partly hidden by the host galaxy.

\subsubsection{Color}
The median $g-r$ color values computed from the colormaps for tidal tails and streams are presented in Table \ref{table:color-all-structures}.
\begin{table*}[!h]
\caption{Median $g-r$ value for tidal tails, streams, shells and halos for each type of MATLAS galaxy, associated with their standard deviation. The number of structures annotated is indicated in parentheses.}            
\label{table:color-all-structures}    
\centering                          
\begin{tabular}{ccccccccccccc}        
\hline \hline
Galaxy type & \multicolumn{2}{c}{Tidal Tails}  & \multicolumn{2}{c}{Streams}& \multicolumn{2}{c}{Shells}& \multicolumn{2}{c}{Halos}\\
&Median $g-r$ & std & Median $g-r$ & std & Median $g-r$ & std & Median $g-r$ & std\\
\hline
MATLAS all galaxies & 0.57 (148) &0.17 & 0.65 (52)&0.16 &  0.60 (217) & 0.25 & 0.62 (221)&0.12 \\
MATLAS ETGs & 0.57 (95)&0.16 & 0.64 (49)&0.15 & 0.60 (198) & 0.25 & 0.63 (170)&0.11  \\
MATLAS LTGs & 0.56 (53)&0.19 & 1.00 (3)&0.21 & 0.70 (19) & 0.20 & 0.57 (51)&0.15    \\
\hline
\end{tabular}
\end{table*}
It must be noted that the colors were computed on the images without any cleaning process: pollutants such as bright ghost reflections or high background may affect the color measurements. As pollutants  are more visible in the $r$-band than in the $g$-band, they will tend to redden all measures. 
Nevertheless, our measure of the color of streams surrounding ETGs, with a median $g-r$ value of 0.64 mag, is in agreement with \cite{SSLS} who determined  a $g-r$ value between 0.5 and 0.8 mag for 24 streams around local galaxies. 

Our analysis tends to show that tidal tails are  bluer than streams by around  0.1 mag. This difference is statistically significant, as Mood's test gives a p-value of 0.003. This could be due to the presence of  young stars along the tails. In gas-rich mergers, gas is expelled along tidal tails, just  like the stars, and may be compressed in the collisional debris and trigger star-formation  \citep[e.g.,][]{star_formation_tails, star_formation_tails2}.
On the contrary, dwarf satellites are usually gas poor, and if they have been stripped the color of their tidal streams will reflect that of their old stellar populations. So the observed difference in colors between tails and streams may be due to age effects: as already argued, tails tend to be observed at a younger age  than streams. 
Taking into account the fact that the color of the old stars of satellites is bluer than  that of the primary galaxy, due to their lower metallicity, the age effect may be even stronger.

\subsubsection{Stream progenitors}
If streams emanate from a disrupted satellite, remnants of the progenitor may still be visible. As matter of fact, the presence of a condensation within a tidal structure was one of our criteria to label it as a possible stream, especially if a sign of an S-shape was present \footnote{Condensations and tidal dwarf galaxies may be present in tidal tails made in major mergers, but being formed in situ, they do not exhibit the S-shape typical of tidally disrupted dwarfs.}.  

Hence, it is interesting to check the percentage of streams that have a progenitor from our annotation database. The progenitor of the stream is defined as follows. If there are no companion or dwarf galaxy in the annotations, the stream is considered orphan. If the stream has an overlap (partial or entire) with a companion or dwarf galaxy, then the progenitor is the companion galaxy. 

The absence of a progenitor in a genuine stream may indicate that the satellite has been totally destroyed, if the stream  was formed long ago,  or that it is  hidden  in the primary galaxy. Statistically,  orphan streams should be older than those having a progenitor still visible.

In our results, when taking all CFIS and MATLAS galaxies with streams into account, about 70\% of all streams are orphan. In comparison, for the Milky Way and M31, although there is no precise census of the percentage of orphan streams, most of the streams originating from companion galaxies do not present a progenitor, excepted for the Sagittarius stream \citep{Ibata_et_al_1994}. Trends are similar for our results even though more progenitors are still visible.
As mentioned before, this percentage of orphan streams could be related to the age of these structures or to projection effects, but it might also be linked to misclassifications between tidal tails and streams. Indeed, the absence of a progenitor in the structure made the identification more complicated, which in some cases might have mislead users during their classification. We do not see any difference between ETG and LTG hosts, but we did not expect the stream properties or its progenitor to depend on the morphological type of the primary galaxy.

\subsection{Shells}
A total of 260 shells have been annotated. Geometrical, color and surface brightness measurements of these features are presented here.

\subsubsection{Concentricity and radii} \label{section:shells}
Shells have been annotated using \textit{Curved lines} and thus measuring their area is irrelevant. Such an annotation faces a major issue: the selection of the beginning and ending of a shell might be different for an expert or novice user. The former may be aware of the well-shaped circular shells in idealized numerical simulations and consider as a single structure a shell that might be divided into several arcs by the less expert users. Nevertheless, interesting metrics can still be computed, such as the concentricity or their radius.

\begin{figure}
   \centering
   \includegraphics[width=\linewidth]{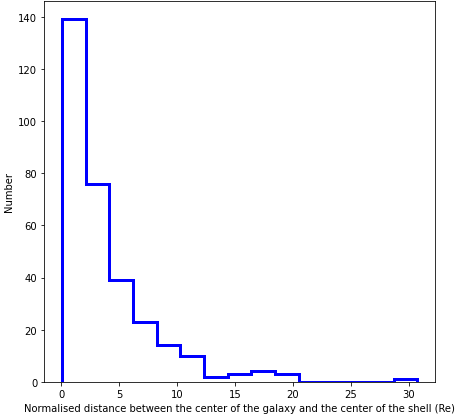}
      \caption{Concentricity test: histogram of the distance in effective radius between the galactic center and the center of the shell. Higher distances correspond to larger deviations from concentricity.}
         \label{fig:hist_center_shells}
\end{figure}

\textbf{Concentricity - }
From numerical simulations, shells are usually formed as concentric structures \citep[e.g.,][]{Pop_et_al_2018,Ebrova_et_al_2021, Bilek_et_al_2021}, a prediction we can directly test with our observations. 
To compute the concentricity, we assume that the \textit{Curved line} defining the shell is a circular arc, and we compute its center\footnote{The center of the shell is hence the center of the circle that passes through the circular arc defining the shell.}. 
Since the \textit{Curved line} is a cubic Bézier curve, it may differ from a circular arc if the user did not draw the shell properly. In that case, we only consider the starting, middle and ending point of the curve and we compute the center of the circle passing through these three points.
We then compute the distance between the shell center and the  center of the host galaxy.
The histogram of theses distances is displayed in Figure \ref{fig:hist_center_shells}. A distance equal to zero means the shell is centered on the galaxy (i.e. concentric), while larger distances indicate a higher deviation from concentricity.

Most often the difference between the galactic center and the center of the shell is less than 10 R$_e$ and the median distance is around 2.5 R$_e$. This means that in general the center of the shell is still located inside the galaxy and we can consider them as relatively concentric. 
Cases for which the relative distance between the centers is large may correspond to bad approximations of the shells as circular structures or to real non-concentric shells such as those of Type III.

\textbf{Radius - }The shell radius is another metric which can easily be  compared to predictions from simulations. It  is computed as the on-sky separation between the point lying in the middle of the \textit{Curved line} defining the shell and  the  center of the galaxy host.

Such a computation needs to take into account the fact that for a given galaxy, multiple users might have annotated the same shells. 
Simply computing  the mean shell radius per galaxy averaged over all the users would introduce biases. Indeed, as mentioned earlier, the number of shells annotated depends on the expertise of the user and on the shape of the shell itself. Clearly defined shells will be annotated as one structure, while less-defined ones will be annotated as several shells. This means that the less well-defined ones will have a higher impact and count for more features.

\begin{figure}
   \centering
   \includegraphics[width=\linewidth]{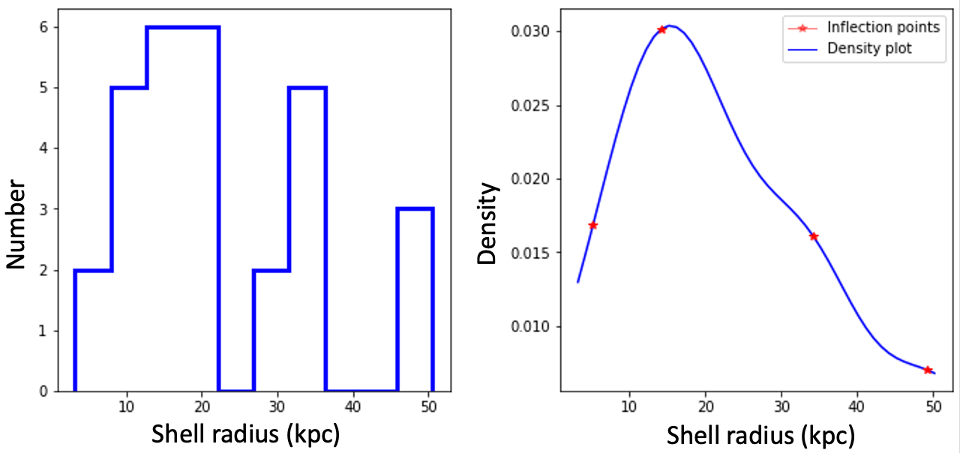}
      \caption{Example of NGC0474. \textit{Left}: Histogram of the shell radii for this galaxy. \textit{Right}: Corresponding density plot with the inflection points of the curve indicated by red stars.}
         \label{fig:density_plot_shell}
\end{figure}

\begin{figure}
  \centering
  \includegraphics[width=\linewidth]{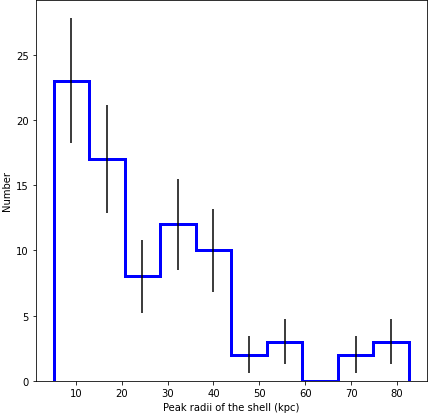}
      \caption{Histogram of the peak radii of shells (in kiloparsecs) for all galaxies, with Poisson uncertainties plotted as the black error bars in each bin.}
         \label{fig:hist_distance_shells}
\end{figure}

To tackle this issue, we have directly plotted for each galaxy the histogram of the radii of shells annotated by all users, and drawn the corresponding density plot. On that  plot, we identified the most representative values, i.e. the radii corresponding to the inflection points, as illustrated in Figure \ref{fig:density_plot_shell}. 
The inflection points, referred here as 'peak radii',  are computed using a Scipy function to find peaks in a 1-D array using a Ricker wavelet transformation\footnote{Scipy find\_peaks\_cwt, \url{https://docs.scipy.org/doc/scipy/reference/generated/scipy.signal.find\_peaks\_cwt.html}}. Their histogram for all galaxies are represented in Figure \ref{fig:hist_distance_shells}.

Most of the shells have a radius smaller than 40 kpc with a few extending to 80 kpc.
For instance the shells of the prototypical galaxy NGC0474 range from 10 to 50 kpc \cite{Bilek_et_al_2021}. We do not observe shells beyond 80 kpc, which seems at odds with  some simulations.  
Whereas they extend to  120 kpc in \cite{shells_distance} and  \cite{shells_distance2}, some of the shells  in \cite{Pop_et_al_2018}  have a radius reaching  150-200 kpc. 
Obviously the comparison is not straightforward as the shell orientation (not well constrained from our annotations of real systems) and differences in surface brightness need to be taken into account.

\subsubsection{Surface brightness} \label{section:SB_shells}
\begin{figure*}
   \centering
   \includegraphics[width=\linewidth]{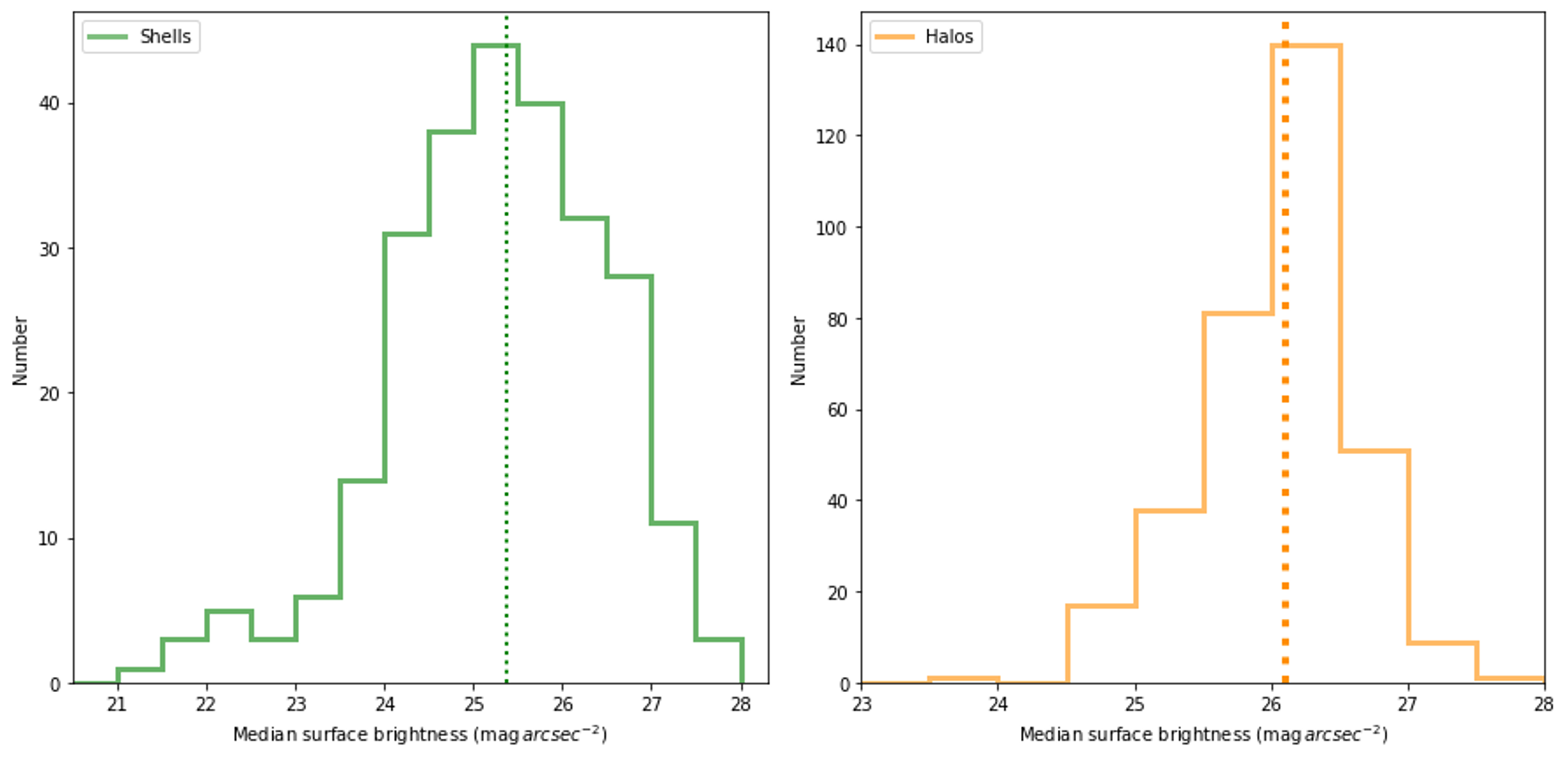} 
      \caption{Histograms of the median surface brightness value in magnitudes per square arcsecond for shells (\textit{left}) and for halos (\textit{right}), in bins of width 0.5 mag\,arcsec$^{-2}$. The medians are represented by the dotted lines. 
      }
         \label{fig:hist_SB_shells_halos}
\end{figure*}

We measured the SB value along the \textit{Curved line} defining the shell annotation.
When considering all galaxies, the median inner SB values for shells is 25.4 mag\,arcsec$^{-2}$; it is of 25.3 mag\,arcsec$^{-2}$ for ETGs and of 25.6 mag\,arcsec$^{-2}$ for LTGs hosts.
The distribution of SB values, presented in Figure \ref{fig:hist_SB_shells_halos}, ranges between 21 and 28 mag\,arcsec$^{-2}$.  Shells are detected with a maximal surface brightness close to the nominal depth of the surveys. This is linked to their shape, as a circular arc is easy to detect and identify on an image: even very faint shells can be visually recognized, while it is more difficult for complicated shapes like tidal tails or streams (see section \ref{section:difference_tails_streams}) 

Note that we did not subtract the stellar halo of the host before our measurement, explaining why the inner shells (i.e. shells that are overlapping with the halo) are apparently brighter than the external ones (i.e. shells further away from the halo).

\subsubsection{Colors} 
The median $g-r$ color values of the shells computed from the MATLAS colormaps are presented in Table \ref{table:color-all-structures}. The median color of shells seems close (slightly redder) than that of other tidal features, but  again the measurements may be polluted by the stellar halo. 

There is a discrepancy of 0.1 mag between the color of the shells of LTGs and  ETGs. Mood's test on the medians of the colors of ETGs and LTGs gives a p-value of $1.8\times 10^{-5}$, which is smaller to 0.05 so we can conclude that the medians of shells for ETGs and LTGs are not the same at a significance of 5\%.

\subsection{Halos}
After our selection process, 340 halos were kept. Here, we present the analysis of their surface brightness, radius and color.

\subsubsection{Surface brightness and radius}\label{section:sb_halos}
We measured the  surface brightness along the external contours of the annotated stellar halos. Figure \ref{fig:hist_SB_shells_halos} plots their distribution for the 340 stellar halos kept after our selection process. It peaks at 26.1 mag\,arcsec$^{-2}$ (median value).

Such a value is clearly much lower than the nominal SB limit of the survey, and that obtained when deriving integrated surface brightness profiles of galaxies. It just reflects the ability of the eye in delineating an external contour on our asinh images. 

\begin{figure}
   \centering
   \includegraphics[width=\linewidth]{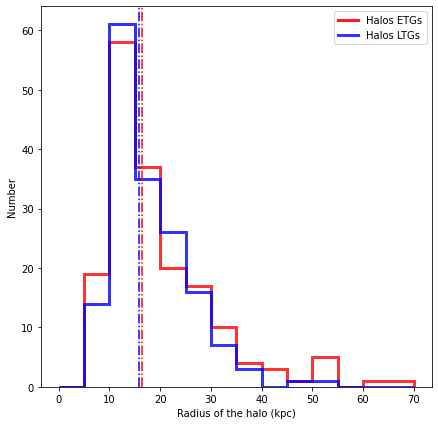}
      \caption{Histogram of the radius of the halos in kiloparsecs as a function of the morphological type, in bins of width 5 kpc. The median of each distribution is represented by the dotted line. }
         \label{fig:radius_halo}
\end{figure}

In addition, from the coordinates of the annotations stored in our database, we can compute the radius of the annotated stellar halos. The  histogram of the values measured for all galaxies is shown  on Figure \ref{fig:radius_halo}.
One can see that the median radius is around 16.5 kpc for ETGs and 15.9 kpc for LTGs, with the majority of the halos having a radius smaller than 30 kpc. A few radii extend up to 70 kpc and correspond mostly to ETGs. However, the distributions are relatively similar for the two morphological types.

We would expect ETGs to have larger radii than LTGs. Indeed, there are more satellite companions around ETGs than LTGs for a given mass, and more companions for more massive galaxies \citep{Kawinwanichakij_et_al_2014}. In addition, the mass growth of present day galaxies seems to be driven by minor mergers \citep{Oser_et_al_2010}, leading to the formation of streams. Hence ETGs should have more material in their outskirts and so larger radii. The fact that we do not observe this could be explained by the low SB of streams. We may not be able to visually recover all the streams, especially the fainter, so we might underestimate the extent of the faint outskirts of the halo. The study of integrated SB profiles of ETGs and LTGs might reveal differences, but this is behind the scope of this paper.

From our annotation database, we cannot directly infer the stellar mass of the halo, an important parameter to constrain the mass assembly of galaxies \citep{Merritt_et_al_2020}. This  estimate requires a lot of processing, in particular to take into account PSF effects \citep{Karabal_et_al_2017}. It will be  the focus of another paper in this series.

\subsubsection{Color}
The median $g-r$ color values computed from the colormaps for halos are presented in Table \ref{table:color-all-structures}.

One can see that there is a 0.06 mag difference between the median $g-r$ color for halos of ETGs and LTGs. It is statistically significant (at a significant level of 5\%), with a p-value of 0.025 from Mood's test. 
Not so surprisingly, the stellar halos of LTGs are bluer than for ETGs, due to (low levels of) star formation occurring there.

\subsection{Covering factor} \label{section:covering_factor}
As mentioned earlier, our annotation database does not provide an estimate of the stellar mass of the LSB structures surrounding their host galaxies. However one proxy of their relative importance is their "covering factor", defined as the percentage of the pixels belonging to one type of structure with respect to a given field of view. For the latter, we considered boxes of side length 20$R_e$ centered on the primary galaxy.

\begin{figure*}
  \centering
  \includegraphics[width=\linewidth]{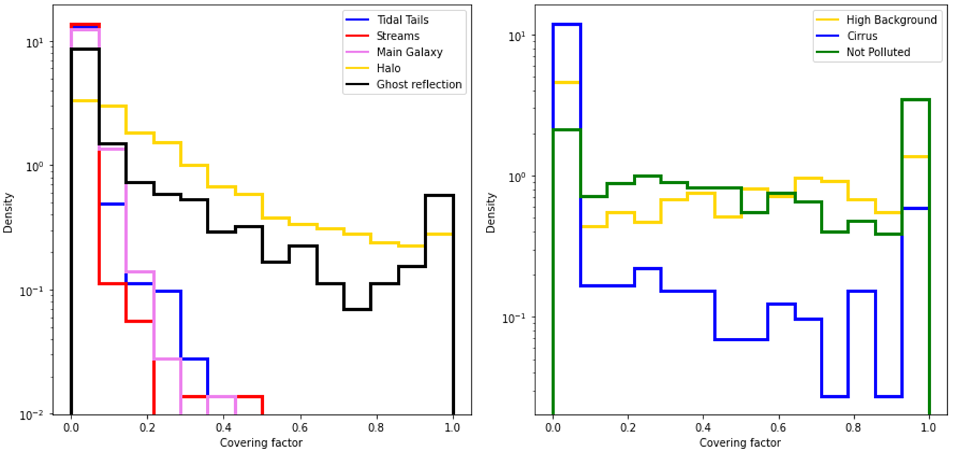}
      \caption{\textit{Left:} histogram of the covering factor for stellar structures and ghost reflections in a selected box of side length 20$R_e$ centered on the primary galaxy. \textit{Right:} histogram of the covering factor for high background, cirrus and clean ("Not Polluted") regions in a field of view of 30.95$\times$30.95' around the primary galaxy.}
         \label{fig:hist_covering_factor_all}
\end{figure*}

For this measurement, we selected the annotations of a given galaxy and a given user. The median values of the covering factor for different types of structures are given in Table \ref{table:covering_factor}. A covering factor of 0 (respectively 1) means that the feature is not present on (respectively entirely covers) the given field of view. Results are shown in Figure \ref{fig:hist_covering_factor_all}. Within the selected boxes, the stellar halo has a median covering factor of 0.16: it is computed for the entire outer ellipse annotation of the halo, without subtracting the main galaxy. This value is to be compared to 0.03 for the central regions ('main galaxy'), 0.04 for tidal tails and 0.02 for streams, considering only galaxies that do have tidal tails or streams.

For  pollutants such as high background and cirrus, it is more relevant  to determine their covering factor for larger fields of view, to get predictions on the contamination levels for other surveys. We considered  a field of view of 31x31'. For regions with high background,  the covering factor has two peaks respectively near 0 and 1, with a relatively uniform distribution between these two values. It is higher than 0.9 (fully polluted images) for 11\% of our galaxies and  lower than 0.1 (clean images) for 34\% of our galaxies. 
Restricting the analysis  to the cirrus, the histogram reveals images that are either fully covered or completely absent with very few intermediate cases. 
 Overall,  regions free of any contamination sources (excepted the ghosted halos), i.e. with a covering  factor higher than 0.9 for the clean pixels, correspond to  about one fourth of our images, while 17\% of our images are almost fully contaminated, i.e. clean regions cover less than 10\%. 
 
\begin{table}
\caption{Median of the percentage of the covering factor in selected boxes of side length 20 $R_e$ around the primary galaxy. For tidal tails and streams, they are counted only when the galaxy exhibits these features.} 
\label{table:covering_factor}    
\centering                
\begin{tabular}{cc}        
\hline \hline
Feature type & Median covering factor \\
\hline
Main Galaxy  & 0.03 \\ 
Halo & 0.17\\ 
Tidal Tails &  0.04\\ 
Streams & 0.02\\ 
Ghost reflections & 0.01\\
\hline
\end{tabular}
\end{table}

\section{Discussion} \label{section:discussion}

The originality of our approach relies on the use of an annotation database of Low-Surface-Brightness features compiled by a group of users who have precisely delineated a large number of individual structures (representing a total of 8441 annotations) directly on displayed images and classified them. The method raises a number of issues partly  posted in the previous sections: 
(1) the difficulty of matching  individual annotations  made by different users (2) the presence of overlapping structures, in particular the contaminants which have a large covering factor, which prevent us from determining with precision the boundaries of some structures  (3) the ambiguity in the classification of the various types of tidal features (4) the fact that we have used images coming from various CFHT surveys with varying depth and surface brightness limits and (5) the reliability of the annotations when considering users with different levels of expertise.
We address all these issues in this discussion.

\subsection{Survey sensitivity to identify tidal features}
One important point to note is the fact that no annotated tidal tails or streams are fainter than 27.5 mag\,arcsec$^{-2}$, even though the nominal depth of the images is at least of 28.3 mag\,arcsec$^{-2}$. This discrepancy can be explained by several factors. First, the nominal survey depth  was estimated from measures done on boxes of 10"x10", while the structures of interest are more extended.
Fluctuations of the SB brightness along the most diffuse structures (possibly above or below the SB limit) make it break into several sub-structures on our images, and its identification and classification as a single genuine stellar feature is very difficult. The presence of artefacts of similar SB as the structures has the same consequence of apparently breaking them into pieces.  

This discrepancy of about 1 mag between the faintest tidal structures that may be identified and classified and the nominal SB limit of the survey must be kept in mind when making comparisons with simulations or estimating their visibility with other facilities. 
Similarly, the outer SB of halos is smaller than the depth of the survey, as the eye is not able to detect the faintest structures compared to what can be obtained with aperture photometry.

On the contrary, shells are more easily identified because of  their characteristic  circular shape, even if they are non-linear, and indeed the SB of the faintest shells are close to the SB limit of the survey (see section \ref{section:SB_shells}).

\subsection{Generalisation to users with different levels of expertise}\label{section:discussion_weights}

In the preliminary study presented in this paper, the annotations were made by four users with a similar level of expertise. However, the annotation tool can be used by anyone and this study can be extended to less expert users. Taking the level of expertise of the user into account in the classification of tidal structures is important, as it reflects the degree of confidence that we can have in the annotation of this user \citep{Bilek_et_al_2020}. Therefore, it is possible with our annotation tool to attribute weights to the users, and these weights would be inherited by the annotations. In that case, the results should take the weights into account, by computing for instance the weighted median and weighted standard deviation instead of simply the median and standard deviation.

Applying weights to the users could also modify our selection method described in section \ref{section:selection_process}. For instance, instead of keeping the largest annotation for halos, tails and streams, one could think of weighting the annotation masks and combining the shapes into a weighted combination of the different annotations.

\subsection{Level of contamination}\label{section:discuss-level-contamination}

\begin{figure}
   \centering
   \includegraphics[width=\linewidth]{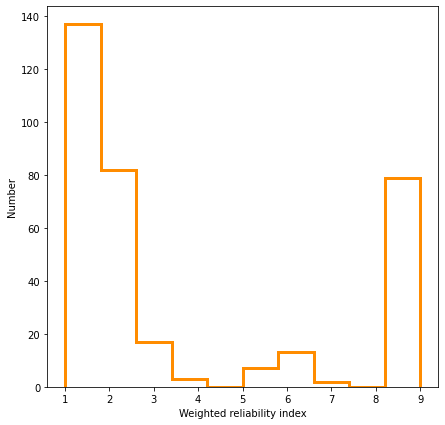}
      \caption{Histogram of the weighted reliability index for the galaxies. A higher index corresponds to a cleaner image around the halo.}
         \label{fig:reliability_index}
\end{figure}

As mentioned in section \ref{section:level_contamination}, we defined a reliability index to take the pollutant sources in the vicinity of the halo of the galaxy into account, as they might lead to biased annotations. The higher the reliability index, the cleaner the image around the galaxy. The histogram of the weighted reliability index can be seen in Figure \ref{fig:reliability_index}.
One can see that only 23\% of the images are completely free of pollutants around the halo, while 65\% are polluted (with an index smaller or equal to 2). This indicates that a majority of our annotations are embedded in a polluted region which might have biased our delineations, but more importantly this could be a major issue for automated classification methods.

\subsection{Similarity between annotations}\label{section:similarity_annotations}
Since several users have annotated the same galaxies, it is necessary to assess their reliability. Following section \ref{section:discussion_weights}, if the users have different levels of expertise, this assessment could be useful to adjust the user weight. Indeed, it is important to characterize whether they annotate like the majority of the group or if their annotations are too different from what is expected (especially for non-expert users). In that case, the weight associated to that user will be lowered in order to avoid outliers in the results due to non-reliable classifications. This step could be part of a methodological process for future studies. We remind that in this paper, all users have a similar level of expertise and no weight was assigned.

A similarity index ranging from 0 (no similarity) to 1 (total similarity) was computed to assess the similarity between the annotations of two users. It depends on the annotation type and is defined as follows:
\begin{itemize}
    \item For halo and main galaxy annotations: the Jaccard index \citep{jaccard_index} is used. It is defined as the area of the intersection of both structures divided by area of their union. If an annotation is missing for one of the user the index is set to zero.
    \item For shells: since their number is relevant, the similarity index is defined as $\frac{min(S_i, S_j)}{max(S_i, S_j)}$, where $S_i$ (respectively $S_j$) is the number of shells annotated by user$_i$ (respectively user$_j$).
    
    \item For tails and streams: two metrics are used. The first one, like for shells, is an index computed over the number of features. However, the area covered is more relevant to characterize them (rather than the number of features, as they range from 0, 1 or 2 for tails and generally 0 or 1 for streams). Hence, a second metric was defined. 
    It relies on the pairing of tidal tails and streams as defined in section \ref{section:selection_process} with the unique identifier. The Jaccard index is computed on the structures annotated by user$_i$ and user$_j$ that share the same unique identifier. 
    \item For contaminants (high background, cirrus): the union of all the contaminant of annotations of the given type is made, as sometimes a user will split a pollutant annotation into two. Then, the Jaccard index is computed between the unions of the contaminants.
\end{itemize}

We found out that the mean similarity index between two of our users reaches about 0.81 for halos and 0.52 for tidal tails and streams sharing the same unique identifier. The relatively high similarity between halo annotations is an indicator that both users annotated in a comparable way, which is important for our study. The lower similarity index for tidal features was expected, since annotating tidal features is not as clear and easy as annotating halos. 

It  is also interesting to determine whether users annotate in a comparable way in the presence of pollutants. When we consider only the cleanest annotations, that is to say annotations with as weighted reliability index as defined in section \ref{section:level_contamination} equal to 9, we obtain a similarity index of 0.85 for halos and 0.57 for tidal tails and streams. When keeping only the most polluted annotations (with a weighted reliability index smaller or equal to 2), the similarity index for halos is 0.8 and 0.53 for tails and streams. One can see that for halos, the cleaner the image, the higher the similarity index, which is likely related to the fact that users annotate in a more similar manner when the image is less polluted. For tails and streams, the trend is not that clear but this might be due to the fact that it is more complicated to precisely delineate these features in a similar manner. Therefore  in general pollutants do  not seem to be the main source of differences in the delineations of tidal debris.

\subsection{Disentangling tidal tails from streams} \label{section:difference_tails_streams}
As mentioned in section \ref{annotation-process}, we defined tidal tail as structures with stellar material  apparently coming from the primary galaxy, while streams  originate from a less-massive smaller companion which may still be visible, hidden or have been destroyed. Most probably the users  have adopted this definition in different ways according to their expertise and used a variety of observables (location, shape, amount of overlap with the closest galaxy, etc.) to assess the classification.   

Getting quantitative measurements from our annotation database, we are able to determine whether these 2 classes of tidal objects show different properties, may thus be really distinguished from our images and retrospectively check  whether the basic initial criterion for disentangling them was relevant.

As presented in detail in section \ref{section:length_width_tails_streams}, we found that streams are narrower than tidal tails, a difference  expected from models, as the width of tidal debris largely depends on the velocity dispersion of the progenitor, itself linked with its  total mass. 
Just considering tidal tails, those found around ETGs appear wider than those associated to LTGs, another result at first order consistent with the expectations, since the velocity dispersion of galaxies decreases with their morphological type. 
In fact this explanation holds only for  ETG-ETG collisions which naturally produce plume-like tidal tails. A merger involving one or two LTGs will produce more narrow tails. Conversely, tails coming from late-type galaxies (that have kept their stellar disk) are necessarily relatively thin. Taking into account all configurations, statistically, there should be more    wide tidal tails around ETGs, as observed. 

Note that the area and length measured (from which we estimated the width) are only based on the projected shape of the structures since it is the only thing that can be annotated. It does not take the inclination and orientation along the line-of-sight of the galaxy into account, so the real intrinsic size of each structure is not known. Some tidal features are likely to be overlooked, especially when the galaxy is seen edge-on or when a tidal tail is hidden behind the galaxy.

We also found a statistically significant difference between the median surface brightness and color of tails and streams. This is also consistent with having progenitors of different masses and ages. 

The fact that we see a statistical distinction in the physical properties of tidal tails and streams is a validation of our definitions of these features (see section \ref{annotation-process}).

However, from visual inspection of the thumbnails with all the individual tidal features shown in Figures \ref{fig:indidivual_stream_tails_LTG} and \ref{fig:indidivual_stream_tails_ETG}, it can be seen that their shapes can vary a lot from one to the other. Some trends are emerging (tidal tails seem broader than streams while streams seem more elongated), but the great variability observed makes difficult a clear visual separation between the two structures. 
This might suggest that an automated classification of streams and tails solely based on their individual shape could be very complicated.
The human expert classifier might have used a number of criteria to classify streams and tails, in addition to  calling his physical intuition on their origin.
Thus results might be better when providing the machine all relevant information, including the properties of the primary galaxy, but still requires reliable labels. 
Deep learning techniques are good candidates to achieve this inclusion of properties thanks to its inherent accounting for visual context.

\subsection{Impact of the depth of the survey}
\begin{figure*}
  \centering
  \includegraphics[width=\linewidth]{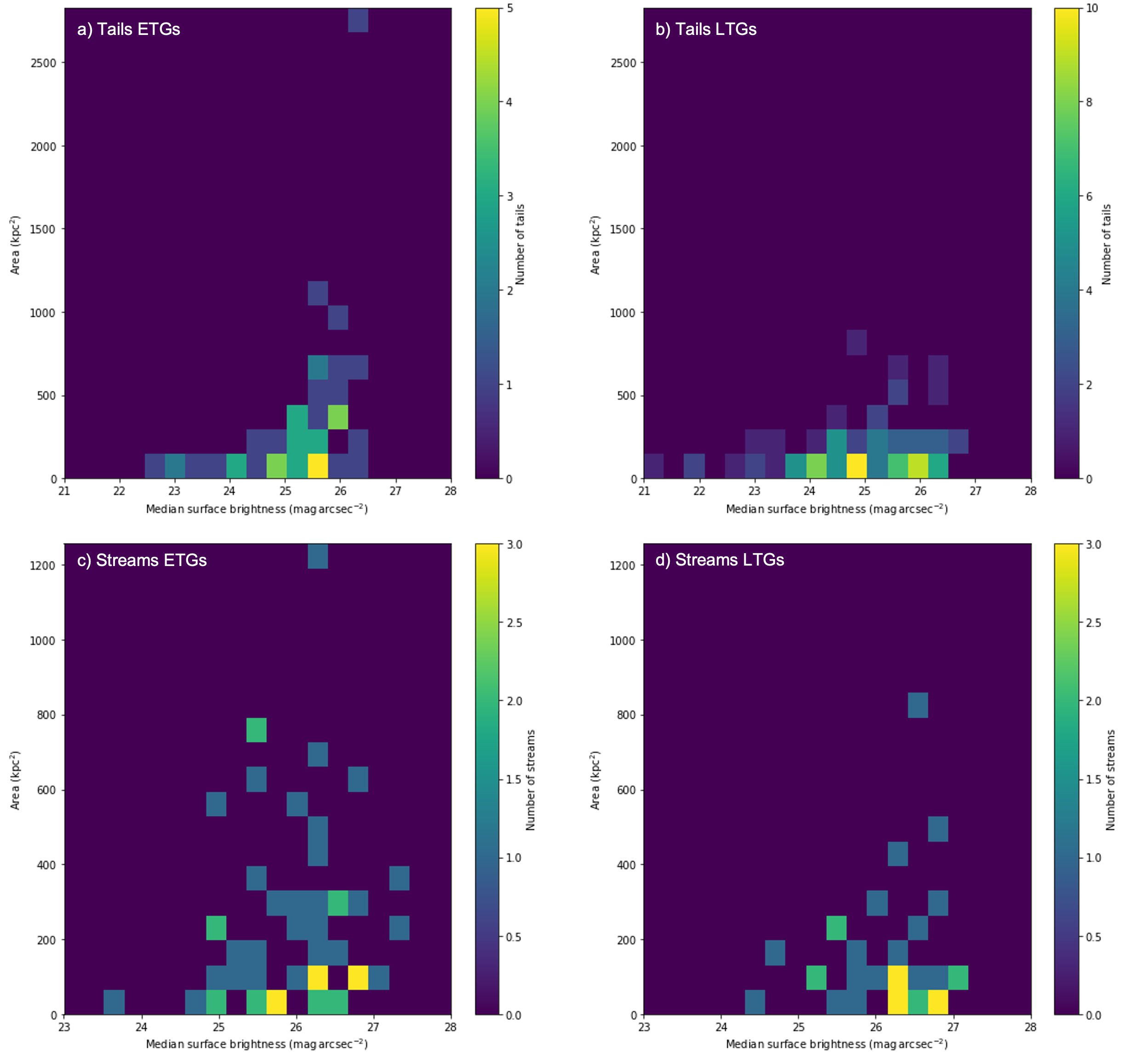}
      \caption{2D-histograms of the area (in square kiloparsecs) of tidal tails and streams as a function of the median surface brightness (in mag\,arcsec$^{-2}$). This 2D-histogram for tidal tails around ETGs is presented in panel \textit{a)} and for LTGs in panel \textit{b)}. For streams, the histogram for ETGs is visible in panel \textit{c)} and in panel \textit{d)} for LTGs. }
         \label{fig:hist2d_area_SB}
\end{figure*}

\begin{figure*}
  \centering
  \includegraphics[width=\linewidth]{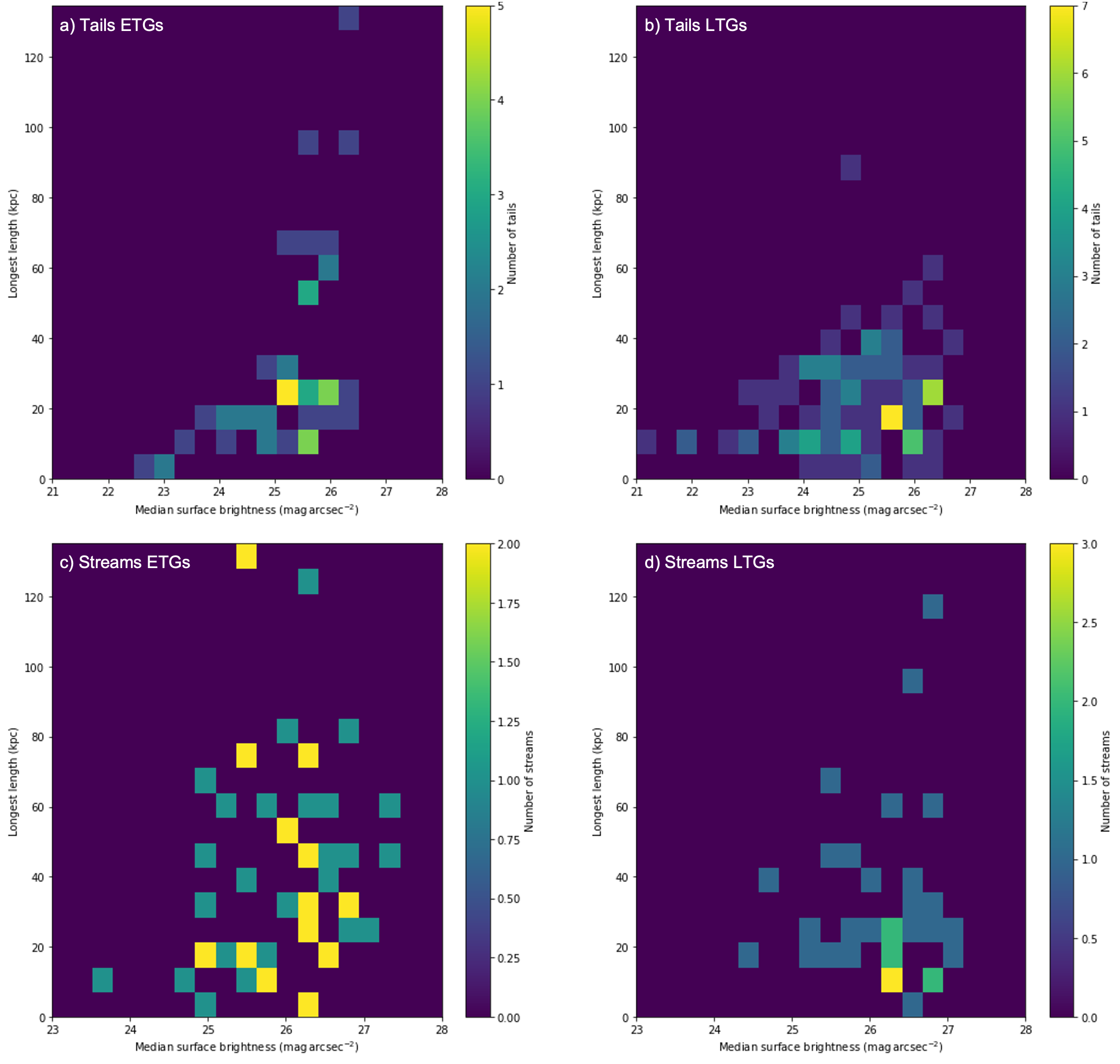}
      \caption{2D-histograms of the length (in kiloparsecs) of tidal tails and streams as a function of the median surface brightness (in mag\,arcsec$^{-2}$). This 2D-histogram for tidal tails around ETGs is presented in panel \textit{a)} and for LTGs in panel \textit{b)}. For streams, the histogram for ETGs is visible in panel \textit{c)} and in panel \textit{d)} for LTGs.}
         \label{fig:hist2d_length_SB}
\end{figure*}

In this paper, we used images from two surveys with different depth, CFIS and the deeper MATLAS. Since most of our ETGs are drawn from MATLAS images and most LTGs from CFIS, we are in principle biased toward finding fainter features in ETGs. In this section, we study the impact of the depth of the survey on tidal features properties like the area or the length. To that end, we plot the 2D-histogram of the SB values of tails and streams as a function of their area (Figure \ref{fig:hist2d_area_SB}), for ETGs and LTGs. A similar 2D-histogram of the length as a function of the SB is represented in Figure \ref{fig:hist2d_length_SB}. One must note that we cannot compare these 2D-histograms between tidal tails and streams as other processes than the depth are involved (see section \ref{section:difference_tails_streams}), and we must keep in mind the fact that we do not have the same number of tidal features around ETGs and LTGs.

For the area (Figure \ref{fig:hist2d_area_SB}) of tidal tails (panels \textit{a)} and \textit{b}), there seems to be a slight trend that fainter structures are larger. We do not observe important differences between ETGs and LTGs, only three tails are larger and fainter in ETGs compared to LTGs.
For the area of streams (panels \textit{c)} and \textit{d}), the trend is less clear. Some streams have similar areas for a large range of SB values. This was expected as streams originate from a companion galaxy and the morphological type of the primary galaxy should not have an influence. Two streams are fainter than 27.2 mag\,arcsec$^{-2}$ around ETGs, contrary to LTGs. 
Hence, for the area, although a few structures are larger and fainter around ETGs compared to LTGs, there is no major difference between ETGs and LTGs that could be attributed to differences in the depth of the surveys.

For the length (Figure \ref{fig:hist2d_length_SB}) of tidal tails (panels \textit{a)} and \textit{b)}, it seems that fainter structures are longer (the trend is clearer for ETGs than for LTGs).
The maximum SB of tails hardly differs: around ETGs it reaches about 26.5 mag\,arcsec$^{-2}$ for four structures, compared to 26.9 mag\,arcsec$^{-2}$ for LTGs for two tails. For the length of streams (panels \textit{c)} and \textit{d)}, for ETGs two streams reach 27.5 mag\,arcsec$^{-2}$, which is 0.2 mag fainter than the faintest stream around LTGs. The longest streams (with a length $ \geq 80 $ kpc) are not the faintest ones, and the majority of streams have comparable length and SB for ETGs and LTGs. Therefore, although there is a difference in the depth of the surveys, we do not observe a major bias in our results for the area or the length of tidal tails and streams.

\subsection{Limits of the study}

We remind here that all our quantitative measurements relied on the analysis of the annotation database. 
In particular, we did not perform aperture photometry in our images. This means that we cannot determine the  flux of the tidal features or of the halos, and hence their stellar mass.
Furthermore, no masking techniques were used to remove the polluting light from foreground or background objects. The SB values that are obtained are useful to compare trends between features, but a more detailed analysis would be needed to precisely assess them.

Like any other visual classification process, our annotations would be more reliable if tens or hundreds of people would have participated. In this paper, we rely on the annotations of 4 users only. However, we realized this process is complicated and the users need to be trained both to recognize LSB structures and to draw their shape with precision, making any citizen science project like Galaxy Zoo difficult to implement. This task is made more complicated by the presence of many pollutants that overlap with the interesting features. The time necessary to annotate is quite important as well (around 10 minutes per galaxy).
Therefore, automated techniques will be needed to analyze larger samples of galaxies with few annotations. The annotations we made could be used to train machine learning algorithms. In fact, it has already been the case with \cite{Richards_et_al_2021}, who used our cirrus annotations to train a new machine learning algorithm to detect cirrus on deep images.
Yet, the small number of annotations is a problem as large datasets are needed to train such algorithms. Appropriate data augmentation or annotations of cosmological simulations may be necessary to have large enough training datasets.

\section{Conclusions} \label{section:conclusion}
The detection and classification of tidal features around galaxies is essential as their characterization gives valuable information about the past assembly history of their host galaxy. In this paper, we present an online annotation tool that enables users to draw the shapes of LSB structures with precision in deep images for a large number of galaxies. 
We also show how we may use our annotation database to estimate some physical quantities, such as the shape, size, surface brightness and colors of tidal tails, streams, shells and stellar halos. Such values were so far not very well constrained on large samples of galaxies. They may be compared to simulations to better understand the type of mergers that took place, and more generally to constrain models of galactic evolution.

Using a customized on-line tool, we have manually identified, delineated  and classified  LSB features including contaminants around 352 nearby massive galaxies from the CFHT MATLAS and CFIS  surveys. Each field has been inspected by up to 4 different users. A database containing 8441 annotations was compiled (Table \ref{table:nb_annotations}). We have developed a methodology to associate common annotations between users and link the associations with their host galaxies. 

With such a project, we aim at differentiating the types of tidal debris and a posteriori determine whether a classification solely based on eye inspection of deep images is reliable. 

Not so surprisingly, the annotated structures show an apparent large variety of shapes and sizes when put and compared together (as visible in Figures \ref{fig:indidivual_stream_tails_LTG} and \ref{fig:indidivual_stream_tails_ETG}). There is a large overlap between tails (tidal structures emanating from the primary galaxy and made during major mergers) and streams (tidal debris from disrupted low-mass companions).
However, a statistical analysis revealed real differences:
 
 \begin{itemize}
    
 \item   Tidal tails are wider than streams, with a typical width of 6 kpc against 4 kpc for streams (Figure \ref{fig:hist_width_length_tails_streams}). This was expected from models: since streams originate from low-mass companions with lower velocity dispersion, their width should be smaller.
 
 \item  Tidal tails are brighter than streams, with a difference between their median surface brightness of almost 1  mag\,arcsec$^{-2}$ (Table \ref{table:sb-measurements-tail-streams}). This may be due to some age effects, with streams having a longer survival (i.e. visibility) time than tails, which are more easily identified as such  when they are young.
 
  \item  Tidal tails are slightly bluer than streams (a difference of  0.1 mag for the g-r color) (Table \ref{table:color-all-structures}), again likely due to an age bias. 

 \end{itemize}
 
These statistical differences comfort the approach we used to specifically distinguish tidal tails from streams. 

\begin{itemize}

 \item The annotated stellar halos have radii extending mostly between 5 and 30 kpc (Figure \ref{fig:radius_halo}) with a  similar distribution for ETGs and LTGs in the environments probed here (outside massive clusters). 

 \item Identified shells have typically radii lower than 40 kpc, with few extending to 80 kpc (Figure \ref{fig:hist_distance_shells}): we do not observe the very external extended shells found in some simulations, even though the comparison is not straightforward.

\end{itemize}

Other tendencies interesting for the analysis for future surveys are worth highlighting:

\begin{itemize}

 \item  The faintest structures identified as tidal tails and streams have a median surface brightness about 1 mag brighter than the nominal limiting surface brightness of the survey.   This should also  be kept in mind when comparing observations to simulations.

 \item  Artefacts or foreground structures occupy a large fraction of the image pixels (more than 50\% in 10 R$_e$ boxes centered on the target galaxy). They may significantly alter our SB and color measurements.  Having the possibility to remove them would also  be a strong asset for future automatic classifications. 
 
 \end{itemize}

Finally, our annotations were all done  manually, which took a non-negligible amount of time of a group of participants. Though feasible in samples of hundreds of galaxies, thanks to our dedicated on-line tool, it cannot be applied to samples of thousands. Machine learning methods will be needed to automatically detect and classify tidal features in large surveys to come, and our annotation database can be used to train such algorithms.

As future prospects, the properties of tidal features with respect to the host galaxy properties will be studied in a future paper. In addition, the amount of stellar mass in the tidal features could be computed. Together with estimates of their lifetime, they can give an estimate of the speed of mass gain of galaxies caused by mergers.

\begin{acknowledgements}
The authors thank Garreth Martin for his feedback on the paper. F.R. was supported by Science and Technology Facilities Council (STFC) grant ST/P006779/1.
This research has made use of the SIMBAD database, operated at CDS, Strasbourg, France, and of "Aladin sky atlas" developed at CDS, Strasbourg Observatory, France.

This work is based on data obtained as part of the Canada-France Imaging Survey, a CFHT large program of the National Research Council of Canada and the French Centre National de la Recherche Scientifique. Based on observations obtained with MegaPrime/MegaCam, a joint project of CFHT and CEA Saclay, at the Canada-France-Hawaii Telescope (CFHT) which is operated by the National Research Council (NRC) of Canada, the Institut National des Science de l’Univers (INSU) of the Centre National de la Recherche Scientifique (CNRS) of France, and the University of Hawaii. This research used the facilities of the Canadian Astronomy Data Centre operated by the National Research Council of Canada with the support of the Canadian Space Agency. This research is based in part on data collected at Subaru Telescope, which is operated by the National Astronomical Observatory of Japan. We are honored and grateful for the opportunity of observing the Universe from Maunakea, which has the cultural, historical and natural significance in Hawaii. Pan-STARRS is a project of the Institute for Astronomy of the University of Hawaii, and is supported by the NASA SSO Near Earth Observation Program under grants 80NSSC18K0971, NNX14AM74G, NNX12AR65G, NNX13AQ47G, NNX08AR22G, YORPD20\_2-0014 and by the State of Hawaii.
\end{acknowledgements}

\bibliographystyle{aa}
\bibliography{main.bbl}

\begin{thebibliography}{109}
\expandafter\ifx\csname natexlab\endcsname\relax\def\natexlab#1{#1}\fi

\bibitem[{{Abraham} {et~al.}(1994){Abraham}, {Valdes}, {Yee}, \& {van den
  Bergh}}]{Abraham_et_al_1994}
{Abraham}, R.~G., {Valdes}, F., {Yee}, H.~K.~C., \& {van den Bergh}, S. 1994,
  \apj, 432, 75

\bibitem[{{Abraham} {et~al.}(1996){Abraham}, {van den Bergh}, {Glazebrook},
  {Ellis}, {Santiago}, {Surma}, \& {Griffiths}}]{Abraham_et_al_1996}
{Abraham}, R.~G., {van den Bergh}, S., {Glazebrook}, K., {et~al.} 1996, \apjs,
  107, 1

\bibitem[{{Abraham} {et~al.}(2003){Abraham}, {van den Bergh}, \&
  {Nair}}]{Abraham_et_al_2003}
{Abraham}, R.~G., {van den Bergh}, S., \& {Nair}, P. 2003, \apj, 588, 218

\bibitem[{{Alabi} {et~al.}(2020){Alabi}, {Romanowsky}, {Forbes}, {Brodie}, \&
  {Okabe}}]{Alabi_et_al_2021}
{Alabi}, A.~B., {Romanowsky}, A.~J., {Forbes}, D.~A., {Brodie}, J.~P., \&
  {Okabe}, N. 2020, \mnras, 496, 3182

\bibitem[{{Antoja} {et~al.}(2020){Antoja}, {Ramos}, {Mateu}, {Helmi}, {Anders},
  {Jordi}, \& {Carballo-Bello}}]{Antoja_et_al_2020}
{Antoja}, T., {Ramos}, P., {Mateu}, C., {et~al.} 2020, \aap, 635, L3

\bibitem[{{Arp}(1966)}]{Arp_1966}
{Arp}, H. 1966, \apjs, 14, 1

\bibitem[{{Atkinson} {et~al.}(2013){Atkinson}, {Abraham}, \&
  {Ferguson}}]{Atkinson_Abraham_Ferguson_2013}
{Atkinson}, A.~M., {Abraham}, R.~G., \& {Ferguson}, A. M.~N. 2013, \apj, 765,
  28

\bibitem[{{Baugh} {et~al.}(2003){Baugh}, {Benson}, {Cole}, {Frenk}, \&
  {Lacey}}]{Baugh_et_al_2003}
{Baugh}, C.~M., {Benson}, A.~J., {Cole}, S., {Frenk}, C.~S., \& {Lacey}, C.
  2003, in The Mass of Galaxies at Low and High Redshift, ed. R.~{Bender} \&
  A.~{Renzini}, 91

\bibitem[{{Bell} {et~al.}(2006){Bell}, {Naab}, {McIntosh}, {Somerville},
  {Caldwell}, {Barden}, {Wolf}, {Rix}, {Beckwith}, {Borch}, {H{\"a}ussler},
  {Heymans}, {Jahnke}, {Jogee}, {Koposov}, {Meisenheimer}, {Peng}, {Sanchez},
  \& {Wisotzki}}]{Bell_et_al_2006}
{Bell}, E.~F., {Naab}, T., {McIntosh}, D.~H., {et~al.} 2006, \apj, 640, 241

\bibitem[{{Belokurov} {et~al.}(2006){Belokurov}, {Zucker}, {Evans}, {Gilmore},
  {Vidrih}, {Bramich}, {Newberg}, {Wyse}, {Irwin}, {Fellhauer}, {Hewett},
  {Walton}, {Wilkinson}, {Cole}, {Yanny}, {Rockosi}, {Beers}, {Bell},
  {Brinkmann}, {Ivezi{\'c}}, \& {Lupton}}]{Belokurov_et_al_2006}
{Belokurov}, V., {Zucker}, D.~B., {Evans}, N.~W., {et~al.} 2006, \apjl, 642,
  L137

\bibitem[{{Bernardi} {et~al.}(2010){Bernardi}, {Shankar}, {Hyde}, {Mei},
  {Marulli}, \& {Sheth}}]{Bernardi_et_al_2010}
{Bernardi}, M., {Shankar}, F., {Hyde}, J.~B., {et~al.} 2010, \mnras, 404, 2087

\bibitem[{{Bertin} {et~al.}(2002){Bertin}, {Mellier}, {Radovich}, {Missonnier},
  {Didelon}, \& {Morin}}]{SWarp}
{Bertin}, E., {Mellier}, Y., {Radovich}, M., {et~al.} 2002, in Astronomical
  Society of the Pacific Conference Series, Vol. 281, Astronomical Data
  Analysis Software and Systems XI, ed. D.~A. {Bohlender}, D.~{Durand}, \&
  T.~H. {Handley}, 228

\bibitem[{{Bezanson} {et~al.}(2012){Bezanson}, {van Dokkum}, \&
  {Franx}}]{Bezanson_et_al_2012}
{Bezanson}, R., {van Dokkum}, P., \& {Franx}, M. 2012, \apj, 760, 62

\bibitem[{{Bickley} {et~al.}(2021){Bickley}, {Bottrell}, {Hani}, {Ellison},
  {Teimoorinia}, {Yi}, {Wilkinson}, {Gwyn}, \& {Hudson}}]{Bickley_et_al_2021}
{Bickley}, R.~W., {Bottrell}, C., {Hani}, M.~H., {et~al.} 2021, \mnras, 504,
  372

\bibitem[{{B{\'\i}lek} {et~al.}(2020){B{\'\i}lek}, {Duc}, {Cuillandre}, {Gwyn},
  {Cappellari}, {Bekaert}, {Bonfini}, {Bitsakis}, {Paudel}, {Krajnovi{\'c}},
  {Durrell}, \& {Marleau}}]{Bilek_et_al_2020}
{B{\'\i}lek}, M., {Duc}, P.-A., {Cuillandre}, J.-C., {et~al.} 2020, \mnras,
  498, 2138

\bibitem[{{B{\'\i}lek} {et~al.}(2021){B{\'\i}lek}, {Fensch}, {Ebrov{\'a}},
  {Nagesh}, {Famaey}, {Duc}, \& {Kroupa}}]{Bilek_et_al_2021}
{B{\'\i}lek}, M., {Fensch}, J., {Ebrov{\'a}}, I., {et~al.} 2021, arXiv
  e-prints, arXiv:2111.14886

\bibitem[{{B{\'\i}lek} {et~al.}(2013){B{\'\i}lek}, {Jungwiert},
  {J{\'\i}lkov{\'a}}, {Ebrov{\'a}}, {Barto{\v{s}}kov{\'a}}, \&
  {K{\v{r}}{\'\i}{\v{z}}ek}}]{Bilek_et_al_2013}
{B{\'\i}lek}, M., {Jungwiert}, B., {J{\'\i}lkov{\'a}}, L., {et~al.} 2013, \aap,
  559, A110

\bibitem[{{Bottrell} {et~al.}(2019){Bottrell}, {Hani}, {Teimoorinia},
  {Ellison}, {Moreno}, {Torrey}, {Hayward}, {Thorp}, {Simard}, \&
  {Hernquist}}]{Bottrell_et_al_2019}
{Bottrell}, C., {Hani}, M.~H., {Teimoorinia}, H., {et~al.} 2019, \mnras, 490,
  5390

\bibitem[{{Bridge} {et~al.}(2010){Bridge}, {Carlberg}, \&
  {Sullivan}}]{Bridge_et_al_2010}
{Bridge}, C.~R., {Carlberg}, R.~G., \& {Sullivan}, M. 2010, \apj, 709, 1067

\bibitem[{{Bullock} \& {Johnston}(2005)}]{Bullock_and_Johnston_2005}
{Bullock}, J.~S. \& {Johnston}, K.~V. 2005, \apj, 635, 931

\bibitem[{{Cappellari} {et~al.}(2011){Cappellari}, {Emsellem}, {Krajnovi{\'c}},
  {McDermid}, {Scott}, {Verdoes Kleijn}, {Young}, {Alatalo}, {Bacon}, {Blitz},
  {Bois}, {Bournaud}, {Bureau}, {Davies}, {Davis}, {de Zeeuw}, {Duc},
  {Khochfar}, {Kuntschner}, {Lablanche}, {Morganti}, {Naab}, {Oosterloo},
  {Sarzi}, {Serra}, \& {Weijmans}}]{Atlas3D}
{Cappellari}, M., {Emsellem}, E., {Krajnovi{\'c}}, D., {et~al.} 2011, \mnras,
  413, 813

\bibitem[{{Casteels} {et~al.}(2013){Casteels}, {Bamford}, {Skibba}, {Masters},
  {Lintott}, {Keel}, {Schawinski}, {Nichol}, \& {Smith}}]{Casteels_et_al_2012}
{Casteels}, K. R.~V., {Bamford}, S.~P., {Skibba}, R.~A., {et~al.} 2013, \mnras,
  429, 1051

\bibitem[{{Chambers} {et~al.}(2016){Chambers}, {Magnier}, {Metcalfe},
  {Flewelling}, {Huber}, {Waters}, {Denneau}, {Draper}, {Farrow}, {Finkbeiner},
  {Holmberg}, {Koppenhoefer}, {Price}, {Rest}, {Saglia}, {Schlafly}, {Smartt},
  {Sweeney}, {Wainscoat}, {Burgett}, {Chastel}, {Grav}, {Heasley}, {Hodapp},
  {Jedicke}, {Kaiser}, {Kudritzki}, {Luppino}, {Lupton}, {Monet}, {Morgan},
  {Onaka}, {Shiao}, {Stubbs}, {Tonry}, {White}, {Ba{\~n}ados}, {Bell},
  {Bender}, {Bernard}, {Boegner}, {Boffi}, {Botticella}, {Calamida},
  {Casertano}, {Chen}, {Chen}, {Cole}, {Deacon}, {Frenk}, {Fitzsimmons},
  {Gezari}, {Gibbs}, {Goessl}, {Goggia}, {Gourgue}, {Goldman}, {Grant},
  {Grebel}, {Hambly}, {Hasinger}, {Heavens}, {Heckman}, {Henderson}, {Henning},
  {Holman}, {Hopp}, {Ip}, {Isani}, {Jackson}, {Keyes}, {Koekemoer}, {Kotak},
  {Le}, {Liska}, {Long}, {Lucey}, {Liu}, {Martin}, {Masci}, {McLean}, {Mindel},
  {Misra}, {Morganson}, {Murphy}, {Obaika}, {Narayan}, {Nieto-Santisteban},
  {Norberg}, {Peacock}, {Pier}, {Postman}, {Primak}, {Rae}, {Rai}, {Riess},
  {Riffeser}, {Rix}, {R{\"o}ser}, {Russel}, {Rutz}, {Schilbach}, {Schultz},
  {Scolnic}, {Strolger}, {Szalay}, {Seitz}, {Small}, {Smith}, {Soderblom},
  {Taylor}, {Thomson}, {Taylor}, {Thakar}, {Thiel}, {Thilker}, {Unger},
  {Urata}, {Valenti}, {Wagner}, {Walder}, {Walter}, {Watters}, {Werner},
  {Wood-Vasey}, \& {Wyse}}]{Chambers_et_al_2016}
{Chambers}, K.~C., {Magnier}, E.~A., {Metcalfe}, N., {et~al.} 2016, arXiv
  e-prints, arXiv:1612.05560

\bibitem[{{Cheng} {et~al.}(2021){Cheng}, {Huertas-Company}, {Conselice},
  {Arag{\'o}n-Salamanca}, {Robertson}, \& {Ramachandra}}]{Cheng_et_al_2021}
{Cheng}, T.-Y., {Huertas-Company}, M., {Conselice}, C.~J., {et~al.} 2021,
  \mnras, 503, 4446

\bibitem[{{Cole} {et~al.}(2000){Cole}, {Lacey}, {Baugh}, \&
  {Frenk}}]{Cole_et_al_2000}
{Cole}, S., {Lacey}, C.~G., {Baugh}, C.~M., \& {Frenk}, C.~S. 2000, \mnras,
  319, 168

\bibitem[{{Conselice}(2009)}]{Conselice_2009}
{Conselice}, C.~J. 2009, \mnras, 399, L16

\bibitem[{{Conselice} {et~al.}(2003){Conselice}, {Bershady}, {Dickinson}, \&
  {Papovich}}]{Conselice_et_al_2003}
{Conselice}, C.~J., {Bershady}, M.~A., {Dickinson}, M., \& {Papovich}, C. 2003,
  \aj, 126, 1183

\bibitem[{{Conselice} {et~al.}(2008){Conselice}, {Rajgor}, \&
  {Myers}}]{Conselice_et_al_2008}
{Conselice}, C.~J., {Rajgor}, S., \& {Myers}, R. 2008, \mnras, 386, 909

\bibitem[{{Cooper} {et~al.}(2010){Cooper}, {Cole}, {Frenk}, {White}, {Helly},
  {Benson}, {De Lucia}, {Helmi}, {Jenkins}, {Navarro}, {Springel}, \&
  {Wang}}]{Cooper_et_al_2010}
{Cooper}, A.~P., {Cole}, S., {Frenk}, C.~S., {et~al.} 2010, \mnras, 406, 744

\bibitem[{{Crain} {et~al.}(2015){Crain}, {Schaye}, {Bower}, {Furlong},
  {Schaller}, {Theuns}, {Dalla Vecchia}, {Frenk}, {McCarthy}, {Helly},
  {Jenkins}, {Rosas-Guevara}, {White}, \& {Trayford}}]{Crain_et_al_2015}
{Crain}, R.~A., {Schaye}, J., {Bower}, R.~G., {et~al.} 2015, \mnras, 450, 1937

\bibitem[{{Dieleman} {et~al.}(2015){Dieleman}, {Willett}, \&
  {Dambre}}]{Dieleman_et_al_2015}
{Dieleman}, S., {Willett}, K.~W., \& {Dambre}, J. 2015, \mnras, 450, 1441

\bibitem[{{Dom{\'\i}nguez S{\'a}nchez} {et~al.}(2018){Dom{\'\i}nguez
  S{\'a}nchez}, {Huertas-Company}, {Bernardi}, {Tuccillo}, \&
  {Fischer}}]{Dominguez-Sanchez_et_al_2018}
{Dom{\'\i}nguez S{\'a}nchez}, H., {Huertas-Company}, M., {Bernardi}, M.,
  {Tuccillo}, D., \& {Fischer}, J.~L. 2018, \mnras, 476, 3661

\bibitem[{{Duc}(2020)}]{Duc_2020}
{Duc}, P.-A. 2020, arXiv e-prints, arXiv:2007.13874

\bibitem[{{Duc} {et~al.}(2015){Duc}, {Cuillandre}, {Karabal}, {Cappellari},
  {Alatalo}, {Blitz}, {Bournaud}, {Bureau}, {Crocker}, {Davies}, {Davis}, {de
  Zeeuw}, {Emsellem}, {Khochfar}, {Krajnovi{\'c}}, {Kuntschner}, {McDermid},
  {Michel-Dansac}, {Morganti}, {Naab}, {Oosterloo}, {Paudel}, {Sarzi}, {Scott},
  {Serra}, {Weijmans}, \& {Young}}]{Duc_et_al_2015}
{Duc}, P.-A., {Cuillandre}, J.-C., {Karabal}, E., {et~al.} 2015, \mnras, 446,
  120

\bibitem[{{Ebrova}(2013)}]{Ebrova_2013}
{Ebrova}, I. 2013, arXiv e-prints, arXiv:1312.1643

\bibitem[{{Ebrov{\'a}} {et~al.}(2019){Ebrov{\'a}}, {B{\'\i}lek}, \&
  {Jungwiert}}]{shells_distance}
{Ebrov{\'a}}, I., {B{\'\i}lek}, M., \& {Jungwiert}, B. 2019, arXiv e-prints,
  arXiv:1909.07393

\bibitem[{{Ebrov{\'a}} {et~al.}(2021){Ebrov{\'a}}, {B{\'\i}lek},
  {Vudragovi{\'c}}, {Y{\i}ld{\i}z}, \& {Duc}}]{Ebrova_et_al_2021}
{Ebrov{\'a}}, I., {B{\'\i}lek}, M., {Vudragovi{\'c}}, A., {Y{\i}ld{\i}z},
  M.~K., \& {Duc}, P.-A. 2021, \aap, 650, A50

\bibitem[{{Elmegreen} {et~al.}(1993){Elmegreen}, {Kaufman}, \&
  {Thomasson}}]{Elmegreen_et_al_1993}
{Elmegreen}, B.~G., {Kaufman}, M., \& {Thomasson}, M. 1993, \apj, 412, 90

\bibitem[{{Ferguson} {et~al.}(2002){Ferguson}, {Irwin}, {Ibata}, {Lewis}, \&
  {Tanvir}}]{Ferguson_et_al_2002}
{Ferguson}, A. M.~N., {Irwin}, M.~J., {Ibata}, R.~A., {Lewis}, G.~F., \&
  {Tanvir}, N.~R. 2002, \aj, 124, 1452

\bibitem[{{Fernique} {et~al.}(2015){Fernique}, {Allen}, {Boch}, {Oberto},
  {Pineau}, {Durand}, {Bot}, {Cambr{\'e}sy}, {Derriere}, {Genova}, \&
  {Bonnarel}}]{hips}
{Fernique}, P., {Allen}, M.~G., {Boch}, T., {et~al.} 2015, \aap, 578, A114

\bibitem[{{Ferrarese} {et~al.}(2012){Ferrarese}, {C{\^o}t{\'e}}, {Cuillandre},
  {Gwyn}, {Peng}, {MacArthur}, {Duc}, {Boselli}, {Mei}, {Erben}, {McConnachie},
  {Durrell}, {Mihos}, {Jord{\'a}n}, {Lan{\c{c}}on}, {Puzia}, {Emsellem},
  {Balogh}, {Blakeslee}, {van Waerbeke}, {Gavazzi}, {Vollmer}, {Kavelaars},
  {Woods}, {Ball}, {Boissier}, {Courteau}, {Ferriere}, {Gavazzi},
  {Hildebrandt}, {Hudelot}, {Huertas-Company}, {Liu}, {McLaughlin}, {Mellier},
  {Milkeraitis}, {Schade}, {Balkowski}, {Bournaud}, {Carlberg}, {Chapman},
  {Hoekstra}, {Peng}, {Sawicki}, {Simard}, {Taylor}, {Tully}, {van Driel},
  {Wilson}, {Burdullis}, {Mahoney}, \& {Manset}}]{Ferrarese_et_al_2012}
{Ferrarese}, L., {C{\^o}t{\'e}}, P., {Cuillandre}, J.-C., {et~al.} 2012, \apjs,
  200, 4

\bibitem[{{Ferreira} {et~al.}(2020){Ferreira}, {Conselice}, {Duncan}, {Cheng},
  {Griffiths}, \& {Whitney}}]{Ferreira_et_al_2020}
{Ferreira}, L., {Conselice}, C.~J., {Duncan}, K., {et~al.} 2020, \apj, 895, 115

\bibitem[{{Gibbons} {et~al.}(2014){Gibbons}, {Belokurov}, \&
  {Evans}}]{warp_stream}
{Gibbons}, S.~L.~J., {Belokurov}, V., \& {Evans}, N.~W. 2014, \mnras, 445, 3788

\bibitem[{{Gilbert} {et~al.}(2012){Gilbert}, {Guhathakurta}, {Beaton},
  {Bullock}, {Geha}, {Kalirai}, {Kirby}, {Majewski}, {Ostheimer}, {Patterson},
  {Tollerud}, {Tanaka}, \& {Chiba}}]{Gilbert_et_al_2012}
{Gilbert}, K.~M., {Guhathakurta}, P., {Beaton}, R.~L., {et~al.} 2012, \apj,
  760, 76

\bibitem[{{Helmi} \& {White}(1999)}]{Helmi_and_White_1999}
{Helmi}, A. \& {White}, S. D.~M. 1999, \mnras, 307, 495

\bibitem[{{Hendel} \& {Johnston}(2015)}]{Hendel_and_Johnston_2015}
{Hendel}, D. \& {Johnston}, K.~V. 2015, \mnras, 454, 2472

\bibitem[{Hennig(2007)}]{jaccard_index}
Hennig, C. 2007, Comput. Statist. Data Analysis, 52, 258

\bibitem[{{Hernquist} \& {Quinn}(1987)}]{Hernquist_et_al_1987}
{Hernquist}, L. \& {Quinn}, P.~J. 1987, \apj, 312, 1

\bibitem[{{Huertas-Company} {et~al.}(2015){Huertas-Company}, {Gravet},
  {Cabrera-Vives}, {P{\'e}rez-Gonz{\'a}lez}, {Kartaltepe}, {Barro}, {Bernardi},
  {Mei}, {Shankar}, {Dimauro}, {Bell}, {Kocevski}, {Koo}, {Faber}, \&
  {Mcintosh}}]{Huertas-Company_et_al_2015}
{Huertas-Company}, M., {Gravet}, R., {Cabrera-Vives}, G., {et~al.} 2015, \apjs,
  221, 8

\bibitem[{{Ibata} {et~al.}(2001){Ibata}, {Irwin}, {Lewis}, {Ferguson}, \&
  {Tanvir}}]{Ibata_et_al_2001}
{Ibata}, R., {Irwin}, M., {Lewis}, G., {Ferguson}, A. M.~N., \& {Tanvir}, N.
  2001, \nat, 412, 49

\bibitem[{{Ibata} {et~al.}(1994){Ibata}, {Gilmore}, \&
  {Irwin}}]{Ibata_et_al_1994}
{Ibata}, R.~A., {Gilmore}, G., \& {Irwin}, M.~J. 1994, \nat, 370, 194

\bibitem[{{Ibata} {et~al.}(2017){Ibata}, {McConnachie}, {Cuillandre}, {Fantin},
  {Haywood}, {Martin}, {Bergeron}, {Beckmann}, {Bernard}, {Bonifacio},
  {Caffau}, {Carlberg}, {C{\^o}t{\'e}}, {Cabanac}, {Chapman}, {Duc}, {Durret},
  {Famaey}, {Fabbro}, {Gwyn}, {Hammer}, {Hill}, {Hudson}, {Lan{\c{c}}on},
  {Lewis}, {Malhan}, {di Matteo}, {McCracken}, {Mei}, {Mellier}, {Navarro},
  {Pires}, {Pritchet}, {Reyl{\'e}}, {Richer}, {Robin}, {S{\'a}nchez-Janssen},
  {Sawicki}, {Scott}, {Scottez}, {Spekkens}, {Starkenburg}, {Thomas}, \&
  {Venn}}]{Ibata_et_al_2017}
{Ibata}, R.~A., {McConnachie}, A., {Cuillandre}, J.-C., {et~al.} 2017, \apj,
  848, 128

\bibitem[{{Iodice} {et~al.}(2021){Iodice}, {Spavone}, {Capaccioli}, {Schipani},
  {Arnaboldi}, {Cantiello}, {D'Ago}, {De Cicco}, {Forbes}, {Greggio},
  {Krajnovi{\'c}}, {La Marca}, {Napolitano}, {Paolillo}, {Ragusa}, {Raj},
  {Rampazzo}, \& {Rejkuba}}]{Iodice_et_al_2021}
{Iodice}, E., {Spavone}, M., {Capaccioli}, M., {et~al.} 2021, The Messenger,
  183, 25

\bibitem[{{Jackson} {et~al.}(2021){Jackson}, {Pasquali}, {La Barbera}, {More},
  \& {Grebel}}]{Jackson_et_al_2021}
{Jackson}, T.~M., {Pasquali}, A., {La Barbera}, F., {More}, S., \& {Grebel},
  E.~K. 2021, arXiv e-prints, arXiv:2102.02241

\bibitem[{{Jarrett} {et~al.}(2006){Jarrett}, {Polletta}, {Fournon}, {Stacey},
  {Xu}, {Siana}, {Farrah}, {Berta}, {Hatziminaoglou}, {Rodighiero}, {Surace},
  {Domingue}, {Shupe}, {Fang}, {Lonsdale}, {Oliver}, {Rowan-Robinson}, {Smith},
  {Babbedge}, {Gonzalez-Solares}, {Masci}, {Franceschini}, \&
  {Padgett}}]{star_formation_tails}
{Jarrett}, T.~H., {Polletta}, M., {Fournon}, I.~P., {et~al.} 2006, \aj, 131,
  261

\bibitem[{{Javanmardi} {et~al.}(2016){Javanmardi}, {Martinez-Delgado},
  {Kroupa}, {Henkel}, {Crawford}, {Teuwen}, {Gabany}, {Hanson}, {Chonis}, \&
  {Neyer}}]{Javanmardi_et_al_2016}
{Javanmardi}, B., {Martinez-Delgado}, D., {Kroupa}, P., {et~al.} 2016, \aap,
  588, A89

\bibitem[{{Ji} {et~al.}(2014){Ji}, {Peirani}, \& {Yi}}]{Ji_et_al_2014}
{Ji}, I., {Peirani}, S., \& {Yi}, S.~K. 2014, \aap, 566, A97

\bibitem[{{Johnston}(1998)}]{Johnston_1998}
{Johnston}, K.~V. 1998, in Astronomical Society of the Pacific Conference
  Series, Vol. 136, Galactic Halos, ed. D.~{Zaritsky}, 365

\bibitem[{{Johnston} {et~al.}(2008){Johnston}, {Bullock}, {Sharma}, {Font},
  {Robertson}, \& {Leitner}}]{Johnston_et_al_2008}
{Johnston}, K.~V., {Bullock}, J.~S., {Sharma}, S., {et~al.} 2008, \apj, 689,
  936

\bibitem[{{Johnston} {et~al.}(1996){Johnston}, {Hernquist}, \&
  {Bolte}}]{Johnston_et_al_1996}
{Johnston}, K.~V., {Hernquist}, L., \& {Bolte}, M. 1996, \apj, 465, 278

\bibitem[{{Kado-Fong} {et~al.}(2018){Kado-Fong}, {Greene}, {Hendel},
  {Price-Whelan}, {Greco}, {Goulding}, {Huang}, {Johnston}, {Komiyama}, {Lee},
  {Lust}, {Strauss}, \& {Tanaka}}]{Kado-Fong_et_al_2018}
{Kado-Fong}, E., {Greene}, J.~E., {Hendel}, D., {et~al.} 2018, \apj, 866, 103

\bibitem[{{Karabal} {et~al.}(2017){Karabal}, {Duc}, {Kuntschner}, {Chanial},
  {Cuillandre}, \& {Gwyn}}]{Karabal_et_al_2017}
{Karabal}, E., {Duc}, P.~A., {Kuntschner}, H., {et~al.} 2017, \aap, 601, A86

\bibitem[{{Karademir} {et~al.}(2019){Karademir}, {Remus}, {Burkert}, {Dolag},
  {Hoffmann}, {Moster}, {Steinwandel}, \& {Zhang}}]{shells_distance2}
{Karademir}, G.~S., {Remus}, R.-S., {Burkert}, A., {et~al.} 2019, \mnras, 487,
  318

\bibitem[{{Kauffmann} {et~al.}(1993){Kauffmann}, {White}, \&
  {Guiderdoni}}]{Kauffman_et_al_1993}
{Kauffmann}, G., {White}, S.~D.~M., \& {Guiderdoni}, B. 1993, \mnras, 264, 201

\bibitem[{Kawinwanichakij {et~al.}(2014)Kawinwanichakij, Papovich, Quadri,
  Tran, Spitler, Kacprzak, Labbé, Straatman, Glazebrook, Allen, \&
  et~al.}]{Kawinwanichakij_et_al_2014}
Kawinwanichakij, L., Papovich, C., Quadri, R.~F., {et~al.} 2014, The
  Astrophysical Journal, 792, 103

\bibitem[{{Kluge} {et~al.}(2020){Kluge}, {Neureiter}, {Riffeser}, {Bender},
  {Goessl}, {Hopp}, {Schmidt}, {Ries}, \& {Brosch}}]{Kluge_et_al_2020}
{Kluge}, M., {Neureiter}, B., {Riffeser}, A., {et~al.} 2020, \apjs, 247, 43

\bibitem[{Koch \& Rosolowsky(2015)}]{filfinder}
Koch, E. \& Rosolowsky, E. 2015, \mnras, 452, 3435

\bibitem[{{Lotz} {et~al.}(2004){Lotz}, {Primack}, \& {Madau}}]{Lotz_2004}
{Lotz}, J.~M., {Primack}, J., \& {Madau}, P. 2004, \aj, 128, 163

\bibitem[{{Lux} {et~al.}(2013){Lux}, {Read}, {Lake}, \&
  {Johnston}}]{Lux_et_al_2013}
{Lux}, H., {Read}, J.~I., {Lake}, G., \& {Johnston}, K.~V. 2013, \mnras, 436,
  2386

\bibitem[{{Malhan} {et~al.}(2018){Malhan}, {Ibata}, \&
  {Martin}}]{Malhan_et_al_2018}
{Malhan}, K., {Ibata}, R.~A., \& {Martin}, N.~F. 2018, \mnras, 481, 3442

\bibitem[{{Mancillas} {et~al.}(2019){Mancillas}, {Duc}, {Combes}, {Bournaud},
  {Emsellem}, {Martig}, \& {Michel-Dansac}}]{Mancillas_et_al_2019}
{Mancillas}, B., {Duc}, P.-A., {Combes}, F., {et~al.} 2019, \aap, 632, A122

\bibitem[{{Mantha} {et~al.}(2019){Mantha}, {McIntosh}, {Ciaschi}, {Evan},
  {Ferguson}, {Fries}, {Guo}, {Koekemoer}, {Landry}, {McGrath}, {Simons},
  {Snyder}, {Thompson}, {Bell}, {Ceverino}, {Hathi}, {Pacifici}, {Primack},
  {Rafelski}, \& {Rodriguez-Gomez}}]{Mantha_et_al_2019}
{Mantha}, K.~B., {McIntosh}, D.~H., {Ciaschi}, C.~P., {et~al.} 2019, \mnras,
  486, 2643

\bibitem[{{Martin} {et~al.}(2022){Martin}, {Bazkiaei}, {Spavone}, {Iodice},
  {Mihos}, \& {others}}]{Martin_et_al_2022}
{Martin}, G., {Bazkiaei}, A., {Spavone}, M., {et~al.} 2022, \mnras

\bibitem[{{Martin} {et~al.}(2020){Martin}, {Kaviraj}, {Hocking}, {Read}, \&
  {Geach}}]{Martin_et_al_2020}
{Martin}, G., {Kaviraj}, S., {Hocking}, A., {Read}, S.~C., \& {Geach}, J.~E.
  2020, \mnras, 491, 1408

\bibitem[{{Martin} {et~al.}(2014){Martin}, {Ibata}, {Rich}, {Collins},
  {Fardal}, {Irwin}, {Lewis}, {McConnachie}, {Babul}, {Bate}, {Chapman},
  {Conn}, {Crnojevi{\'c}}, {Ferguson}, {Mackey}, {Navarro}, {Pe{\~n}arrubia},
  {Tanvir}, \& {Valls-Gabaud}}]{Martin_et_al_2014}
{Martin}, N.~F., {Ibata}, R.~A., {Rich}, R.~M., {et~al.} 2014, \apj, 787, 19

\bibitem[{{Martinez-Delgado} {et~al.}(2021){Martinez-Delgado}, {Cooper},
  {Roman}, {Pillepich}, {Erkal}, {Pearson}, {Moustakas}, {Laporte}, {Laine},
  {Akhlaghi}, {Lang}, {Makarov}, {Borlaff}, {Donatiello}, {Pearson},
  {Miro-Carretero}, {Cuillandre}, {Dominguez}, {Roca-Fabrega}, {Frenk},
  {Schmidt}, {Gomez-Flechoso}, {Guzman}, {Libeskind}, {Dey}, {Weaver},
  {Schlegel}, {Myers}, \& {Valdes}}]{SSLS}
{Martinez-Delgado}, D., {Cooper}, A.~P., {Roman}, J., {et~al.} 2021, arXiv
  e-prints, arXiv:2104.06071

\bibitem[{{Mart{\'\i}nez-Delgado} {et~al.}(2010){Mart{\'\i}nez-Delgado},
  {Gabany}, {Crawford}, {Zibetti}, {Majewski}, {Rix}, {Fliri},
  {Carballo-Bello}, {Bardalez-Gagliuffi}, {Pe{\~n}arrubia}, {Chonis}, {Madore},
  {Trujillo}, {Schirmer}, \& {McDavid}}]{Martinez_Delgado_et_al_2010}
{Mart{\'\i}nez-Delgado}, D., {Gabany}, R.~J., {Crawford}, K., {et~al.} 2010,
  \aj, 140, 962

\bibitem[{{Masters} {et~al.}(2021){Masters}, {Krawczyk}, {Shamsi}, {Todd},
  {Finnegan}, {Bershady}, {Bundy}, {Cherinka}, {Fraser-McKelvie}, {Krishnarao},
  {Kruk}, {Lane}, {Law}, {Lintott}, {Merrifield}, {Simmons}, {Weijmans}, \&
  {Yan}}]{Masters_et_al_2021}
{Masters}, K.~L., {Krawczyk}, C., {Shamsi}, S., {et~al.} 2021, \mnras, 507,
  3923

\bibitem[{{McConnachie} {et~al.}(2009){McConnachie}, {Irwin}, {Ibata},
  {Dubinski}, {Widrow}, {Martin}, {C{\^o}t{\'e}}, {Dotter}, {Navarro},
  {Ferguson}, {Puzia}, {Lewis}, {Babul}, {Barmby}, {Bienaym{\'e}}, {Chapman},
  {Cockcroft}, {Collins}, {Fardal}, {Harris}, {Huxor}, {Mackey},
  {Pe{\~n}arrubia}, {Rich}, {Richer}, {Siebert}, {Tanvir}, {Valls-Gabaud}, \&
  {Venn}}]{McConnachie_et_al_2009}
{McConnachie}, A.~W., {Irwin}, M.~J., {Ibata}, R.~A., {et~al.} 2009, \nat, 461,
  66

\bibitem[{{McIntosh} {et~al.}(2008){McIntosh}, {Guo}, {Hertzberg}, {Katz},
  {Mo}, {van den Bosch}, \& {Yang}}]{McIntosh_et_al_2008}
{McIntosh}, D.~H., {Guo}, Y., {Hertzberg}, J., {et~al.} 2008, \mnras, 388, 1537

\bibitem[{{Merritt} {et~al.}(2020){Merritt}, {Pillepich}, {van Dokkum},
  {Nelson}, {Hernquist}, {Marinacci}, \& {Vogelsberger}}]{Merritt_et_al_2020}
{Merritt}, A., {Pillepich}, A., {van Dokkum}, P., {et~al.} 2020, \mnras, 495,
  4570

\bibitem[{{Mihos}(1995)}]{Mihos_1995}
{Mihos}, J.~C. 1995, \apjl, 438, L75

\bibitem[{{Mihos} {et~al.}(2015){Mihos}, {Durrell}, {Ferrarese}, {Feldmeier},
  {C{\^o}t{\'e}}, {Peng}, {Harding}, {Liu}, {Gwyn}, \&
  {Cuillandre}}]{Mihos_et_al_2015}
{Mihos}, J.~C., {Durrell}, P.~R., {Ferrarese}, L., {et~al.} 2015, \apjl, 809,
  L21

\bibitem[{{Morales} {et~al.}(2018){Morales}, {Mart{\'\i}nez-Delgado}, {Grebel},
  {Cooper}, {Javanmardi}, \& {Miskolczi}}]{Morales_et_al_2018}
{Morales}, G., {Mart{\'\i}nez-Delgado}, D., {Grebel}, E.~K., {et~al.} 2018,
  \aap, 614, A143

\bibitem[{{Olson} \& {Kwan}(1990)}]{star_formation_tails2}
{Olson}, K.~M. \& {Kwan}, J. 1990, \apj, 361, 426

\bibitem[{{Oser} {et~al.}(2010){Oser}, {Ostriker}, {Naab}, {Johansson}, \&
  {Burkert}}]{Oser_et_al_2010}
{Oser}, L., {Ostriker}, J.~P., {Naab}, T., {Johansson}, P.~H., \& {Burkert}, A.
  2010, \apj, 725, 2312

\bibitem[{{Pawlik} {et~al.}(2016){Pawlik}, {Wild}, {Walcher}, {Johansson},
  {Villforth}, {Rowlands}, {Mendez-Abreu}, \& {Hewlett}}]{Pawlik_et_al_2016}
{Pawlik}, M.~M., {Wild}, V., {Walcher}, C.~J., {et~al.} 2016, \mnras, 456, 3032

\bibitem[{{Pearson} {et~al.}(2019){Pearson}, {Wang}, {Trayford}, {Petrillo}, \&
  {van der Tak}}]{Pearson_et_al_2019}
{Pearson}, W.~J., {Wang}, L., {Trayford}, J.~W., {Petrillo}, C.~E., \& {van der
  Tak}, F.~F.~S. 2019, \aap, 626, A49

\bibitem[{{Peng} {et~al.}(2002){Peng}, {Ho}, {Impey}, \&
  {Rix}}]{Peng_et_al_2002}
{Peng}, C.~Y., {Ho}, L.~C., {Impey}, C.~D., \& {Rix}, H.-W. 2002, \aj, 124, 266

\bibitem[{{Pillepich} {et~al.}(2018){Pillepich}, {Springel}, {Nelson}, {Genel},
  {Naiman}, {Pakmor}, {Hernquist}, {Torrey}, {Vogelsberger}, {Weinberger}, \&
  {Marinacci}}]{Pillepich_et_al_2018}
{Pillepich}, A., {Springel}, V., {Nelson}, D., {et~al.} 2018, \mnras, 473, 4077

\bibitem[{{Pop} {et~al.}(2018){Pop}, {Pillepich}, {Amorisco}, \&
  {Hernquist}}]{Pop_et_al_2018}
{Pop}, A.-R., {Pillepich}, A., {Amorisco}, N.~C., \& {Hernquist}, L. 2018,
  \mnras, 480, 1715

\bibitem[{{Prieur}(1990)}]{Prieur_1990}
{Prieur}, J.~L. 1990, {Status of shell galaxies.}, ed. R.~{Wielen}, 72--83

\bibitem[{{Quinn}(1984)}]{Quinn_1984}
{Quinn}, P.~J. 1984, \apj, 279, 596

\bibitem[{{Richards} {et~al.}(2020){Richards}, {Paiement}, {Xie}, {Sola}, \&
  {Duc}}]{Richards_et_al_2021}
{Richards}, F., {Paiement}, A., {Xie}, X., {Sola}, E., \& {Duc}, P.-A. 2020,
  arXiv e-prints, arXiv:2011.11734

\bibitem[{{Schaye} {et~al.}(2015){Schaye}, {Crain}, {Bower}, {Furlong},
  {Schaller}, {Theuns}, {Dalla Vecchia}, {Frenk}, {McCarthy}, {Helly},
  {Jenkins}, {Rosas-Guevara}, {White}, {Baes}, {Booth}, {Camps}, {Navarro},
  {Qu}, {Rahmati}, {Sawala}, {Thomas}, \& {Trayford}}]{Schaye_et_al_2015}
{Schaye}, J., {Crain}, R.~A., {Bower}, R.~G., {et~al.} 2015, \mnras, 446, 521

\bibitem[{{Spindler} {et~al.}(2021){Spindler}, {Geach}, \&
  {Smith}}]{Spindler_et_al_2021}
{Spindler}, A., {Geach}, J.~E., \& {Smith}, M.~J. 2021, \mnras, 502, 985

\bibitem[{{Spitler} {et~al.}(2019){Spitler}, {Longbottom}, {Alvarado-Montes},
  {Bazkiaei}, {Caddy}, {Gee}, {Horton}, {Lee}, \& {Prole}}]{Spitler_et_al_2019}
{Spitler}, L.~R., {Longbottom}, F.~D., {Alvarado-Montes}, J.~A., {et~al.} 2019,
  arXiv e-prints, arXiv:1911.11579

\bibitem[{{Stringer} \& {Benson}(2007)}]{Stringer_and_Benson_2007}
{Stringer}, M.~J. \& {Benson}, A.~J. 2007, \mnras, 382, 641

\bibitem[{{Tal} {et~al.}(2009){Tal}, {van Dokkum}, {Nelan}, \&
  {Bezanson}}]{Tal_et_al_2009}
{Tal}, T., {van Dokkum}, P.~G., {Nelan}, J., \& {Bezanson}, R. 2009, \aj, 138,
  1417

\bibitem[{{Tohill} {et~al.}(2021){Tohill}, {Ferreira}, {Conselice}, {Bamford},
  \& {Ferrari}}]{Tohill_et_al_2021}
{Tohill}, C., {Ferreira}, L., {Conselice}, C.~J., {Bamford}, S.~P., \&
  {Ferrari}, F. 2021, \apj, 916, 4

\bibitem[{{Toomre} \& {Toomre}(1972)}]{Toomre_and_Toomre_1972}
{Toomre}, A. \& {Toomre}, J. 1972, \apj, 178, 623

\bibitem[{{Uzeirbegovic} {et~al.}(2020){Uzeirbegovic}, {Geach}, \&
  {Kaviraj}}]{Uzeirbegovic_et_al_2020}
{Uzeirbegovic}, E., {Geach}, J.~E., \& {Kaviraj}, S. 2020, \mnras, 498, 4021

\bibitem[{{van der Walt} {et~al.}(2014){van der Walt}, {Sch{\"o}nberger},
  {Nunez-Iglesias}, {Boulogne}, {Warner}, {Yager}, {Gouillart}, {Yu}, \&
  {scikit-image contributors}}]{scikit-image}
{van der Walt}, S., {Sch{\"o}nberger}, J.~L., {Nunez-Iglesias}, J., {et~al.}
  2014, arXiv e-prints, arXiv:1407.6245

\bibitem[{{van Dokkum} {et~al.}(2014){van Dokkum}, {Abraham}, \&
  {Merritt}}]{van_Dokkum_et_al_2014}
{van Dokkum}, P.~G., {Abraham}, R., \& {Merritt}, A. 2014, \apjl, 782, L24

\bibitem[{{Vega-Ferrero} {et~al.}(2021){Vega-Ferrero}, {Dom{\'\i}nguez
  S{\'a}nchez}, {Bernardi}, {Huertas-Company}, {Morgan}, {Margalef}, {Aguena},
  {Allam}, {Annis}, {Avila}, {Bacon}, {Bertin}, {Brooks}, {Carnero Rosell},
  {Carrasco Kind}, {Carretero}, {Choi}, {Conselice}, {Costanzi}, {da Costa},
  {Pereira}, {De Vicente}, {Desai}, {Ferrero}, {Fosalba}, {Frieman},
  {Garc{\'\i}a-Bellido}, {Gruen}, {Gruendl}, {Gschwend}, {Gutierrez},
  {Hartley}, {Hinton}, {Hollowood}, {Honscheid}, {Hoyle}, {Jarvis}, {Kim},
  {Kuehn}, {Kuropatkin}, {Lima}, {Maia}, {Menanteau}, {Miquel}, {Ogando},
  {Palmese}, {Paz-Chinch{\'o}n}, {Plazas}, {Romer}, {Sanchez}, {Scarpine},
  {Schubnell}, {Serrano}, {Sevilla-Noarbe}, {Smith}, {Suchyta}, {Swanson},
  {Tarle}, {Tarsitano}, {To}, {Tucker}, {Varga}, \&
  {Wilkinson}}]{Ferrero_et_al_2020}
{Vega-Ferrero}, J., {Dom{\'\i}nguez S{\'a}nchez}, H., {Bernardi}, M., {et~al.}
  2021, \mnras, 506, 1927

\bibitem[{{Venhola} {et~al.}(2017){Venhola}, {Peletier}, {Laurikainen}, {Salo},
  {Lisker}, {Iodice}, {Capaccioli}, {Verdois Kleijn}, {Valentijn}, {Mieske},
  {Hilker}, {Wittmann}, {van de Ven}, {Grado}, {Spavone}, {Cantiello},
  {Napolitano}, {Paolillo}, \& {Falc{\'o}n-Barroso}}]{Venhola_et_al_2017}
{Venhola}, A., {Peletier}, R., {Laurikainen}, E., {et~al.} 2017, \aap, 608,
  A142

\bibitem[{{Walmsley} {et~al.}(2019){Walmsley}, {Ferguson}, {Mann}, \&
  {Lintott}}]{Walmsley_et_al_2018}
{Walmsley}, M., {Ferguson}, A. M.~N., {Mann}, R.~G., \& {Lintott}, C.~J. 2019,
  \mnras, 483, 2968

\bibitem[{{Wen} {et~al.}(2014){Wen}, {Zheng}, \& {An}}]{Wen_et_al_2014}
{Wen}, Z.~Z., {Zheng}, X.~Z., \& {An}, F.~X. 2014, \apj, 787, 130

\bibitem[{{Wilkinson} {et~al.}(1987){Wilkinson}, {Sparks}, {Carter}, \&
  {Malin}}]{Wilkinson_et_al_1987}
{Wilkinson}, A., {Sparks}, W.~B., {Carter}, D., \& {Malin}, D.~A. 1987, in
  Structure and Dynamics of Elliptical Galaxies, ed. P.~T. {de Zeeuw}, Vol.
  127, 465

\end{thebibliography}

\begin{appendix}
\section{Individual shapes of tidal tails and streams}\label{section:appendix_tail_streams}
\begin{figure*}
   \centering
   \includegraphics[width=\linewidth]{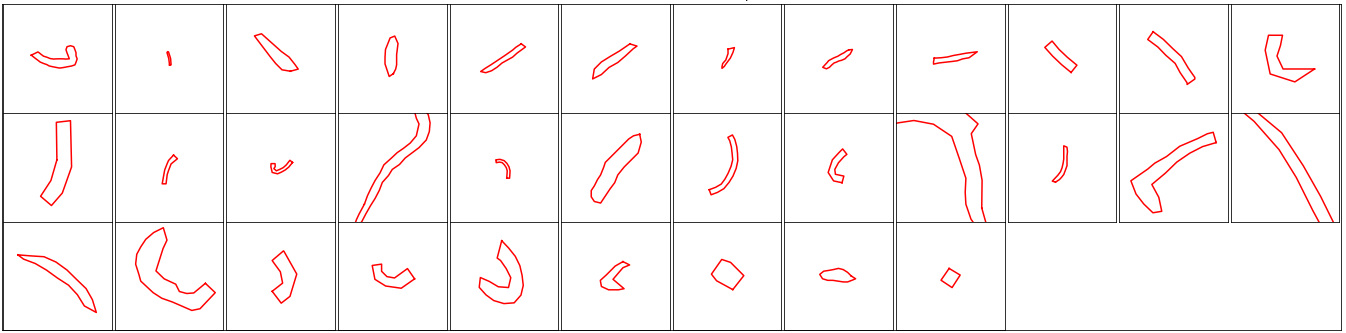}
    \hfill
    \includegraphics[width=\linewidth]{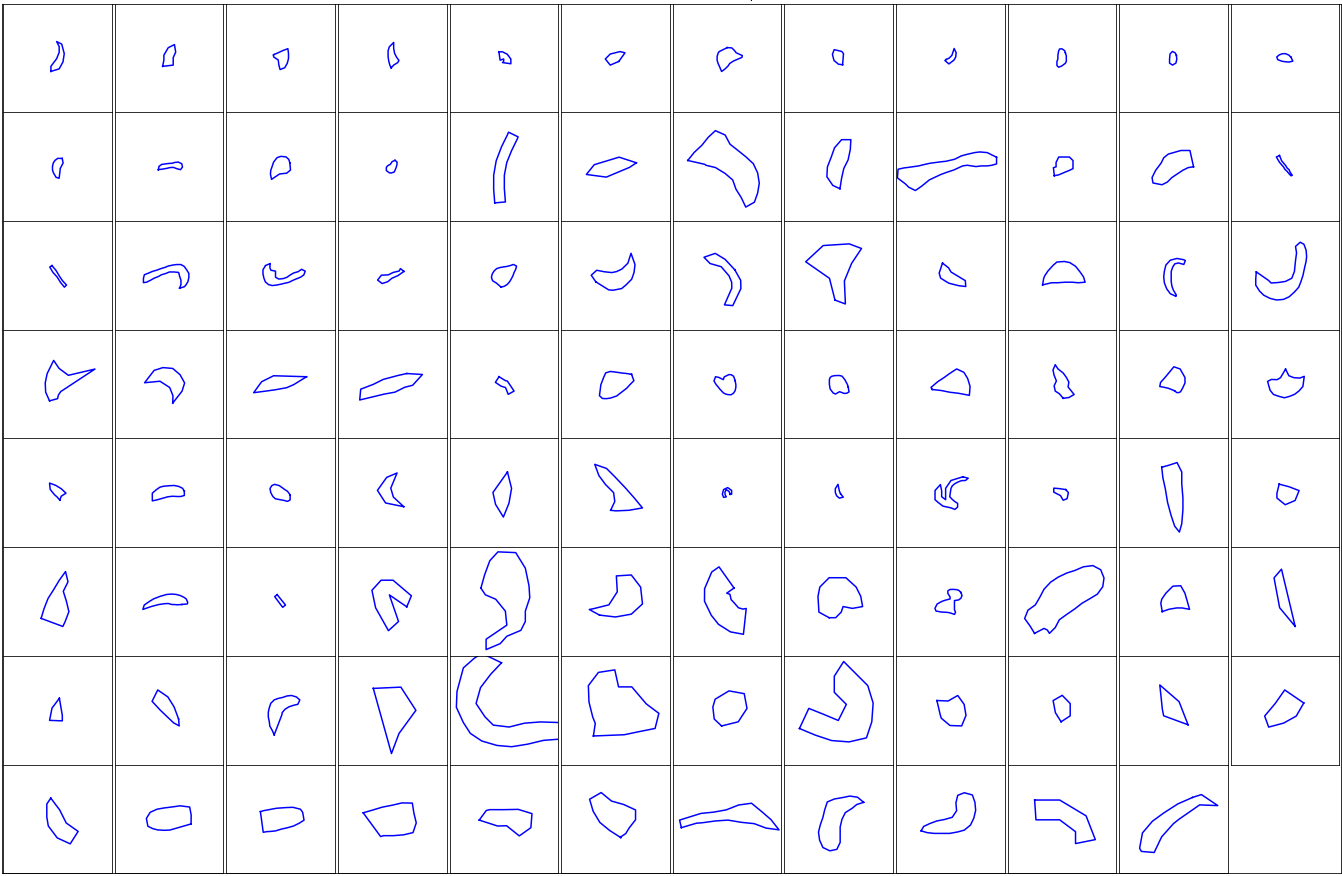}
      \caption{Thumbnails of the streams (\textit{top, in red}) and tidal tails (\textit{bottom, in blue}) identified around the LTGs, plotted in boxes of $50\times50$ kpc. They are sorted by increasing mass of the host galaxy, starting from the top-left  for the lightest LTG to the bottom-right for the most massive LTG.}
         \label{fig:indidivual_stream_tails_LTG}
\end{figure*}

\begin{figure*}
   \centering
   \includegraphics[width=\linewidth]{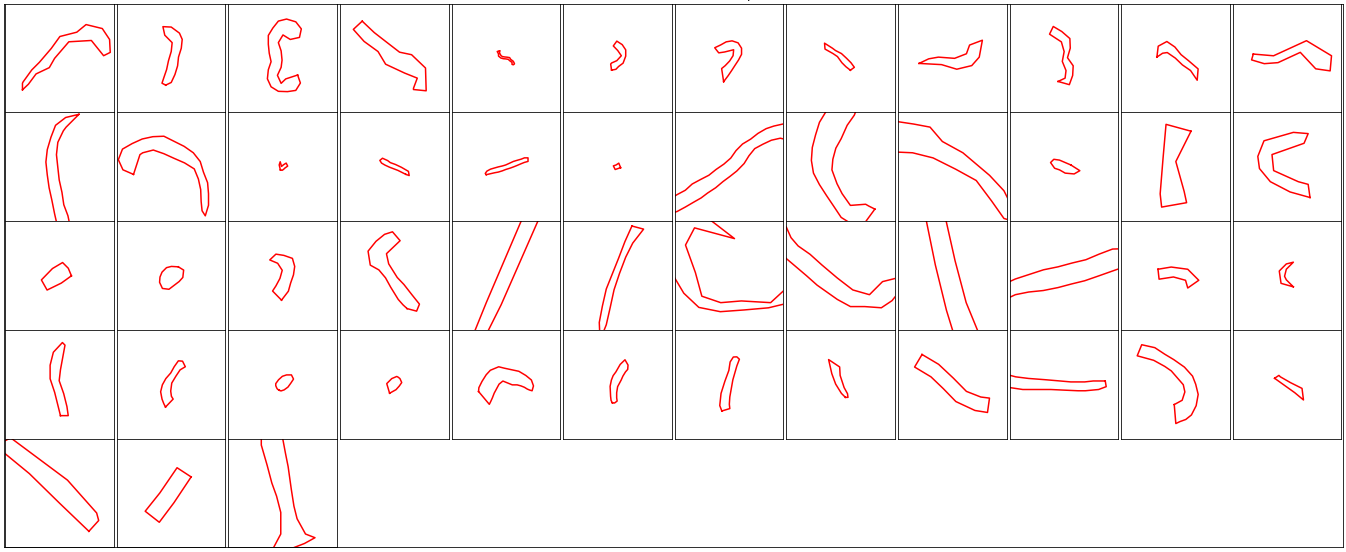}
    \hfill
    \includegraphics[width=\linewidth]{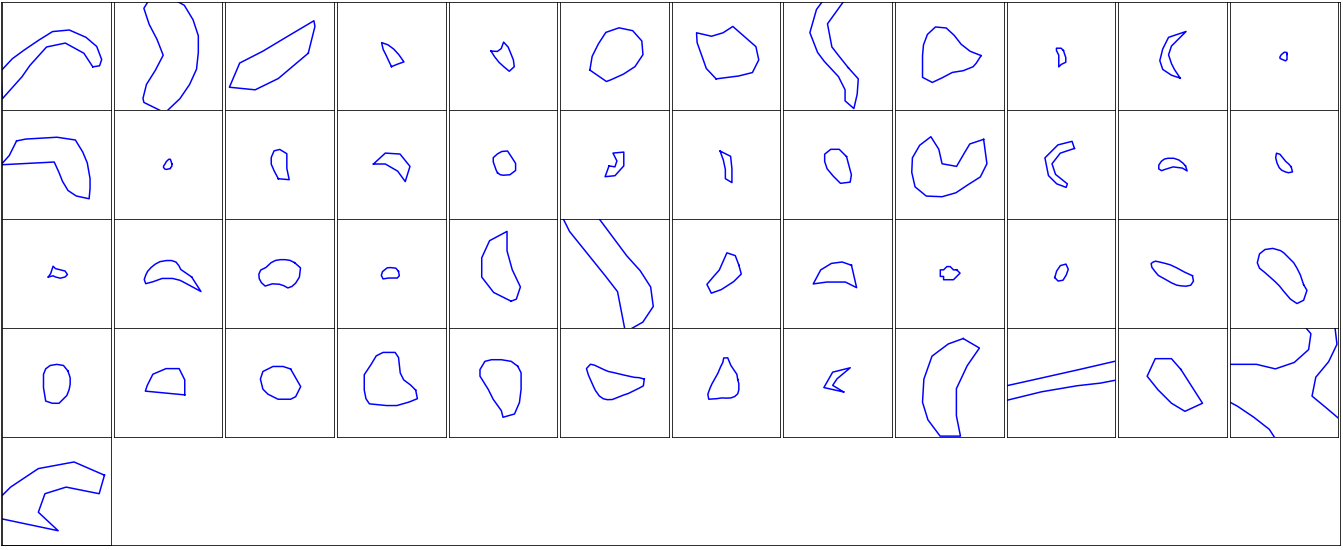}
      \caption{Thumbnails of the streams (\textit{(top, in red)}) and tidal tails (\textit{bottom, in blue}) identified around the ETGs, plotted in boxes of $50\times50$ kpc. They are sorted by increasing mass of the host galaxy, starting from the top-left  for the lightest ETG to the bottom-right for the most massive ETG.}
         \label{fig:indidivual_stream_tails_ETG}
    
\end{figure*}

In this section, we present the individual thumbnails of tidal tails (in blue) and streams (in red) identified around LTGs (Figure \ref{fig:indidivual_stream_tails_LTG}) and ETGs (Figure \ref{fig:indidivual_stream_tails_ETG}). The galaxies are sorted by increasing mass, from the top-left for the less massive one to the bottom-right for the most massive one.

For the streams of LTGs (Figure \ref{fig:indidivual_stream_tails_LTG}), the mean galaxy mass per each row is respectively $1.5\times10^{10}$, $4.6\times10^{10}$ and $1.1\times10^{11} M_\odot$  for the first, second and third rows. For the tails of LTGs, the mean mass per row is: $8.5\times10^{9}$, $1.4\times10^{10}$, $2.3\times10^{10}$, $3.2\times10^{10}$, $4.9\times10^{10}$, $6.9\times10^{10}$, $7.8\times10^{10}$  and finally $1.1\times10^{11} M_\odot$.

Likewise, the mean galaxy mass per row for the streams of ETG (Figure \ref{fig:indidivual_stream_tails_ETG}) are $1.3\times10^{10}$, $3.1\times10^{10}$, $6.2\times10^{9}$, $1.0\times10^{11}$ and $1.8\times10^{11}M_\odot$. For the tails of ETGs, this progression per row is: $1.9\times10^{10}$, $3.3\times10^{10}$, $5.3\times10^{10}$, $1.1\times10^{11}$ and $1.5\times10^{11}M_\odot$. 

\section{Normalized length and width of tidal tails} \label{section:normalized_length_width_tails}

In this section, we present the histograms of the length and width of tidal tails normalized by the effective radius of their host galaxy, and as a function of the morphological type. One can see that the trends are similar to the ones presented in section \ref{section:length_width_tails_streams}, i.e. the distribution of the width is flatter for ETGs than for LTGs. The length of the tidal tails for ETGs is slightly longer than for LTGs.

\begin{figure*}
   \centering
   \includegraphics[width=\linewidth]{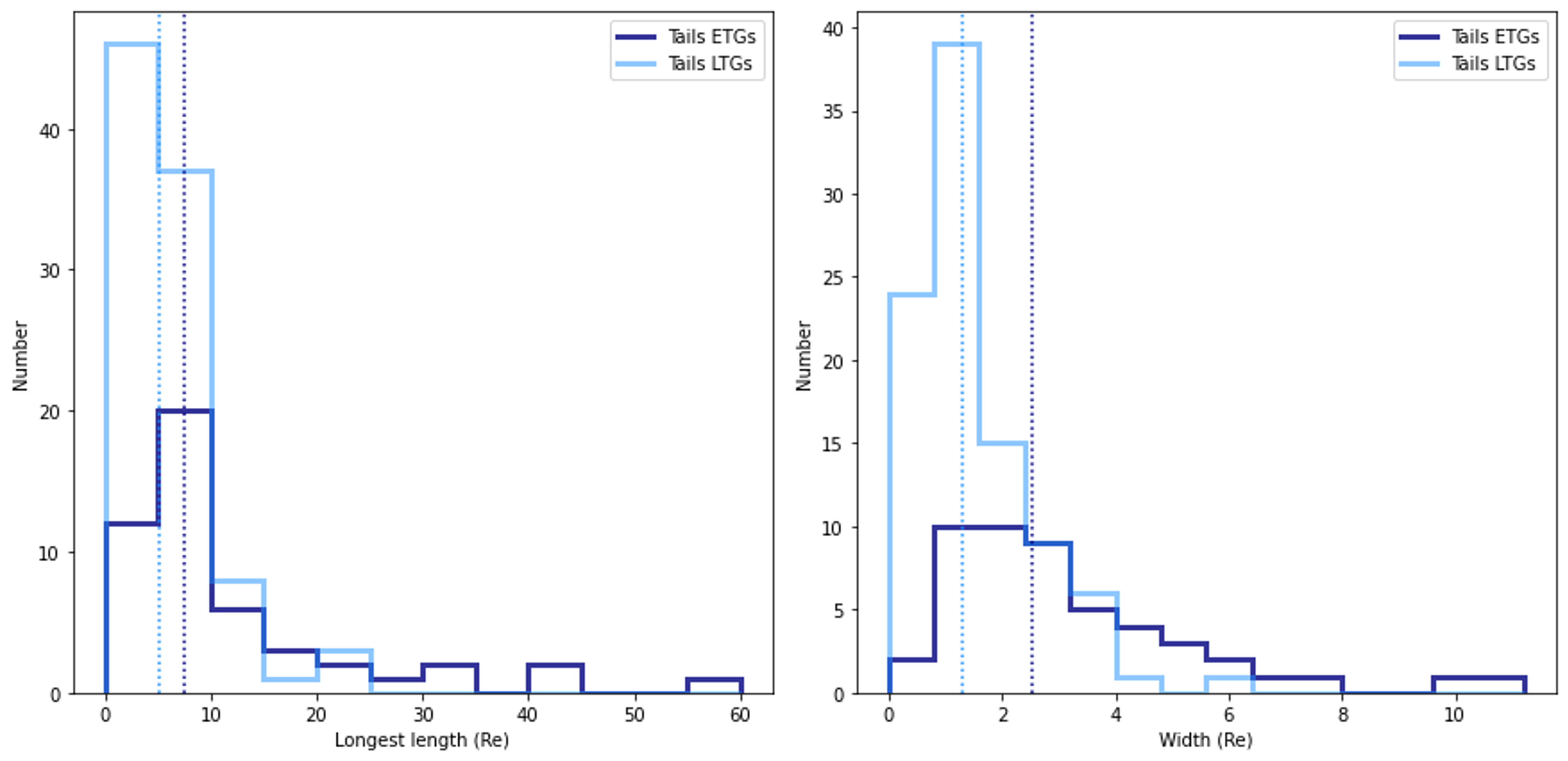}
     \caption{Histogram of the length (\textit{left}) and of the width (\textit{right}) of tidal tails normalized by the effective radius of the host galaxy, as a function of the morphological type. ETGs are represented by darker shades than LTGs. The median of each distribution is represented by the dotted lines. The histogram of the length has bins width of 5 R$_e$ and of 0.8 R$_e$ for the width.}
         \label{fig:tails_width_length_ETG_LTG}
\end{figure*}

\section{Precision on the annotations kept}
\begin{figure*}
   \centering
   \includegraphics[width=\linewidth]{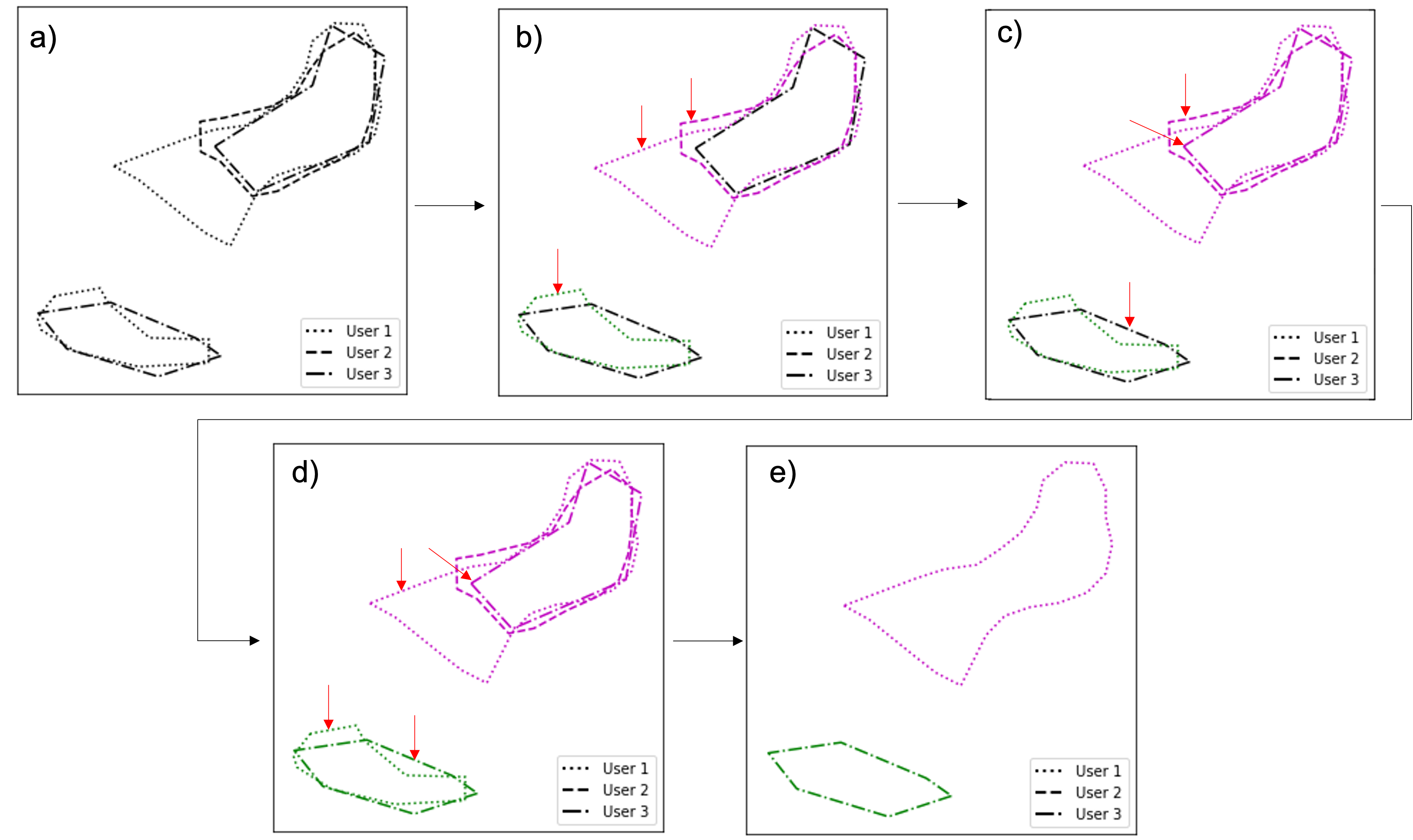}
      \caption{Illustration of the selection process for tidal tails and streams for one galaxy. Black annotations indicate that the unique identifier has not been attributed yet, while each color corresponds to one unique identifier. The linestyles correspond to the different users. The red arrows indicate which annotations are currently compared. \textit{a)} Initialization: all the annotations of tidal tails and streams are considered. \textit{b)} First iteration: the annotations of User 2 and User 1 are compared. \textit{c)} Second iteration: comparison of the annotations of User 3 and User 2. \textit{d)} Third iteration: comparison of the annotations of User 3 and User 1. \textit{e)} Final step: for the annotations sharing the same unique identifier, only the one with the largest area is kept.}
         \label{fig:unique_id}
\end{figure*}

In this appendix, we describe in more detail the selection process for tidal tails and streams seen in section \ref{section:selection_process}. The aim is to keep the most representative features by attributing a unique identifier to the annotations, and then keeping, for annotations sharing the same unique identifier, the one with the largest area.
This iterative process is illustrated in Figure \ref{fig:unique_id} for one galaxy. First, all the tidal tails and streams from all users are considered (panel \textit{a)}. The first iteration starts (panel \textit{b)}: the annotations of User 2 and User 1 are compared. The red arrows outline which annotations are currently compared.  If their intersection score is high enough, they are paired and share the same unique identifier (e.g. here the magenta annotations); otherwise a new unique identifier is given (e.g. here the green annotation). During the second iteration (panel \textit{c)}, the annotations of User 3 and User 2 are compared with the same method. During the third iteration (panel \textit{d)}, the annotations of User 3 and User 1 are compared. At the end of this step, all the annotations have been attributed a unique identifier. Finally, for the annotations sharing the same unique identifier, only the one with the largest area is kept (panel \textit{e)}.

\section{The Next Generation Virgo Cluster Survey}
In addition to the images from CFIS and MATLAS, we have also annotated the LSB structures visible in the deep images from the Next Generation Virgo Cluster Survey (NGVS\footnote{NGVS, \url{http://astrowww.phys.uvic.ca/~lff/NGVS/Home.html}}). This CFHT Large program surveyed 104 squared degrees in the Virgo cluster in four bands, with a depth of 29 mag\,arcsec$^{-2}$ in the $g$-band \citep{Ferrarese_et_al_2012}. 
Unfortunately the $r$-band image is not available for this survey, and no direct comparison could be directly done with  CFIS and MATLAS. We therefore decided not to include the NGVS annotations in our survey but to summarize our analysis in this Appendix. 

\begin{figure}
   \centering
   \includegraphics[width=\linewidth]{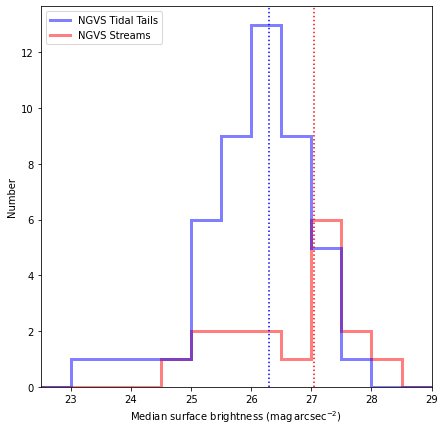}
      \caption{For NGVS only: histogram of the median surface brightness values in magnitudes per square arcsecond for tidal tails (in blue) and streams (in red), in bins of 0.5 mag\,arcsec$^{-2}$. The median of each distribution is represented by the dotted lines.}
         \label{fig:hist_SB_tails_streams_NGVS}
\end{figure}

\begin{figure}
   \centering
   \includegraphics[width=\linewidth]{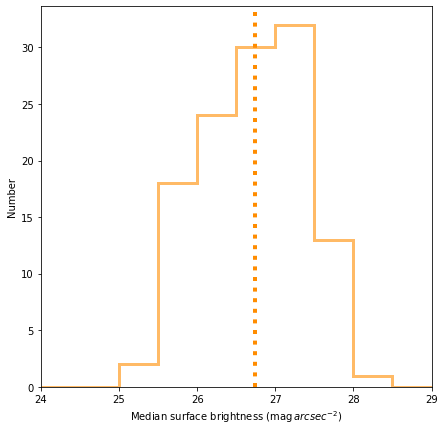}
      \caption{For NGVS only: histogram of the outer median surface brightness values in magnitudes per square arcsecond for halos, in bins of 0.5 mag\,arcsec$^{-2}$. The median is represented by the dotted line.}
         \label{fig:hist_sb_halo_ngvs}
\end{figure}

Two users annotated a total of 2217 features (among which 1898 have been kept after our selection process) around 58 ETGs and 65 LTGs from the Virgo Cluster. Figure \ref{fig:hist_SB_tails_streams_NGVS} displays the median SB of tails and streams, and Figure \ref{fig:hist_sb_halo_ngvs} the median value of the outer contour of the annotated stellar halos. 

\begin{figure}
   \centering
   \includegraphics[width=\linewidth]{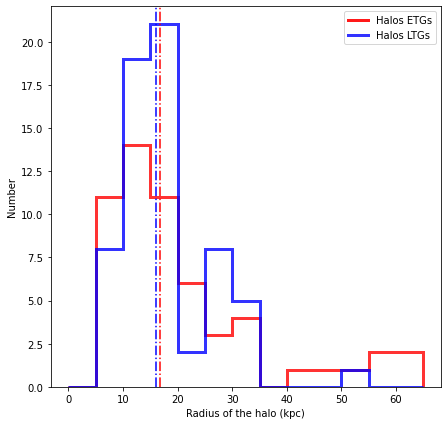}
      \caption{For NGVS only: histogram of the radius of the halos, in bins of 5 kpc, as a function of the morphological type. The median of each distribution is represented by the dotted line.}
         \label{fig:radius_halo_ngvs}
\end{figure}

The maximal outer SB value (28.5 mag\,arcsec$^{-2}$) in the $g$-band is closer to the nominal depth of the survey than the maximal outer SB value in the $r$-band for MATLAS. One possible reason is that  the $g$-band is less sensitive to artefacts such as ghost reflections or high background regions and therefore less contaminated, which enables to detect fainter isophotes than in the $r$-band. Figure \ref{fig:radius_halo_ngvs} displays the radius of the annotated halos for the Virgo LTGs and ETGs. 

\end{appendix}

\end{document}